\documentclass[fleqn,usenatbib]{mnras}
\usepackage[utf8x]{inputenc}
\usepackage[T1]{fontenc}
\usepackage{ae,aecompl}
\usepackage[english]{babel}
\usepackage{lmodern}
\usepackage{enumitem}
\usepackage[squaren, Gray, cdot]{SIunits}
\usepackage{multirow}
\usepackage{color}
\usepackage{array}
\usepackage{setspace}
\usepackage{graphicx}
\usepackage{amsmath}
\usepackage{amssymb}
\usepackage{bm}
\newcommand \red[1] {{#1}}

\title[Convergence in cones]{Game of cones: A nulling strategy for modelling lensing convergence in cones with large deviation theory}

\author[A. Barthelemy et al.]{
A. Barthelemy$^{1}$\thanks{E-mail: alexandre.barthelemy@iap.fr},
S. Codis$^{1}$, C. Uhlemann$^{2,3}$, F. Bernardeau$^{1,4}$, 
R. Gavazzi$^{1}$
\\ 
$^{1}$CNRS \& Sorbonne Universit\'e, UMR 7095, Institut d'Astrophysique de Paris, 75014, Paris, France\\
$^{2}$Centre for Theoretical Cosmology, DAMTP, University of Cambridge, Cambridge CB3 0WA, UK\\
$^{3}$Fitzwilliam College, University of Cambridge, Cambridge CB3 0DG, UK\\
$^{4}$Institut de Physique Th\'eorique, Universit\'e Paris-Saclay,
CEA, CNRS, UMR 3681, 91191 Gif-sur-Yvette, France
}

\date{Accepted 2020 January 6. Received 2020 January 6; in original form July 26.}

\pubyear{2019}

\begin{document}

\label{firstpage}
\pagerange{\pageref{firstpage}--\pageref{lastpage}}
\maketitle

\begin{abstract}
\red{The distribution of the cosmic convergence field is modeled using a large-deviation principle where all non-Gaussian contributions are computed from first principles}. The geometry of the past light cone is accounted for by constructing the total weak-lensing signal from contributions of the matter density in thin disk slices. The prediction of this model is successfully tested against numerical simulation with ray tracing, and found to be accurate within at least 5 per cent in the tails at redshift 1 and opening angle of 10~arcmin and even more so with increasing source redshift and opening angle. An accurate analytical approximation to the theory is also provided for practical implementation. The lensing kernel that mixes physical scales along the line-of-sight tends to reduce the domain of validity of this theoretical approach compared to the three dimensional case of cosmic densities in spherical cells. This effect is shown to be avoidable if a nulling procedure is implemented in order to localise the lensing line-of-sight integrations in a tomographic analysis. Accuracy in the tails is thus achieved within a percent for source redshifts between 0.5 and 1.5 and an opening angle of 10~arcmin. Applications to future weak-lensing surveys like Euclid and the specific issue of shape noise are discussed.
\end{abstract}


\section{Introduction}
Light rays coming from distant sources propagate through the inhomogeneous distribution of baryonic and dark matter and are scattered many times which results in lensed galaxies being (de)magnified in brightness and distorted from their intrinsic shape. This gravitational lensing provides information on the gravitational potential that rays go through. Weak gravitational lensing, where slight distortions of galaxy shapes are used, is a powerful tool for precision cosmology (see for example a review in \cite{kilbinger15}). By essence, the weak nature of this effect renders viable its exploitation only in large area surveys observing many millions to billions of galaxies such as the Dark Energy Survey (DES) \citep{DES}, the Kilo-Degree Survey (KiDS) \citep{kids} or the hyper suprime cam (HSC) \citep{HSC} for the ongoing surveys, and in the near future the Euclid satellite \citep{Euclid} and Large Synoptic Survey Telescope (LSST) \citep{LSST}. This huge quantity of data thus renders mandatory the use of accurate statistical probes able to infer the matter distribution between us and the sources. 
From a theoretical point of view, since weak-lensing by the large-scale structure of the Universe is an integrated effect along the line-of-sight, accurate predictions for the matter distribution at both linear and non-linear scales are required to model weak-lensing observables. 
Already at intermediate scales, the highly non-linear couplings between scales become important, 
not to mention the impact of baryonic physics on small-scales which could significantly affect lensing observables like the power spectrum \citep[and references therein]{Celine,2019JCAP...03..020S} and higher order statistics. As an illustration, recently, \cite{BaryonEffect} suggested for instance a cut at 16~arcmin for a Euclid-like survey in order to safely ignore the physics of baryons.

Many works have focused on the information contained in the power spectrum or equivalently its real-space counterpart the two-point correlation function. Unfortunately this observable contains only complete statistical information for Gaussian random fields, a prescription valid with extremely good accuracy
to describe primordial metric perturbations visible in the cosmic microwave background as shown by \cite{planck}. However, starting from Gaussian initial conditions, the subsequent {\it non-linear} time-evolution of density fluctuations by means of the gravitational instability develops significant non-Gaussianities, in particular for small scales and late times. In this non-linear regime of structure formation, we observe both an increase in power in the power spectrum measurements (relative to linear evolution) and a generation of distinct non-Gaussianities in the late-time density field.
While empirically, the one-point distributions of matter and tracer densities are found to be close to lognormal \citep{ColesJones91,Kayo01,Bel16,Hurtado-Gil17,Repp18}, similarly to the weak-lensing distribution \citep{Taruya02,Hilbert11,Clerkin17}, lognormal models are fundamentally limited in jointly modelling the two \citep{Xavier16}. Even at the level of the one-point distribution, the lognormal model does not capture the detailed behaviours of the field evolution. Several tentative models have been considered to circumvent this issue with skewed lognormal \citep{Colombi94}, generalised normal distributions \citep{Shin17} or double exponential cutoffs \citep{2018MNRAS.481.4588K} for instance.

More physically motivated approaches have also been developed over the course of the last decades. In the quasi-linear regime, perturbation theory can be used for the computation of cumulants and probability distribution functions (hereafter PDF) of the density field via Edgeworth or Gamma expansions, the latter allowing for a strictly positive definite PDF \citep{Gaztanaga00}. 
In order to get the full hierarchy of cumulants and a meaningful PDF, \citep{Bernardeau1992,Valageas,seminalLDT} suggested to use the spherical collapse model in the context of large-deviation theory. In this approach, the matter PDF is predicted with very good accuracy in the mildly non-linear regime \citep{2014PhRvD..90j3519B}. \cite{saddle} showed that an analytical approximation could be used with the same accuracy. \red{\citep{encircling} applied this technique to show, as a proof of principle, how to extract cosmological information from the matter non-Gaussianities, namely by getting constrains on the growth factor through the variance in different redshift bins and thus on the dark energy equation of state, using the full PDF to measure the variance instead of a suboptimal sample variance estimate. This was shown to provide tighter cosmological constraints}.

Recently, non-perturbative effects on the dark matter PDF have been quantified analytically in one dimension \citep{vanderWoude017} and estimated using a 
path-integral approach based on perturbation theory and a renormalisation procedure for small-scale physics \citep{Ivanov19}.
Additionally, one-point PDFs of the thermal and kinetic Sunyaev-Zel'dovich signals have been extracted from simulations \citep{Dolag16} and predicted from a halo-model based approach for the thermal Sunyaev-Zel'dovich effect in maps of the cosmic microwave background \citep{Thiele18}. These can be used to create cross-correlation statistical tools with convergence maps as was done in \cite{Munshi14SZ+WL} where the joint two-point probability distribution function for smoothed thermal Sunyaev-Zel’dovich and convergence maps is expressed in terms of individual one-point PDFs. 

Such PDFs can be used to generate mock catalogues for large surveys \citep{Baratta19} and  might provide valuable cosmological information once applied to the observed galaxy distribution (although the subtle effect of galaxy bias might be an issue \citep{bias}) or weak-lensing.
A Fisher analysis based on fast simulations in \cite{Patton17} demonstrated that the weak-lensing convergence PDF provides information complementary to the cosmic shear two-point correlation. This is in line with reports of an increase in the lensing figure of merit through higher order convergence moments \citep{Vicinanza18} and an improvement by a factor of two when adding moments over shear power spectrum tomography alone \citep{Petri16cos}. The same conclusions were reached by \cite{Petri13} demonstrating the additional information gained on cosmological parameters (in particular the equation of state of dark energy, amplitude of fluctuations and total matter fraction) by including higher order moments compared to an analysis of the sole power spectrum of weak-lensing data. In addition, numerical simulations suggest that the lensing convergence PDF contains signatures of massive neutrinos \citep{Liu19} amongst other things beyond $\Lambda$CDM effects and that including non-Gaussian information is key to break degeneracies in modified gravity models \citep{Peel18}.

Cosmic shear experiments like DES and KiDS are sensitive to the matter distribution itself and can be used to extract the weak-lensing signal around shear peaks \citep{KacprzakDES16}, galaxy troughs and ridges \citep{Gruen16troughs,Brouwer18troughs} or more general density split statistics \citep{FriedrichDES17,GruenDES17}. In particular, \cite{FriedrichDES17} used the cumulant generating function to construct the one-point PDF of galaxy densities in cones and the weak-lensing convergence profile around line-of-sight with given galaxy density. This density-split statistics from counts and lensing in cells can yield cosmological constraints competitive with the two-point function measurements \citep[see figure~10 in][]{GruenDES17}. Potentially, the lensing profile around special density environments like voids, as suggested for example by \cite{voids}, can also be used to test gravity \citep{Baker18}.

The approach chosen in this work is to study the one-point PDF of the lensing convergence PDF from first principles using large deviation theory in continuation of the work done using this theory.
More precisely, we focus on the convergence field, here filtered in top-hat windows in position space, which can be reconstructed from weak-lensing observations \citep{KaiserSquires,aski} and reflects the projected matter density between the observer and the sources. Some works in this direction have already been performed by \cite{paolo} for the aperture mass and focusing on the reduced-shear statistics although without considering the geometrical effects of the cone, nor building the projected density from the underlying 3D density distribution nor comparing the predictions to numerical simulations, which is the topic of the work proposed here. 
Even before that, computation of the convergence PDF was performed by \cite{BernardeauValageas,Munshi00,Valageas1,Valageas2,Munshi14} relying on different hierarchical ansatz.

The outline of the paper is as follows. 
In section \ref{Convergence}, we review the basics of the weak-lensing convergence, introduce relevant statistical quantities and describe how cone-projected densities can be determined from previous 3D computations using large deviation theory. Section \ref{PDF} aims at testing our approach against numerical N-body simulations at redshifts of interest for the next upcoming photometric galaxy surveys. Section \ref{nulling} introduces a nulling procedure to avoid the contribution from poorly understood small scales and considerably improves theoretical predictions. Finally, section \ref{discussion} discusses more realistic setups as well as perspectives for future works.

\section{Statistics of convergence maps} \label{Convergence}
\subsection{Definition of the convergence}
The convergence $\kappa$ can be interpreted as a line-of-sight projection of the matter density distribution between the observer and the source. More quantitatively, under Born-approximation and neglecting lens-lens coupling, it can be written as \citep{kappadef}
\begin{equation}
    \kappa(\bm{\theta}) = \int_0^{R_s} {\rm d}R \, \omega(R,R_s) \, \delta(R,D\bm{\theta}),
    \label{def-convergence}    
\end{equation}
where $R$ is the comoving radial distance -- $R_s$ radial distance of the source -- that depends on the cosmological model, $D$ is the comoving angular distance ($K$ is the constant space curvature -- $K = 0$ in our case but we recall the quantities in a general FLRW framework)
\begin{equation}
    D(R) \equiv\left\{
    \begin{aligned}{\frac{\sin (\sqrt{K} R)}{\sqrt{K}}}  {\text { for } K>0} \\ {R \qquad}  {\text { for } K=0} \\ {\frac{\sinh (\sqrt{-K} R)}{\sqrt{-K}}}  {\text { for } K<0}
    \end{aligned}\right. ,
\end{equation}
and thus the weight function $\omega$ is defined as
\begin{equation}
\label{eq:weight}
    \omega(R,R_s) = \frac{3\,\Omega_m\,H_0^2}{2\,c^2} \, \frac{D(R)\,D(R_s-R)}{D(R_s)}\,(1+z(R)).
\end{equation}
We define the projected density $\delta_{\text{proj}}$ as
\begin{equation}
\label{eq:cone}
    \delta_{\text{proj}}(\bm{\theta}) = \frac{\kappa}{|\kappa_{min}|} = \int_0^{R_s} {\rm d}R \, F(R) \, \delta(R,D\bm{\theta}),
\end{equation}
with $F(R) = \omega(R,R_s)/|\kappa_{min}|$ and where $\kappa_{min}$ is the convergence in an \lq empty\rq \ beam (when $\delta = -1$). Doing so is not mandatory for the formalism itself but the projected density has the advantage of having behaviours similar to its 3D counterpart, notably cumulants of the same order of magnitude and being able to use the value of the variance as a probe of non-linearity: the smaller it is the more the linear regime applies.

Let us include filtering effects on the convergence maps, in our case a 2D top-hat window  and define the smoothed projected density as
\begin{equation}
    \delta_{\text{proj},\theta} = \int {\rm d}^2\bm{\theta}' \, u_{\theta}(\bm{\theta}') \, \delta_{\text{proj}}(\bm{\theta}' - \bm{\theta}).
\end{equation}
This relation is general enough to account for not only the 2D top-hat filter used in this work but any other 2D filter and compute for example the aperture mass.

Note that equation~(\ref{def-convergence}) assumed no lens-lens couplings and no geodesic perturbations (Born approximation). Although they could affect the high-order cumulants we are computing to get to the convergence PDF, the dominant effect in our case would be on the skewness since it is the dominant correction to the Gaussian. In this case, \cite{Bernardeau1997} showed in their equation~(82) with some simplifying assumptions that the effect for a top-hat opening angle of 10~arcmin, with a power-law power spectrum of spectral index between -1 and -1.5, was subtractive and of order unity for the convergence skewness, independently of the source redshift. As will be shown throughout this work and more specifically in figure~\ref{S}, the skewnesses we consider are much larger than unity so that this correction is of the order of a few percents at most, and its impact on the PDF is small enough that when comparing our model to numerical simulations the effect is completely within the error bars. 
However, for CMB lensing this could have a more drastic impact making our predictions too skewed and one would need to take this effect into account not to overpredict the skewness.
Let us mention that fully taking into account those effects in numerical simulations to quantify them as was done by \cite{Born} also showed that the post-Born corrections (lens-lens couplings and geodesic perturbations) have only a few percent effect on the convergence skewness in weak-lensing studies for the scales of interest. 

\subsection{From cumulant generating function to PDF}
Throughout this work, we make use of different statistical quantities that we briefly introduce here for clarity.
From the PDF ${\mathcal P}_X$ of some continuous random variable $X$ (in our case the cosmic matter density or the weak-lensing convergence) one can define the moment generating function as the Laplace transform of the PDF
\begin{equation}
    M_X(y) =\mathrm{E}\left(e^{y X}\right) =  \int_{-\infty}^{+ \infty} e^{y x} {\mathcal P}_X(x) {\rm d}x,
    \label{laplace}
\end{equation}
or equivalently as the expectation value\footnote{Note that we make use throughout this work of the ergodicity hypothesis where one assumes that ensemble averages are equivalent to spatial averages ($E(.)\rightarrow \left\langle.\right\rangle$) over one realisation of a random field at one fixed time. This requires that spatial correlations decay sufficiently rapidly with separation such that one has access to many statistically independent volumes in one realisation.} of the random variable $e^{y X}$. The moment generating function, as its name implies, can be used to find the moments of the distribution as can be seen from the series expansion of the expectation of $e^{y X}$,
\begin{equation}
\begin{aligned}
    M_{X}(y)&\!\!=\!\!\mathrm{E}\left(e^{y X}\right)\!\!=\!\!1\!\!+\!\!y \mathrm{E}(X)\!\!+\!\!\frac{y^{2} \mathrm{E}\left(X^{2}\right)}{2 !}\!\! +\!\!\frac{y^{3} \mathrm{E}\left(X^{3}\right)}{3 !}\!\!+\!\cdots 
    \\
    &= \sum_{n = 0}^{+ \infty} \frac{y^{n} \mathrm{E}\left(X^{n}\right)}{n !},
\end{aligned}  
\end{equation}
so that the $n$-th derivative of the moment generating function in zero is equal to the moment of order $n$, $\mathrm{E}\left(X^{n}\right)$.
The logarithm of the moment generating function is the cumulant generating function (CGF)
\begin{equation}
    \phi_X(y) = \log(M_X(y)) = \sum_{n=1}^{+ \infty} k_{n} \frac{y^{n}}{n !}
    \label{log}
\end{equation}
where $k_n$ are the cumulants (i.e the connected moments) of the distribution. It turns out that the quantities
\begin{equation}
    S_n = \frac{k_n}{k_2^{n-1}},
    \label{Sp}
\end{equation}
called reduced cumulants and where $k_2$ is the variance of the distribution, are of importance in our context \footnote{In cosmology, $S_n$ of the matter density field were indeed shown to be independent of the variance down to quite small scales \citep{Peebles,1995MNRAS.274.1049B}.} and we thus also define the scaled cumulant generating function (SCGF) as
\begin{equation}
    \varphi_X(y) = \lim_{k_2 \rightarrow 0} \sum_{n=0}^{+\infty} S_n \, \frac{y^n}{n!},
    \label{defscgf}
\end{equation}
that we will in our context extend to non-zero values of the variance. Finally, if one is able to compute the CGF, one can then reconstruct the PDF for the random variable $X$ as an inverse Laplace transform (inverting equation~(\ref{laplace})) given by
\begin{equation}
    {\mathcal P}_X(x) = \int_{-i\infty}^{+i\infty} \frac{{\rm d}y}{2\pi i} \, \text{exp}\left(-y x + \phi_X(y)\right).
    \label{eq::laplace}
\end{equation}

\subsection{From 3D densities to projections in cones}
\label{relating}

\begin{figure}
    \centering
    \includegraphics[width=\columnwidth]{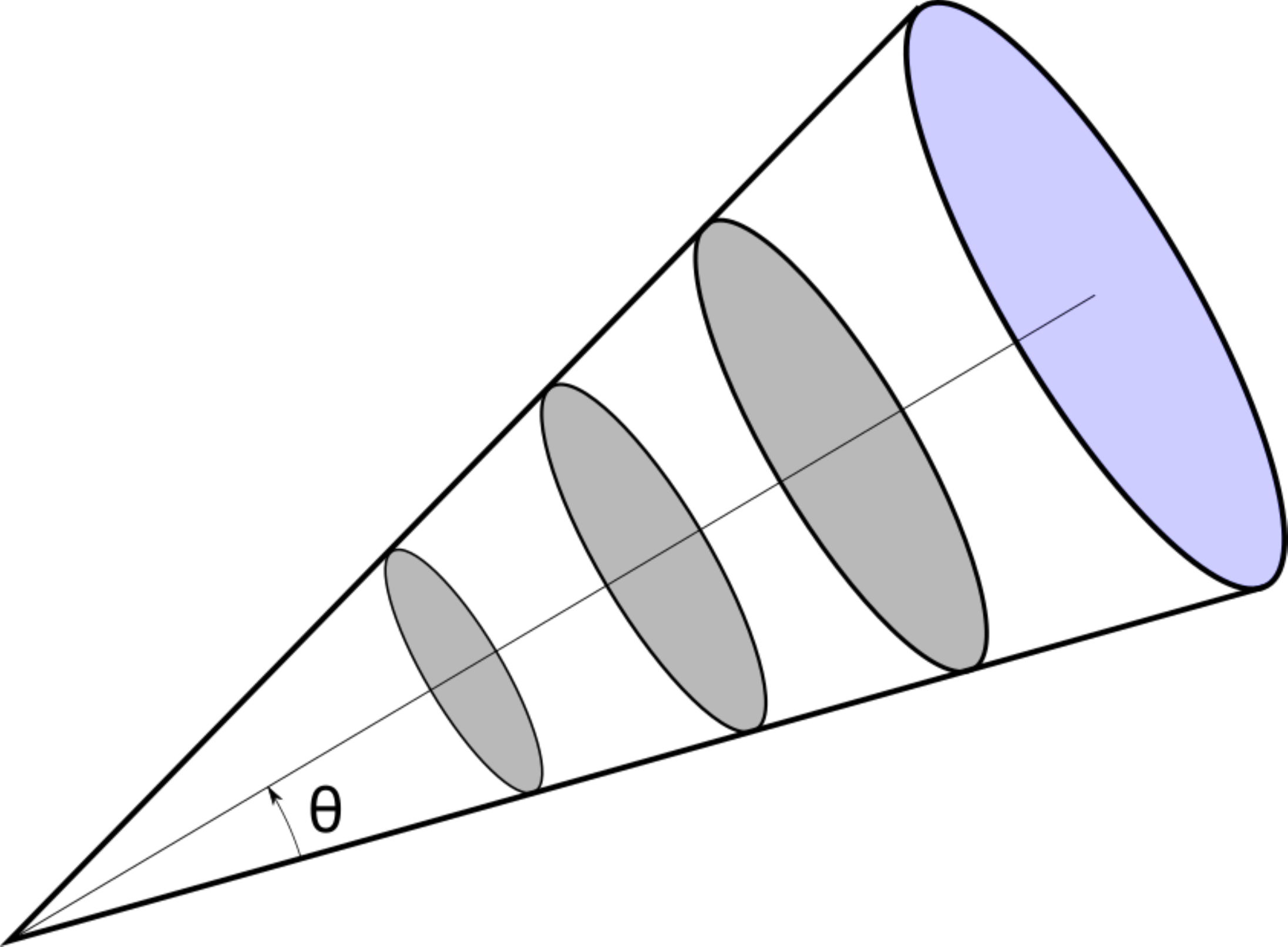}
    \caption{Schematic description of our line-of-sight integral for the convergence. It is seen as a superposition of independent thin slices of 3D matter density in a cone of opening angle $\theta$.}
    \label{dessin}
\end{figure}

From equation~(\ref{def-convergence}), the convergence can be viewed as a superposition of (supposedly) independent layers of the 3D cosmic matter field, as illustrated in figure~\ref{dessin}. 
Thus the computation of the one-point PDF would involve an infinite number of convolution products. Alternatively, one can more conveniently consider the cumulants and thus the cumulant generating function which simply adds the superposed layers as a consequence of equations (\ref{laplace}) and (\ref{log}). This is the path we follow in this paper, the result being explicitly shown in equation~(\ref{cumulant}) below.

Hence, following \cite{BernardeauValageas}, we relate the projected cumulant generating function to the one for the 3D density field\footnote{Note that this derivation is akin to the Limber approximation for the p-point correlation function \citep[see for example][]{Peebles}.}. To do so, let us start by writing explicitly the p-point correlation functions of the projected density field. From equation~(\ref{eq:cone}), denoting $D_i\equiv D(R_i)$ and  using the subscript $c$ for cumulants (i.e the connected part of the moments), we get
\begin{multline}
    \langle \delta_{\text{proj}}(\bm{\theta}_1) \ ... \ \delta_{\text{proj}}(\bm{\theta}_p)\rangle_c = \int_0^{R_s} \prod_{i=1}^{p} {\rm d}R_i \, F(R_i) \\ \times \, \langle\delta(R_1,D_1\bm{\theta}_1) \ ... \ \delta(R_p,D_p\bm{\theta}_p) \rangle_c .
\end{multline}
This quantity can be computed by making the change of variable $R_i = R_1 + r_i$. 
Since the correlation length (beyond which the p-point correlation functions are negligible) is much smaller than the Hubble scale $c/H(z)$, only values of $r_i$ which obey $|r_i| \ll c/H(z)$ contribute to the integral over $r_i$. 
Thus the integral boundaries over $r_i$ can be pushed to infinity or to any value greater or equal to the values for which to correlation is negligible.
Also we get $F(R_1 + r_i) \simeq F(R_1)$ and $D(R_1 + r_i) \simeq D(R_1)$ such that
\begin{multline}
\label{eq:ppoint}
    \langle \delta_{\text{proj}}(\bm{\theta}_1) \ ... \ \delta_{\text{proj}}(\bm{\theta}_p)\rangle_c = \int_0^{R_s}  F(R_1)^p \, {\rm d}R_1 \\ \hspace{-0.2cm} \times \int_{-\infty}^{\infty} \prod_{i=2}^{p} {\rm d}r_i \, \times \, \langle\delta(R_1,D_1\bm{\theta}_1) \ ... \ \delta(R_1 + r_p,D_1\bm{\theta}_p) \rangle_c .
\end{multline}
The p-point correlation function appearing in equation~(\ref{eq:ppoint}) is translation-invariant at constant $z$ and thus depends on $R_1$ only through $D_1$. We again assume that $\delta(R_1 + r_i,D_1\bm{\theta}_i) \simeq \delta(R_1,D_1\bm{\theta}_i)$, take filtering effects into account and are finally led to
\begin{equation}
    \langle \delta_{\text{proj},\theta}^p \rangle_c = \int_0^{R_s} {\rm d}R \, F^p(R) \, \langle \delta_{D\theta,\text{cyl}}^p \rangle_c \, L^{p-1},
\label{cumulant}
\end{equation}
where $\langle \delta_{D\theta,\text{cyl}}^p \rangle_c$ are the cumulants of the 3D density contrast filtered in a cylinder of transverse size $D(R)\theta$ and length $L$\footnote{The length $L$ of the cylinders is only a dimension parameter that will cancel out with terms in the cylindrical variance (equation~(\ref{cylindrevariance})) since we consider arbitrarily long cylinders and are working within the small-angle approximation. Moreover, the cylindrical collapse given in equation~(\ref{collapse}) corresponds to the 2D collapse of a thin slice of the cone. Thus our formalism is the same considering the 3D density in long cylinders or in thin slices which was hinted in figure~\ref{dessin}.}.

In the quasi-linear regime,
high-order cumulants follows -- at least at tree order -- the scaling $\langle \delta^p\rangle_c \propto \langle \delta^2\rangle_c^{p-1}$ so that reduced cumulants are commonly defined as in equation~(\ref{Sp}) and are therefore redshift-independent at tree order \citep{BernardeauReview}. Thus using equation~(\ref{defscgf}) we get the relation between the SCGF of the projected density and the SCGF of the density filtered in cylinders
\begin{equation}
    \varphi_{\text{proj},\theta}(y) = \int_0^{R_s} \frac{{\rm d}R}{\Psi_{\theta}(R)} \, \varphi_{\rm cyl}(F(R)\Psi_{\theta}(R)  y,R),
    \label{phiproj}
\end{equation}
with
\begin{equation}
    \Psi_{\theta}(R) = \frac{\langle\delta^2_{D\theta,\text{cyl}}\rangle}{\langle\delta^2_{\text{proj},\theta}\rangle}  L.
\end{equation}
In practice, we will implement the projected cumulant generating function as shown in equation~(\ref{log}) using 
\begin{equation}
    \phi_{\text{proj},\theta}(y) = \int_0^{R_s} {\rm d}R \, \phi_{\rm cyl}(F(R) y,R).
    \label{CGF}
\end{equation}

\subsection{Cumulants in cylinders from theory} \label{LDT3D}

The relation given by equation~(\ref{phiproj}) is important in the context of the computation of the PDFs for weak-lensing convergence maps. Indeed it relates the SCGF of the projected density to the one used for 3D density in cylinders, where the latter is easily obtained in the context of large deviation theory (LDT) as we will show in this section. 

To start with, let us recall some of the generic results of LDT before applying them to densities in cylinders. For more details, we refer the reader to \cite{seminalLDT} and \cite{cylindres}.
LDT is a mathematical theory \citep{touchette} which allows us to quantify the asymptotic exponential shape of the PDF of a series of random variables when some driving parameter goes to infinity. 
A typical example in this context is the case of the mean value of a dice after a large number $N$ of draws ($N$ is the driving parameter). For us, the random variable will be the matter density and the driving parameter the inverse variance $1/\sigma^2$.

More specifically, a random variable $\rho$ (more precisely its PDF ${\mathcal P}_{\rho}$) satisfies a large deviation principle if the following limit exists
\begin{equation}
    \Psi_{\rho}(\rho) = - \lim_{\sigma^2 \rightarrow 0} \sigma^2 \log({\mathcal P}_{\rho}(\rho))
    \label{LDP}
\end{equation}
and defines the rate function $\Psi_{\rho}$ which characterises the exponential decay of the PDF. 
In general, the existence of a large deviation principle for $\rho$ implies
that the SCGF $ \varphi_{\rho}$ is given through Varadhan's theorem as the Legendre-Fenchel transform of the rate function $\Psi_{\rho}$
\begin{equation}
    \varphi_{\rho}(y) = \sup_{\rho} \,[y\rho - \Psi_{\rho}(\rho)],
    \label{varadhan}
\end{equation}
where the Legendre-Fenchel transform reduces to a simple Legendre transform when $\Psi_{\rho}$ is convex. In that case, 
\begin{equation}
    \varphi_{\rho}(y) =  y\rho - \Psi_{\rho}(\rho),
    \label{Legendre}
\end{equation}
where $\rho$ is a function of $y$ through the following stationary condition
\begin{equation}
    y = \frac{\partial \Psi_{\rho}}{\partial \rho}.
    \label{stationnary}
\end{equation}
Another consequence of the large-deviation principle, which is very useful in the cosmological context, 
is the so-called contraction principle.
This principle states that if we consider a random variable $\tau$ related to $\rho$ through the continuous map $f$ then its rate function can be computed as
\begin{equation}
    \Psi_{\rho}(\rho) = \inf_{\tau:f(\tau) = \rho} \Psi_{\tau}(\tau).
    \label{contraction}
\end{equation}
This formula is called the contraction principle because $f$ can be many-to-one. 
In other words, there might be many $\tau$ such that $\rho = f(\tau)$, 
in which case we are {\it contracting}  information about the rate function of $\tau$ down to $\rho$. 
In physical terms, this formula is interpreted by saying that an improbable fluctuation of $\rho$ is brought about by the most probable of all improbable fluctuations of $\tau$.

Thus the rate function of the late-time density field can be computed from the initial conditions if the most likely mapping between the two is known, 
that is if one is able to identify the leading field configuration that will contribute to this infimum. 
In cylindrically symmetric configurations (with transverse size $D\theta$ and length $L$), one could conjecture \citep{Valageas} that the most likely mapping between initial and final conditions is cylindrical collapse
(similarly to spherical collapse being the most likely dynamics for 3D density fluctuations). 
Then starting from Gaussian initial conditions\footnote{Primordial non-Gaussianities could also straightforwardly be accounted for in this formalism as shown by \cite{NonGaussianities}.} the rate function is 
\begin{equation}
    \Psi_{\rm cyl}(\rho) = \frac{\sigma^2_{l}(D\theta,z)}{2\sigma^2_{l}(r_{ini},z)} \, \tau^2(\rho),
    \label{psicyl}
\end{equation}
where $1/\sigma^2_{l}(D\theta,z)$ is the inverse variance in the cylinder and plays the role of the driving parameter, $r_{ini}$ is related to the dimensions of the cylinder through mass conservation and $\tau$ is the initial linear density. In the spirit of the approximations we have developed until now, we will assume sufficiently long cylinders such that mass conservation is expressed through
\begin{equation}
    r_{ini} = D\theta \cdot \rho^{1/2}
\end{equation}
and the densities in cylinders are expressed through 2D spherical collapse, for which an accurate parametrisation is given by\footnote{\red{This parametrisation was first proposed by \cite{Bernardeau1995}. and can be shown to provide a very accurate approximation to the real spherical collapse dynamics so that the effect on the PDF for the 3D matter density field is much smaller than the difference between the theory as it is and the measurement in simulations (Codis, Uhlemann, Wang in prep.).}}
\begin{equation}
    \zeta(\tau) = \left(1 - \frac{\tau}{\nu} \right)^{-\nu}.
    \label{collapse}
\end{equation}
In the spirit of previous work done involving the density filtered in spherical cells, the parametrisation of $\zeta(\tau)$ in $\nu$ is taken so as to reproduce the value of the tree-order skewness for cylindrical symmetry as computed with perturbation theory. We shall then take $\nu = 1.4$ \citep{cylindres}. See Appendix~\ref{app:theory} for more details.

Finally, the rate function given by equation~(\ref{psicyl}) is also the rate function of any monotonic transformation of $\rho$, such that for the density contrast $\delta = \rho - 1$, we have $\Psi_{\text{cyl},\delta}(\delta) = \Psi_{\text{cyl},\rho}(\rho(\delta))$. 

From equations~(\ref{Legendre}-\ref{psicyl}), we can now define the SCGF, at tree order, of the 3D density contrast in a cylindrical filter of transverse size $D\theta$ and (long) length $L$
\begin{equation}
    \varphi_{\rm cyl}(y) \!=\! \sup_{\delta}  \left[y\delta \!-\! \frac{\sigma^2_{l}(D\theta, L, z)}{2\sigma^2_{l}(D\theta(1+\delta)^{1/2}, L, z)}  \tau^2(1+\delta) \right].
    \label{phicyl}
\end{equation}

\subsection{Parametrisation of the linear variance} 
\label{var}

In the distant observer approximation, let us define the cylindrical variance that is needed to compute the SCGF
\begin{multline}
    \sigma^2_{l}(D\theta, L, z) \!=\! \int_0^{\infty}\!\frac{{\rm d}k_{||}}{2\pi} \!\int \! \frac{{\rm d}^2\bm{k}_{\perp}}{(2\pi)^2} \, P_l(k,z) \\ W_{||}\left(k_{||}L\right)^2 W(D\theta k_{\_})^2,
\end{multline}
where $\bm{k}_{\perp}$ and $k_{||}$ are the components of the wave vector $\bm{k}$ orthogonal and parallel to the line-of-sight, and where $W_{||}$ and $W$ are respectively longitudinal and transverse top-hat-windows. The radial component is of the order of $1/L$ and the transverse $1/D\theta$. Thus when $L$ is large the radial part is negligible which leads to
\begin{equation}
    \sigma^2_{l}(D\theta, L, z) = \frac{1}{L} \,\int \frac{{\rm d}^2\bm{k}_{{\perp}}}{(2\pi)^2} P_l(k_{{\perp}},z) \, W(D\theta k_{\perp})^2,
    \label{cylindrevariance}
\end{equation}
where $W(l) = 2J_1(l)/l$ and $J_1$ is the Bessel function of first order. 
This expression is valid for any linear power spectrum. However, the Laplace transform in equation~(\ref{eq::laplace}) requires to have an analytic expression of the integrand so as to be able to perform an analytic continuation in the complex plane. To do so, one needs to have an analytical approximation of the numerical power spectrum.
For a power-law power spectrum,
\begin{equation}
    P_l(k,z) = P_0(z) \, \left(\frac{k}{k_0}\right)^{n},
\end{equation}
the cylindrical variance at scale $D\theta$ reads
\begin{equation}
    \sigma^2_{l}(D\theta, L, z) = \sigma^2_{l}(D_p\theta, L, z) \left(\frac{D}{D_p}\right)^{-n-2},
\end{equation}
where $D_p\theta$ is some pivot scale. However, this approximation is not accurate enough in our case, especially because the lensing kernel integrates over many different scales. We therefore introduce a more sophisticated parametrisation for the variance that accounts for the running of the spectral index following \cite{2014PhRvD..90j3519B}
\begin{equation}
    \sigma^2_{l}(D\theta, L, z) = \frac{2\,\sigma^2_{l}(D_p\theta, L, z) }{\left(D/D_p\right)^{n_1+2}+\left(D/D_p\right)^{n_2+2}},
    \label{variance}
\end{equation}
where $D_p\theta$ is still a pivot scale, $n_1$ and $n_2$ are parameters chosen to reproduce the correct variance by fitting equation~\ref{cylindrevariance} (in our case $D_p=4.5$~Mpc~$h^{-1}$, $n_1=-0.89$ and $n_2=-1.97$). Note that the dependence on the cosmological parameters is entirely contained in the values of $n_1$ and $n_2$ and the pivot variance. This form was chosen since it ensures the analyticity of the variance mandatory for the continuation to the complex plane that is needed in equation~\ref{laplacePDF} below. It is also a natural extension for the form of the variance in the case of a power-law power spectrum. We checked that this parametrization was very close to its value as computed from the full power-spectrum corresponding to the cosmology considered in the manuscript. The cosmology dependence of the variance was for instance used in \citep{encircling} in order to extract constraint on the dark energy equation-of-state.

\subsection{Cumulants of projected density/convergence} 
\label{CGFproj}

We can now construct the SCGF (and CGF) of the projected density field plugging equation~(\ref{phicyl}) into (\ref{phiproj}) once the variance is modelled via equation~(\ref{variance}).
Up to this point, we focused on the SCGF and modelled it with LDT in the asymptotic $\sigma\rightarrow 0$ regime, which boils down to tree order perturbation theory. 
In the spirit of previous works done with LDT we extrapolate the SCGF to non zero values of the projected variance and take the latter as a free parameter of the theory in order to model the CGF
\begin{equation}
    \begin{aligned} \phi_{\text{proj},\theta,\text{nl}}(y) &=\frac{1}{\left(\sigma_{\mathrm{nl}}^{\mathrm{proj}}(\theta)\right)^{2}}\ \varphi_{\text{proj},\theta}\left(\left(\sigma_{\mathrm{nl}}^{\mathrm{proj}}(\theta)\right)^{2} y\right) \\ &=\!\left(\frac{\sigma_{\mathrm{l}}^{\mathrm{proj}}(\theta)}{\sigma_{\mathrm{nl}}^{\mathrm{proj}}(\theta)}\right)^{2} \phi_{\text{proj},\theta}\left(\left(\frac{\sigma_{\mathrm{nl}}^{\mathrm{proj}}(\theta)}{\sigma_{\mathrm{l}}^{\mathrm{proj}}(\theta)}\right)^{2} y\right)\!. \end{aligned}
    \label{NLPhi}
\end{equation}
This is motivated by the fact that reduced cumulants (i.e the SCGF) are well-described by tree order perturbation theory (the redshift dependence is small in the regime we are describing here) while the variance is not and requires a better modelling accounting for additional non-linear corrections (including tidal effects beyond the spherical collapse) already at play in the mildly non-linear regime.

\subsection{Non-linear PDF}

From the non-linear CGF, the non-linear PDF for the projected density is expressed via an inverse Laplace transform as
\begin{equation}
    {\mathcal P}(\hat{\delta}_{\text{proj}}, \sigma_{\mathrm{nl}}^{\mathrm{proj}})\! =\!\!\! \int_{-i\infty}^{+i\infty}\!\!\! \frac{{\rm d}y}{2\pi i} \, \text{exp}\left(-y\hat{\delta}_{\text{proj}}\! +\! \phi_{\text{proj},\theta,\text{nl}}(y)\right).
    \label{laplacePDF}
\end{equation}

To perform this computation, one has to rely on a numerical integration in the complex plane which can be performed accurately and rather quickly. We implement Newton-Cotes formula of 3$^{\text{rd}}$ order to compute both the integrals along the line-of-sight and along the imaginary axis. For the  integral in the complex plane, the path chosen induces a highly oscillatory behaviour of the imaginary part and thus both the step and the highest $y$ (which should technically go to infinity along the imaginary axis) are chosen so as to ensure convergence of the integral. Methods of steepest descent were also considered and tried but were not as fast as the straightforward approach although maybe more elegant. 

Note that if the obtained projected rate function were convex, the PDF could be accurately predicted using a simple saddle-point approximation and written analytically in terms of the projected rate function as was done in \cite{saddle} for the 3D case. 
At this stage, this is not imperative because the projected CGF (and hence rate function) have to be obtained through a numerical integration in any case.
As it might prove useful for practical applications of our results, we present an approximate analytical formula for the convergence PDF in Appendix~\ref{analytical_shortcut}, given by
\begin{equation}
\label{eq:PDFapprox}
\mathcal P_{\rm proj}^{\rm approx}(\hat\delta_{\rm proj})\!=\!\frac{\tau_{\rm SC}'(1+\hat\delta_{\rm proj})}{\sqrt{2\pi}\sigma_{\rm proj,nl}(\theta)}\exp\!{\left[\!-\!\frac{\tau^2_{\rm SC}(1+\hat\delta_{\rm proj})}{2\sigma_{\rm proj,nl}^2(\theta)}\!\right]}\!,
\end{equation}
where $\tau_{\rm SC}(\rho)$ is defined as $\zeta^{-1}(\rho)$.
The domain of validity of this approximation is thoroughly studied in Appendix~\ref{analytical_shortcut}, together with a proper scheme to assure its normalisation and zero mean. We found that this approximation is accurate at 5 per cent level in the 2-sigma region around the peak for opening angles and redshifts considered here and outperforms the lognormal approximation.

\section{Implementation and validation} \label{PDF}

\subsection{Convergence maps simulations} \label{simu}

\begin{figure}
    \centering
    \includegraphics[width=\columnwidth]{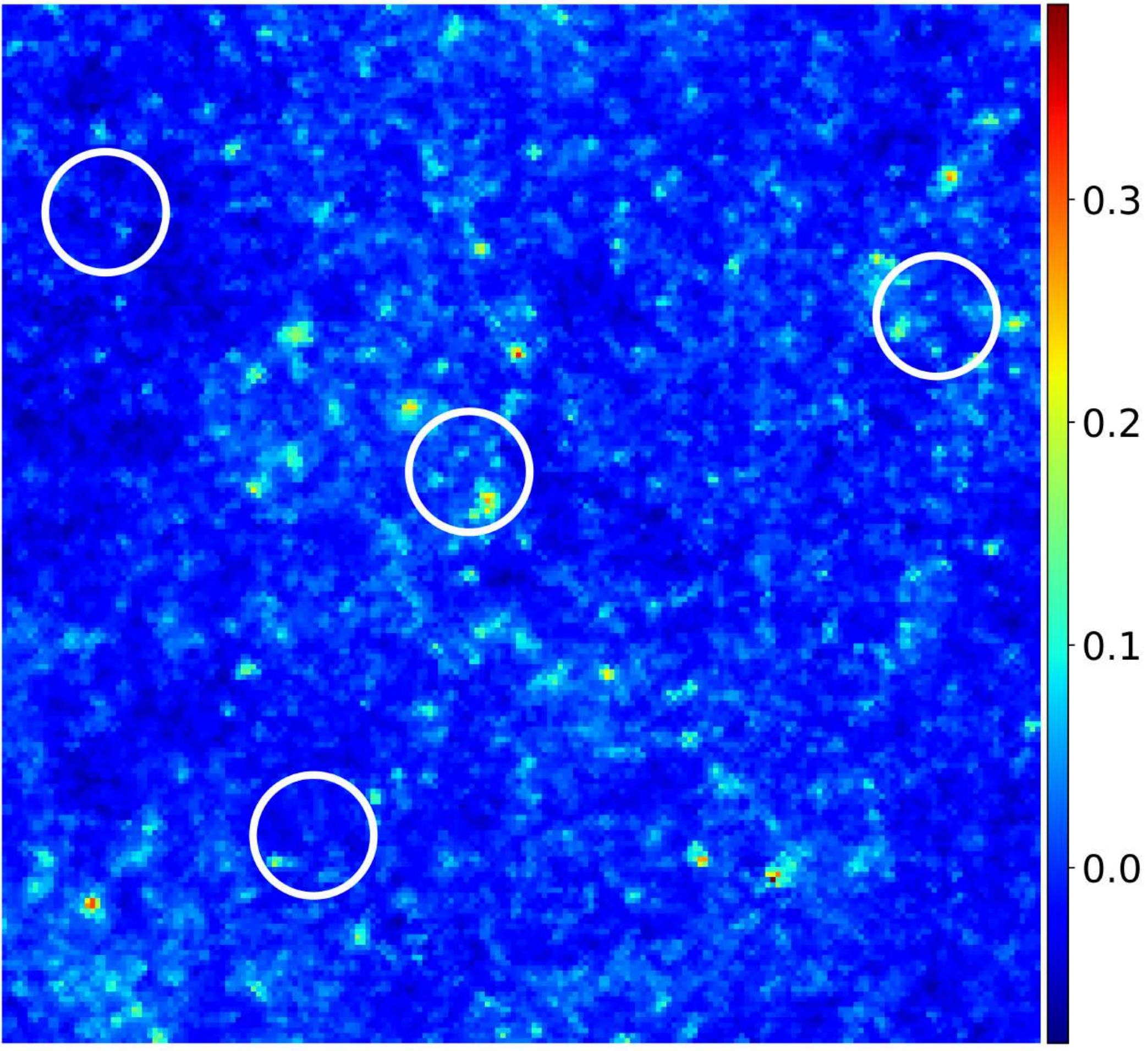}
    \caption{Gnomonic projection of a 172$\times$172~arcmin$^2$ piece of the convergence field at redshift 2. The resolution per pixel is 0.86~arcmin. The white circles show an example of a top-hat smoothing we apply to the maps in this paper (10~arcmin here).}
    \label{gnomview}
\end{figure}

Let us now compare our theoretically-predicted one-point PDFs to the ones taken from numerical simulations.

To do so, 108 full-sky gravitational lensing simulation data sets were generated by \cite{Simulation} performing multiple-lens plane ray-tracing through high-resolution cosmological N-body simulations: a system of nested cubic simulation boxes were prepared to reproduce the mass distribution in the Universe and placed around a fixed vertex representing the observer’s position. They were evolved in a periodic cosmological N-body simulation following the gravitational evolution of dark matter particles without baryonic processes where initial conditions were based on the second-order Lagrangian perturbation theory with the initial linear power spectrum calculated using the Code for Anisotropies in the Microwave Background ({\sc camb}, \cite{camb}). The number of particles for each box was $2048^3$, making the mass and spatial resolutions better for the inner boxes. It was checked that the matter power spectra agreed with theoretical predictions of the revised Halofit and ray-tracing was performed using the public code {\sc graytrix} which follows the standard multiple-lens plane algorithm in spherical coordinates using the {\sc healpix} algorithm. The data sets include full-sky convergence maps from redshifts $z= 0.05$ to 5.3 at intervals of 150~Mpc~$h^{−1}$ comoving radial distance and are freely available for download\footnote{\url{http://cosmo.phys.hirosaki-u.ac.jp/takahasi/allsky_raytracing/}}. The adopted cosmological parameters are consistent with the WMAP-9 year result as shown in Table \ref{table1}. The pixelization of the full-sky maps follows the {\sc healpix} ring scheme with resolution of about 0.86~arcmin.

\begin{table}
    \centering
    \bgroup
    \def\arraystretch{1.5}
    \begin{tabular}{|c|c|c|c|c|c|c|}
   \hline
    $\Omega_m$ & $\Omega_{\Lambda}$ & $\Omega_{cdm}$ & $\Omega_b$ & h & $\sigma_8$ & $n_s$  \\
    \hline
    0.279 & 0.721 & 0.233 & 0.046 & 0.7 & 0.82 & 0.97 \\
    \hline
    \end{tabular}
    \egroup
    \caption{Cosmological parameters used to run the simulations used in this paper.}
    \label{table1}
\end{table}

In this work, \red{ and mostly because of a lack of ressources to process them all,} we pick only one of the 108 realisations. A glimpse of a convergence map at redshift 2 from this realisation is shown in figure~\ref{gnomview}.

To convolve this map with a top-hat window of the desired angular radius (for different source redshifts), we use the \textit{query\_disc} function of {\sc healpy} to find all pixels whose centres are located within a disk centred at one specific pixel $p$. Then we reassign the value of $p$ as being the mean of all the pixels inside the disk. This real-space method was favoured over a Fourier-space method where ringing effects were too large to be satisfactory, especially for the tails of the PDF where the effect has the largest impact. More details on the filtering can be found in Appendix~\ref{top-hat filter}. The error bars on the measured PDFs are estimated via the error on the mean amongst eight subvolumes. \red{ More details on the justification of this procedure can be found in Appendix~\ref{justify}.}

For each source redshift we consider, the non-linear variance is measured in the simulated convergence maps but note that no significant differences were found when computing the non-linear variance from the measured PDF. 
Before focusing on the result for the PDF, let us start by showing the CGF.

\subsection{Validating the cumulant generating function} \label{CGFvalid}

\begin{figure}
    \centering
    \includegraphics[width=\columnwidth]{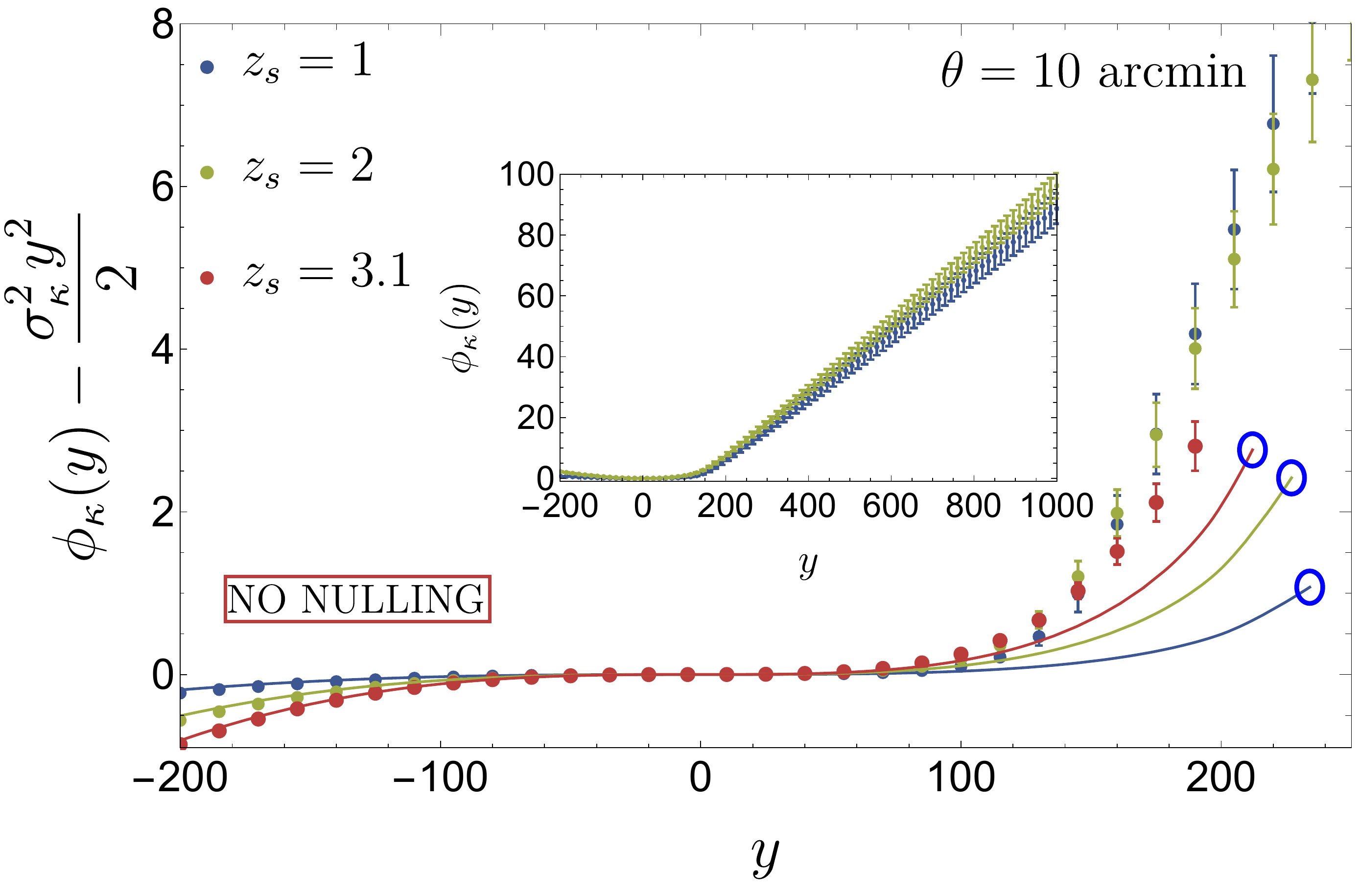}
    \caption{CGF of the convergence for different source redshifts and an opening angle of $\theta = 10$~arcmin. Data points were measured on the simulation described in section \ref{simu}. Discrepancy between theory and observations is discussed in detail in section \ref{compare}. Blue circles indicate the critical points along the real-axis. The small panel inside shows the linear asymptotic behaviour of the measured CGF.}
    \label{CGFplot}
\end{figure}

The stationary condition (link between $y$ and $\delta_{\textrm proj}$ given in equation~(\ref{stationnary}) at each redshift slice) makes the (S)CGF only dependent on its argument $y$, the source redshift $z_s$ and cosmology 
by means of the normalised lensing kernel $F$, the angular distance $D$, the amplitude of the linear power spectrum $P_0$ and the two parameters used to approximate its shape $n_1$ and $n_2$. Note that the (S)CGF is an observable on its own as shown in equation~(\ref{laplace}).

figure~\ref{CGFplot} displays the CGF for three source redshifts and one opening angle. Note that we subtract on this plot the Gaussian (quadratic) contribution to the CGF as the mean is imposed by definition to be zero and the variance is chosen as a free parameter to match the data. Hence, we are only comparing contributions coming from the skewness, kurtosis and higher orders.
Similarly to the 3D case, this approach leads to the emergence of a critical value (blue circles on figure~\ref{CGFplot}) along the real axis for the projected SCGF/CGF which arises directly from the Legendre transform of the projected rate function. In principle, those critical points should also be visible on the simulated data, they manifest as a drastic rise of the error bars to a point where the measured signal does not make sense anymore, as was shown in the 3D case in \cite{2014PhRvD..90j3519B}. This is because above the critical value, ensemble averages of $\exp(y\delta_{\rm proj})$ formally diverge. In practice they are finite in a finite sample but dominated by the rarest events available. More precisely the maximum value of $\kappa$ available in the sample, $\kappa_{\rm max}$, gives the CGF a linear asymptotic behaviour of slope $\kappa_{\rm max}$ and a linear asymptotic for its r.m.s value of slope given by the r.m.s. value of $\kappa_{\rm max}$ in subsamples.

Overall,
we observe a rather good agreement between theory and simulated data close to zero, however the agreement degrades towards the tails when the error on the skewness starts to be visible together with the addition  of higher order contributions.
This is due to
the lensing kernel that mixes all scales and in particular the
small ones at the tip of the cone that cannot be well modelled
by tree-order perturbation theory. This is discussed in more detail in the next section when we compare theory and simulation in terms of PDFs. A way to circumvent these issues by means of a nulling procedure will be presented in section~\ref{nulling}.

\subsection{Weak-lensing convergence PDFs} \label{compare}

\begin{figure}
    \centering
    \includegraphics[width=\columnwidth]{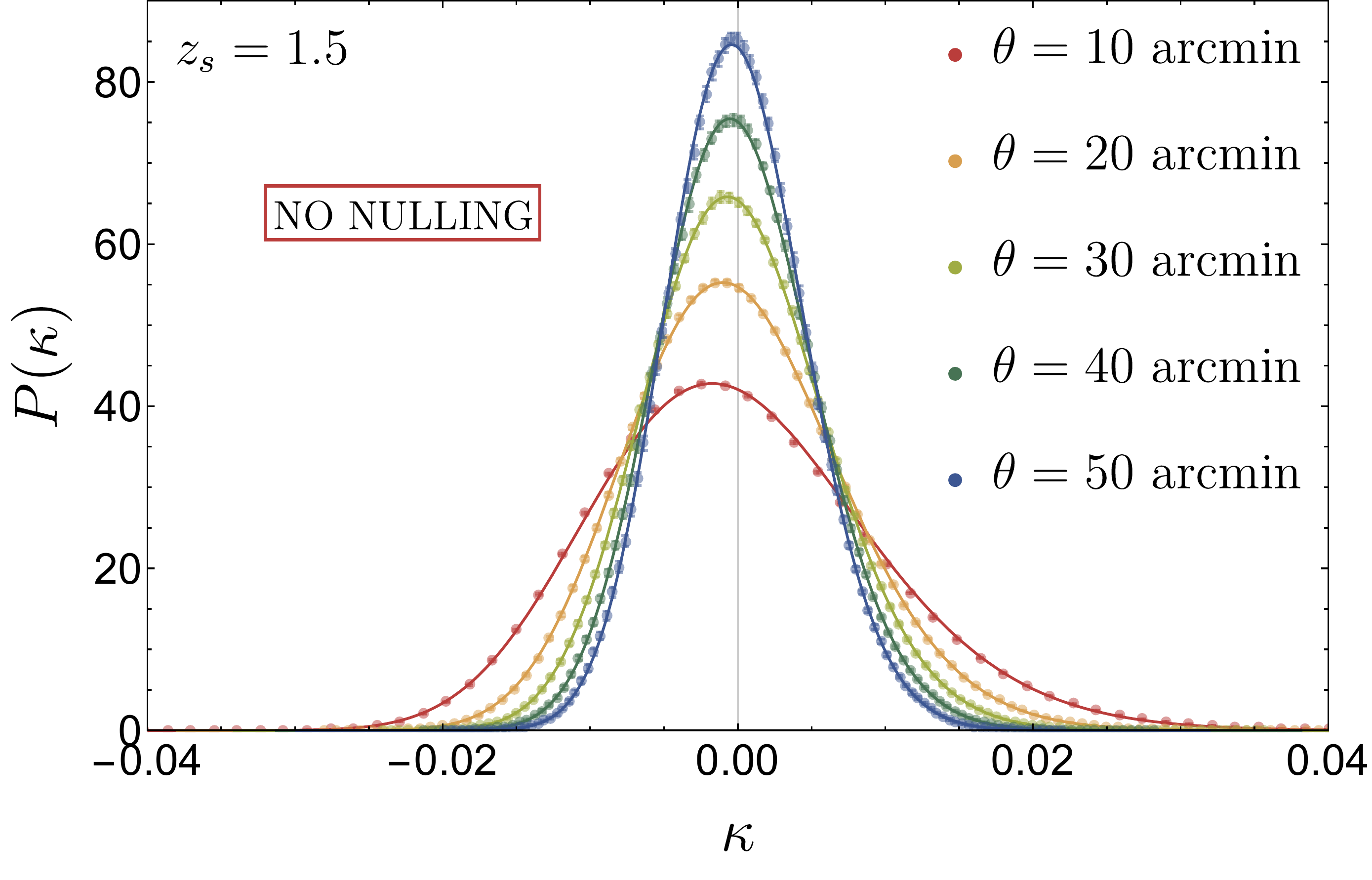}\\
    \includegraphics[width=\columnwidth]{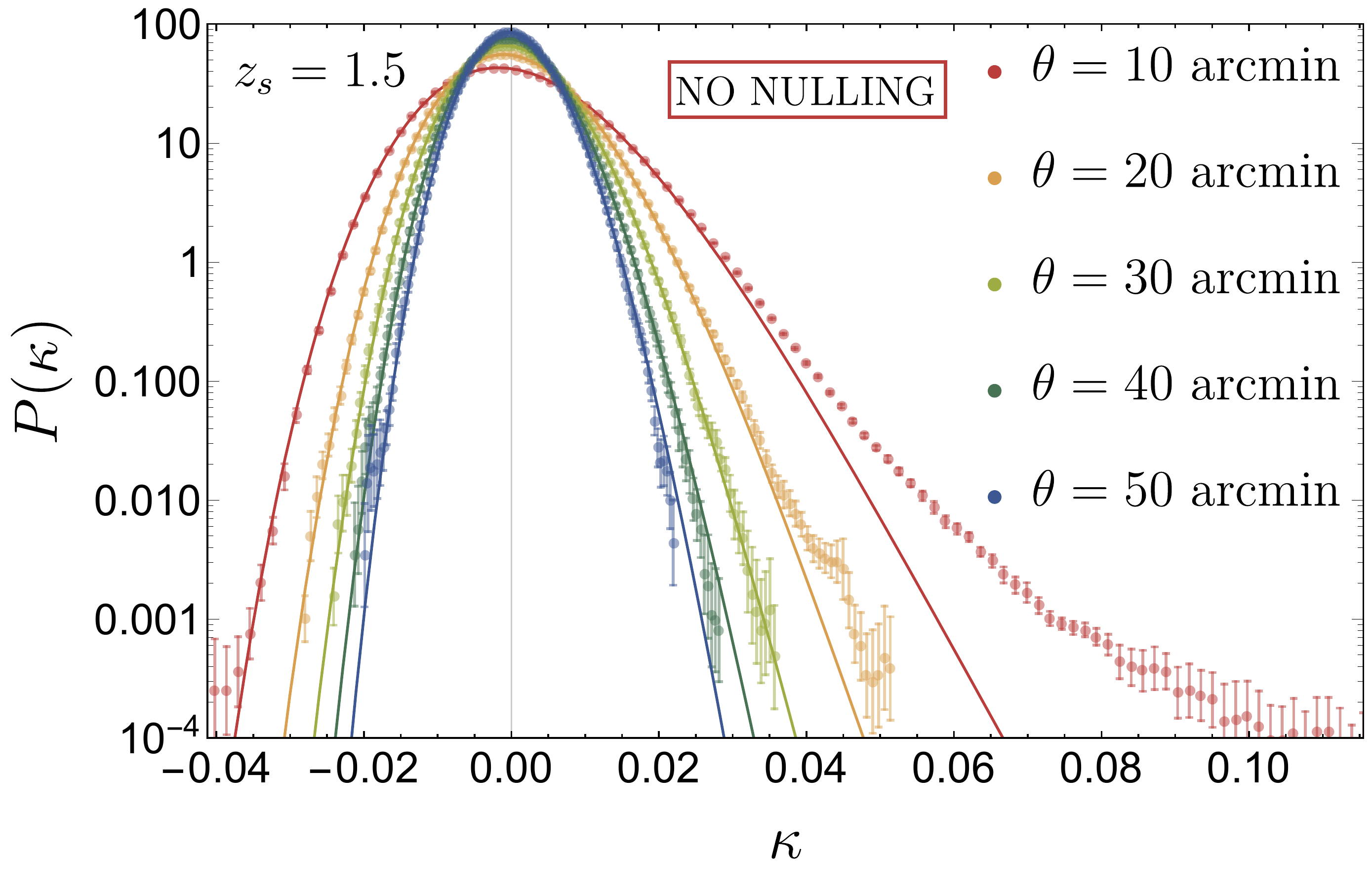}\\
    \includegraphics[width=\columnwidth]{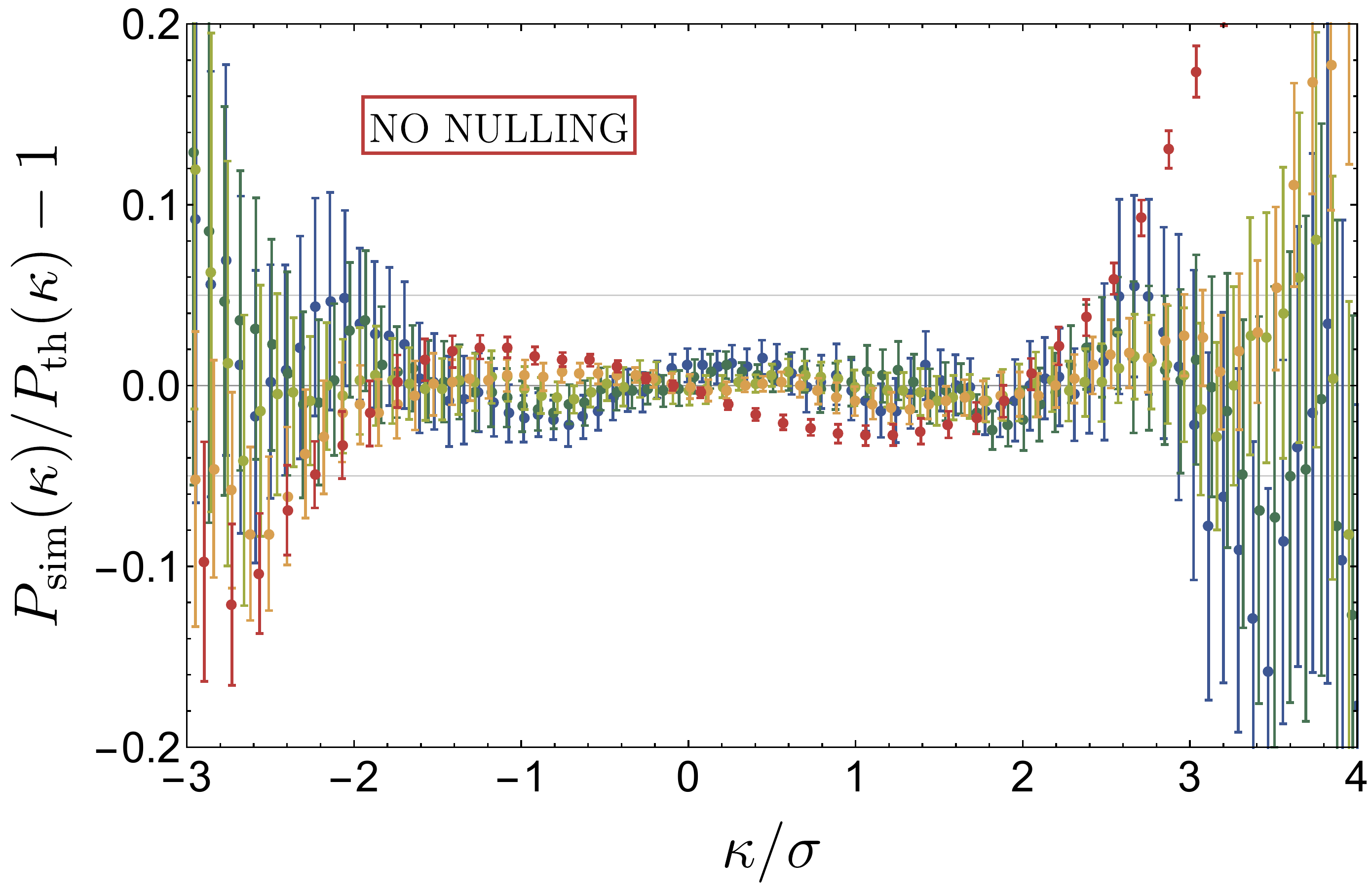}
    \caption{One point PDF of the weak-lensing convergence for different opening angles from 10~arcmin (red) to 50~arcmin (blue) as labelled. The source redshift is fixed here to $z_s$ = 1.5. Solid lines display the LDT predictions given by equation~(\ref{laplacePDF}) while the measurements on the simulated sky is shown with error bars. Top panel: PDF in linear scale. Middle panel: Same as top panel in log scale to better display the tails. Bottom panel: residuals of the simulated data compared to the prediction.}
    \label{angles_pdf}
\end{figure}

\begin{figure}
    \centering
    \includegraphics[width=\columnwidth]{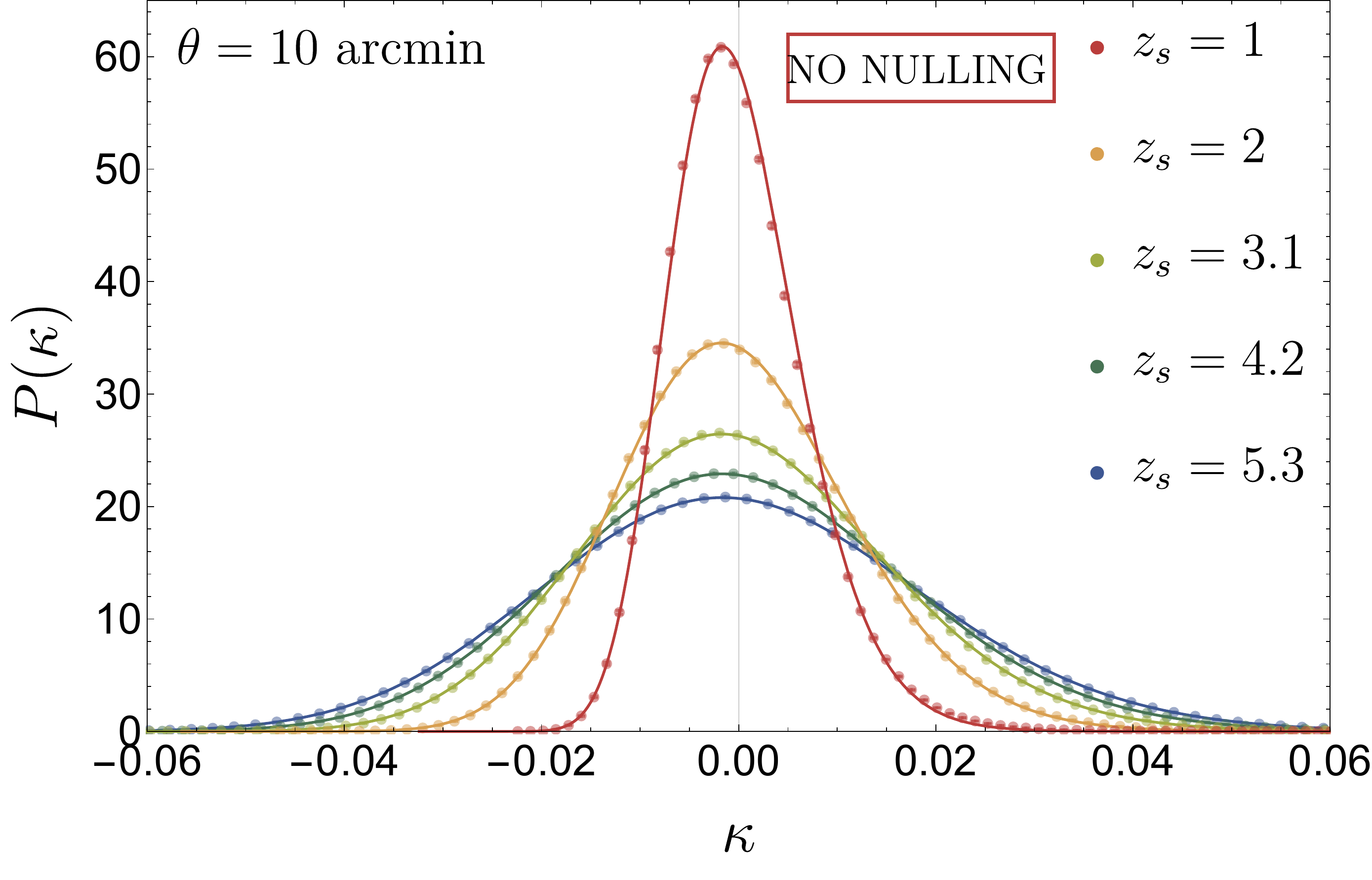}\\
    \includegraphics[width=\columnwidth]{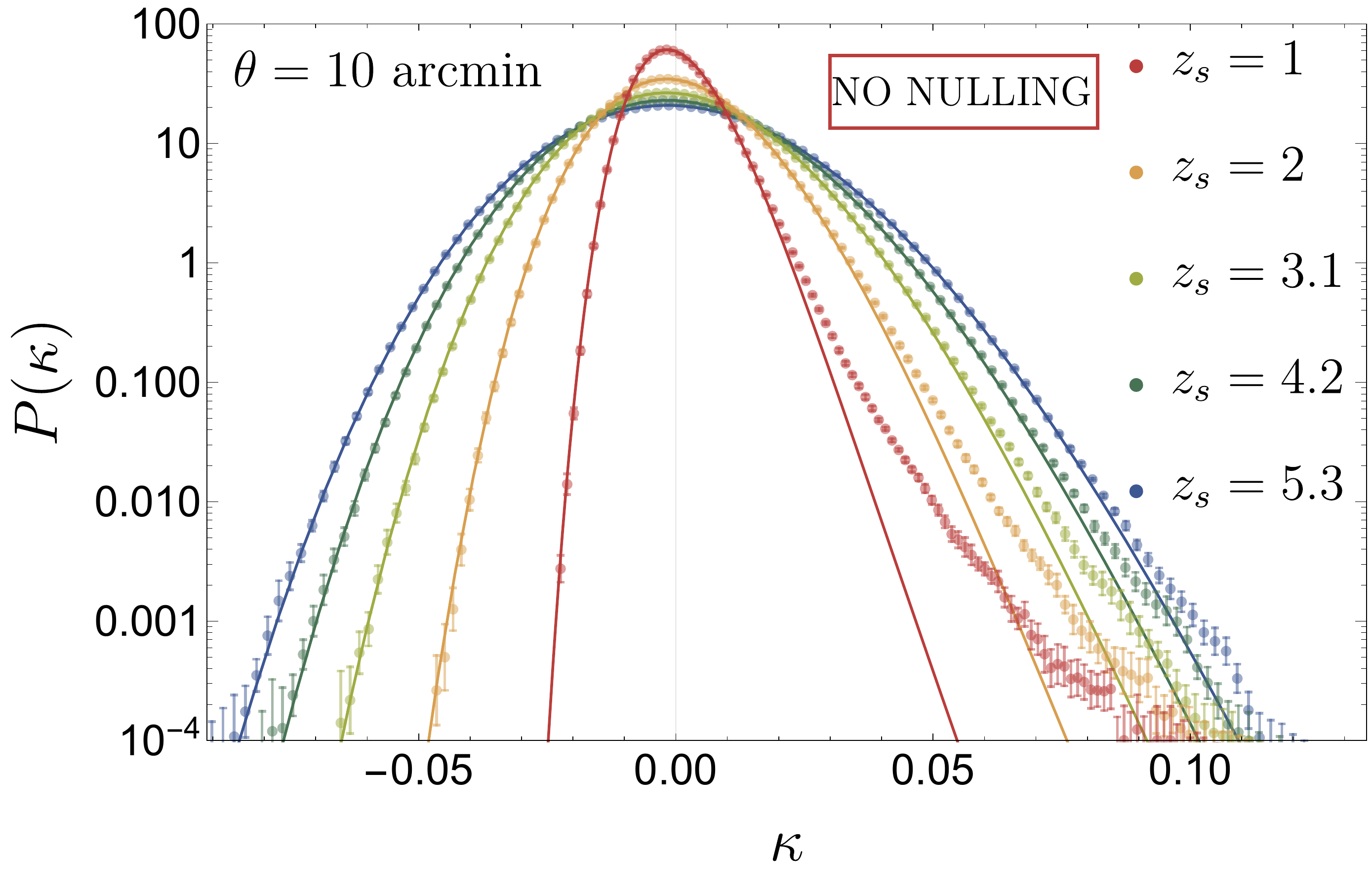}\\
    \includegraphics[width=\columnwidth]{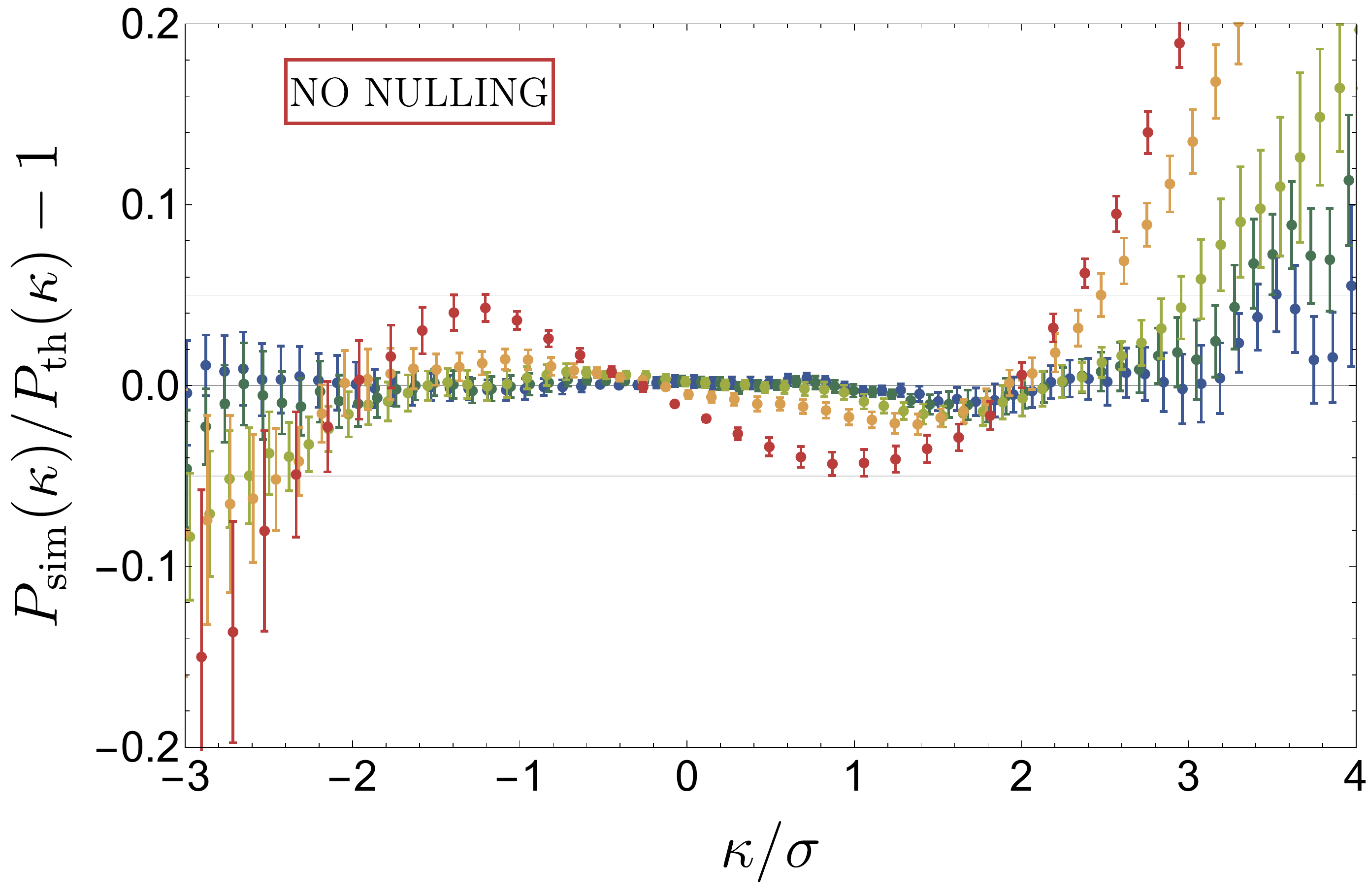}
    \caption{One point PDF of the weak-lensing convergence for different redshifts from 1 (red) to 5.3 (blue) as labelled. The opening angle is fixed here to $\theta$ = 10~arcmin. Solid lines display the LDT predictions given by equation~(\ref{laplacePDF}) while the measurements on the simulated sky are shown with error bars. Top panel: PDF in linear scale. Middle panel: Same as top panel in log scale to better display the tails. Bottom panel: residuals of the simulated data compared to the prediction.}
    \label{redshift_pdf}
\end{figure}

We now show in figure~\ref{angles_pdf}, for a fixed redshift of the source plane of $z_s = 1.5$ and for different opening angles from $\theta = 10$ to 50~arcmin, our theoretical PDFs compared to the ones measured in the simulation.
The upper panel shows the PDFs with a linear scale to emphasize the behaviour around the maximum of the PDF while the middle panel displays the PDFs with a log-scale to highlight the exponential decay in the tails of the distributions (ie for large deviations from the mean convergence). In addition, the lower panel shows the relative difference between the theoretical and measured PDF in the 3 $\sigma$-region around the peak.
Alternatively, figure~\ref{redshift_pdf} shows the same comparison when fixing the opening angle and varying the redshift of the source plane. \red{We also give in Tables~\ref{table2}-\ref{table3} the result of the different fitted variances.}

\begin{table}
    \centering
    \bgroup
    \def\arraystretch{1.5}
    \begin{tabular}{|c|c|c|c|c|c|}
    \hline
       $\theta$ (arcmin) & 10 & 20 & 30 & 40 & 50 \\
    \hline
       $\sigma^2_{\kappa} \ (10^{-5})$  & 9.1 & 5.4 & 3.7 & 2.8 & 2.3  \\
    \hline
    \end{tabular}
    \egroup
    \caption{\red{Variance of the convergence field fitted from the simulated data for various opening angles as labelled and associated with figure~\ref{angles_pdf}. The source redshift is $z_s = 1.5$.}}
    \label{table2}
\end{table}

\begin{table}
    \centering
    \bgroup
    \def\arraystretch{1.5}
    \begin{tabular}{|c|c|c|c|c|c|}
    \hline
       $z_s$ & 1 & 2 & 3.1 & 4.2 & 5.3 \\
    \hline
       $\sigma^2_{\kappa} \ (10^{-4})$  & 0.47 & 1.4 & 2.3 & 3.1 &  3.7 \\
    \hline
    \end{tabular}
    \egroup
    \caption{\red{Variance of the convergence field fitted from the simulated data for various source redshifts as labelled and associated with figure~\ref{redshift_pdf}. The opening angle is $\theta = 10$ arcmin.}}
    \label{table3}
\end{table}

In both cases, we observe that, as expected, the theoretical PDF becomes more and more accurate as one approaches the linear regime (higher and higher redshifts for a fixed opening angle or larger scales at fixed source redshift). 
Still, in this regime (towards the blue curves) the LDT prediction provides us with a better description than the linear Gaussian case, especially in the tails where the departure from a pure Gaussian is clearly seen and well reproduced by LDT. 
When diving into a more non-linear regime (towards the red curves), the distribution clearly gets more skewed towards low convergences, the LDT prediction captures relatively well this non-linear evolution and remains withing 5 per cent from the measured distribution in the 2-$\sigma$ region around the mean convergence. 
Further away in the (rare event) tails, the agreement between our prediction and the simulations gets worse as expected since tree order cumulants are not accurate enough. This is clearly seen on the residuals where a typical $H_3$ modulation by the skewness is visible, showing that higher order correction to the skewness becomes necessary. 
Indeed, let us remind here that the skewness enters the Edgeworth expansion of the PDF at the first non-Gaussian correction order and multiplies a third order Hermite polynomial of the convergence field as follows
\begin{equation}
{\mathcal P}(\kappa)={\cal G}(\kappa)\left[1+\sigma \frac{S_{3,\kappa}}{3!} H_3\left(\frac{\kappa}{\sigma}\right) +{\cal O}(\sigma^2)\right],
\end{equation}
where $H_3(x)=x^3-x$.
Let us emphasize here that the prediction used in this paper and based on LDT does not use or assume an Edgeworth expansion.
Interestingly though, one can show that our approach is equivalent to having an infinite Edgeworth series \citep{BernardeauKofman} -- that is to say with no truncation -- but with reduced cumulants given by spherical collapse which boils down to tree-order in perturbation theory (for more details, the reader is referred to Appendix~\ref{app:theory}).
This infinite series is the main advantage of the LDT based formalism as it allows us to get accurate (and physical) predictions for the tails of the PDF. 
This is to be contrasted with a truncation at a given order in the Edgeworth expansion which by construction -- if it captures correctly the vicinity of the maximum -- would get very inaccurate and nonphysical in the tails (the truncated PDF becoming negative for some values of the convergence and not normalised). 

\begin{figure}
    \centering
    \includegraphics[width=\columnwidth]{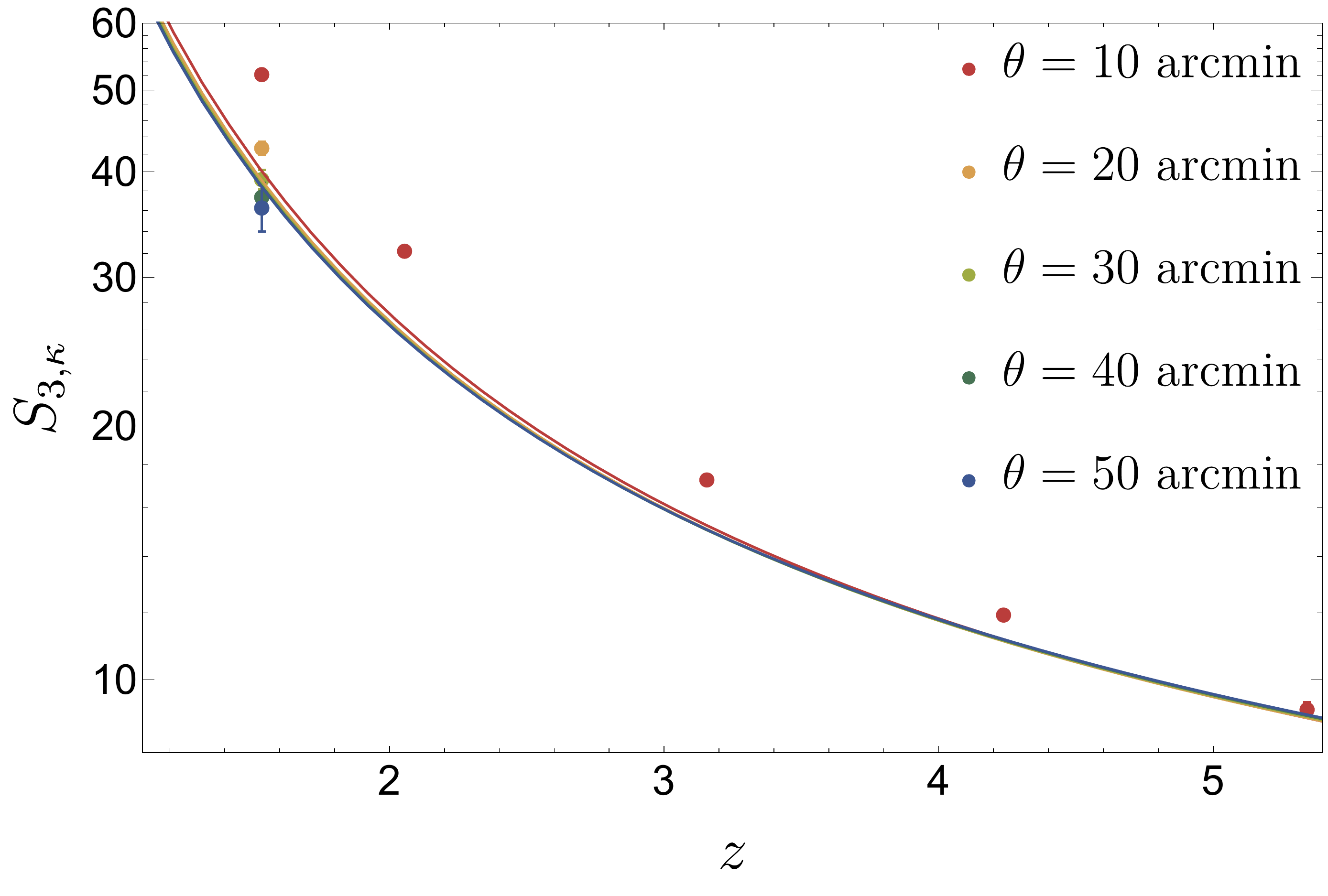}\\
    \includegraphics[width=\columnwidth]{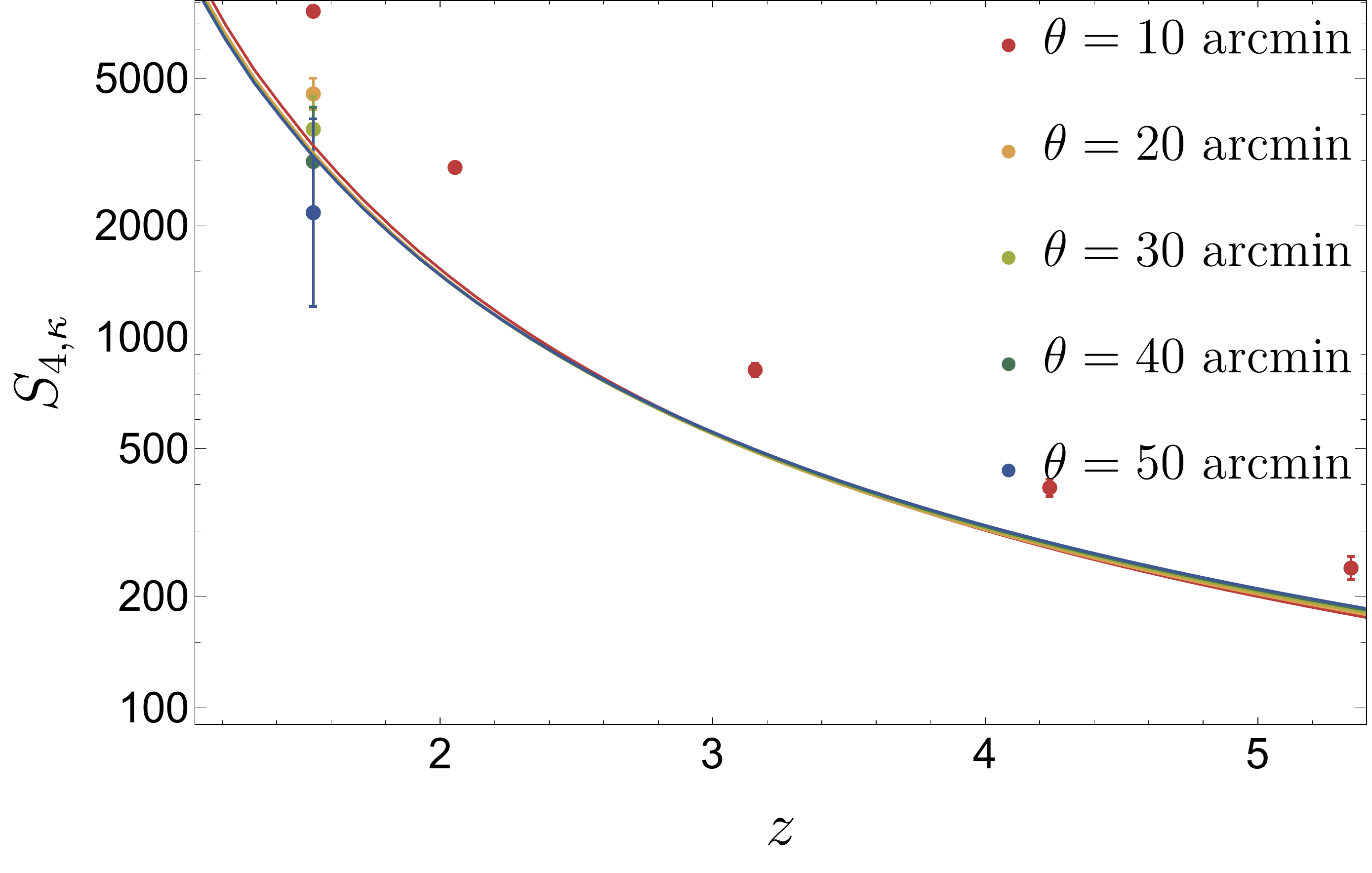}
    \caption{High order cumulants of the weak-lensing convergence as a function of redshift and for different opening angles from 10 (blue) to 50~arcmin (red). The solid lines display the LDT predictions given by equation~(\ref{NLPhi}) while the measurements on the simulated sky are shown with error bars. Top panel: Comparison of the skewness $S_3$ as a function of opening angle and redshift. Bottom panel: Same thing but with the kurtosis $S_4$.}
    \label{S}
\end{figure}

A comparison of the reduced third and fourth order cumulants (skewness and kurtosis) with their tree-order predictions is shown in figure~\ref{S} for various redshifts and opening angles. 
A very good agreement is found in the weakly non-linear regime when the source redshift or the opening angle is large. As one goes towards a more non-linear regime, a clear departure is observed, the prediction systematically underestimating the measured skewness and kurtosis. 

Overall, it is found that -- unsurprisingly -- theoretical predictions of the cumulants and PDF are valid in a somewhat narrower regime than for the 3D densities: while the region around the maximum is quite well captured, the high density tails is heavily affected. 
This is due to the lensing kernel that mixes all scales and in particular the small ones at the tip of the cone that cannot be well modelled by tree-order perturbation theory. 
One way to circumvent this issue is presented in section~\ref{nulling} below by means of nulling.

\section{Multi-source planes and nulling} \label{nulling}

As was shown when naively applying the LDT formalism to weak-lensing statistics, projection effects tend to mix large and small scales which undoubtedly degrades the quality of the theoretical predictions. 
However, there exists a method to make the contribution of lenses null in a range of redshifts thus avoiding scale mixing and allowing for a better theoretical description of lensing observables even when little is known about the details of the small-scale physics (including non-linearities, baryons, etc).
The first implementation of this nulling technique for cosmic shear maps was done in \cite{Nulling} \red{ where it was used in the context of modelling the angular convergence power spectrum from perturbation theory,} and was recently used in \cite{k-cut} to make an explicit link between the angular scale $l$ and the structure scale $k$, thus removing the influence of small scale in the matter power spectrum.
We will first recall the ideas behind nulling before applying it to the convergence PDF.

\subsection{Nulling strategy}

\begin{figure}
    \centering
    \includegraphics[width=\columnwidth]{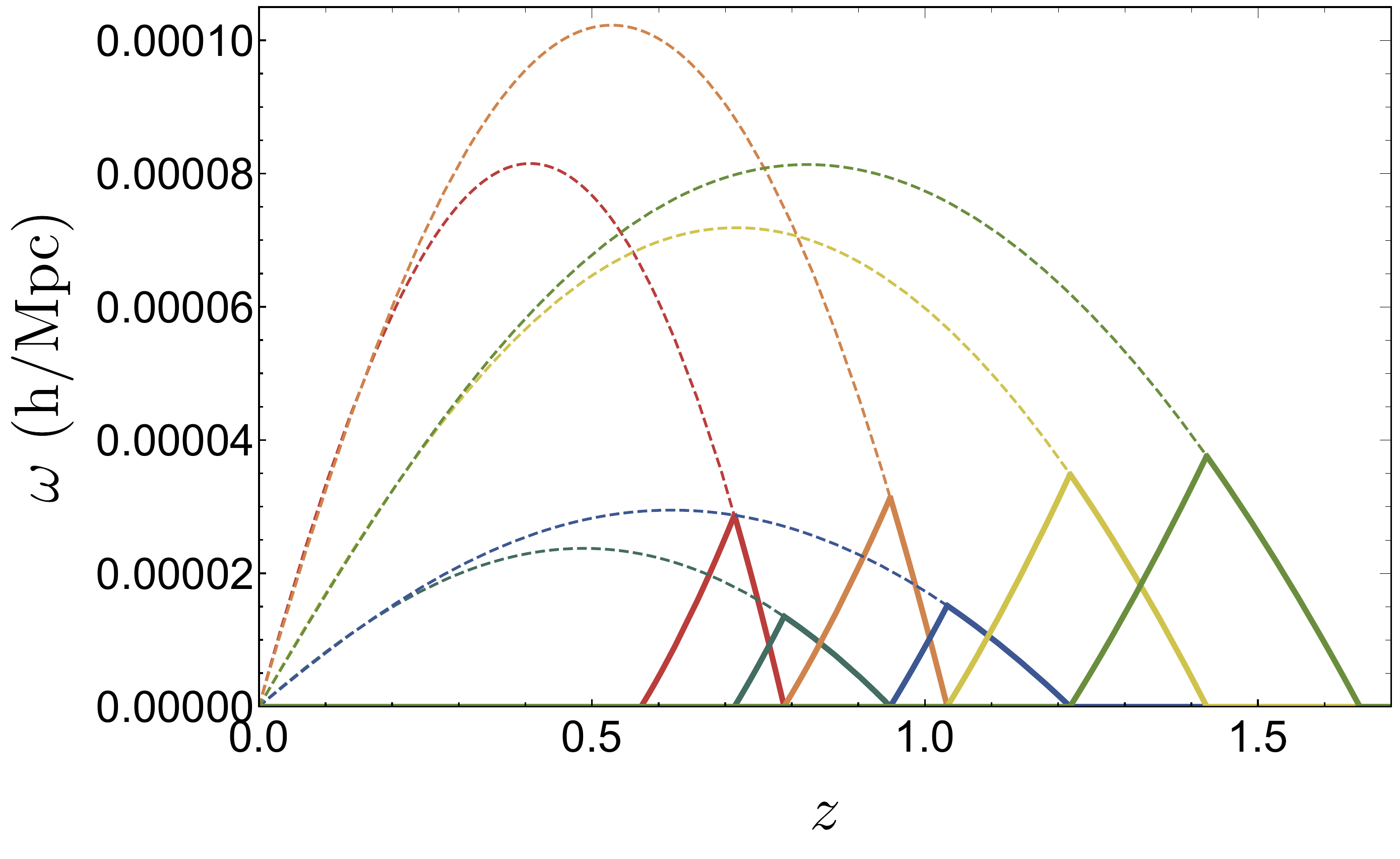}
    \caption{Lens distribution (lensing kernel) for the 6 nulled convergence maps we constructed with weights as shown in equations~(\ref{weights}) and (\ref{matrix}). Dashed lines are the kernel that would be used with no nulling applied on the 3$^{\text{rd}}$ planes of each subset multiplied by the appropriate $p_3$. We roughly follow the redshift binning of the Euclid photometric survey.}
    \label{kernel}
\end{figure}

The principle of nulling is to combine linearly successive convergence maps so as to define new observables that still correspond to weighted line-of-sight integration of the 3D density but we adjust the coefficients in front of each map in order to localise their effective lensing kernel to small redshift ranges (which do not overlap as much as possible). 
This procedure will in particular allow us to avoid having many physical scales contributing to one fixed angular scale as is the case with weak-lensing observables when such a nulling strategy is not implemented.

Starting from several $\kappa$-maps at redshifts $z_i$, the nulled convergence is defined by 
\begin{equation}
    \kappa(\bm{\theta})_{\rm null} = \sum_i p_i \kappa_i,
\end{equation}
where $p_i$ are dimensionless weights whose values will later be chosen so as to reach the desired effect and the lensing kernel $\omega_i$ associated to $\kappa_i$ were given in equation~(\ref{eq:weight}). 
Let us now define the nulled lensing kernel so that the nulled convergence matches the usual convergence definition given in equation~(\ref{def-convergence}),
\begin{equation}
    \omega(R)_{\rm null} \!=\!\!\! \sum_{i,R_{s,i}>R}\!\! p_i \frac{3\,\Omega_m\,H_0^2}{2\,c^2} \, \frac{D(R)\,D(R_{s,i}-R)}{D(R_{s,i})}\,(1+z).
\end{equation}
The game now amounts to finding a set of $p_i$ so that the built nulled convergence map is only sensitive to lenses confined in a certain range of distances for which an exquisite knowledge of the small-scale physics is not necessary and our theoretical prediction are accurate.

Fortunately, such solutions for a set of discrete planes exist. They are unique for sets of 3 source planes up to a normalisation constant. The uniqueness of the solution is not an issue, the general solution could easily be obtained by taking the linear combinations of the three-plane solution for different sets of planes. 
Let us now assume three source planes at distances $R_{s,i}$, $i = 1,2,3$. 
The expression of $\omega(R)_{\rm null}$ can be re-written using trigonometric identity of $\sin$ and $\sinh$ function as
\begin{equation}
    \omega(R)_{\rm null} \!=\! D^{2}(R)\left[\frac{1}{g(R)}\! \sum_{i, R_{s,i}>R}\! p_{i}\!-\!\sum_{i, R_{s,i}>R}\! \frac{p_{i}}{g\left(R_{s,i}\right)}\right]
\end{equation}
where we left out the term $(1+z)(3\Omega_m H_0^2)/(2 c^2)$ to clarify the equations and where $g$ is defined by
\begin{equation}
g(R) \equiv\left\{\begin{aligned}{\frac{\tan (\sqrt{K} R)}{\sqrt{K}}} & {\text { for } K>0} \\ {R \qquad} & {\text { for } K=0} \\ {\frac{\tanh (\sqrt{-K} R)}{\sqrt{-K}}} & {\text { for } K<0}\end{aligned}\right. .
\end{equation}
Now, if the $p_i$ weights satisfy the two following conditions
\begin{equation}
\label{eq:nulling-condition}
\left\{
\begin{aligned}
& \quad \sum_{i=1}^{3} p_{i}=0, \\ 
& \quad \sum_{i=1}^{3} \frac{p_{i}}{g\left(R_{s,i}\right)}=0 ,
\end{aligned}
\right.
\end{equation}
the nulled lensing kernel $\omega(R)_{\rm nulled}$ associated with our nulled convergence will be zero for $R<R_{s,1}$ confining the lenses between $R_{s,1}$ and $R_{s,3}$ as required.

The  conditions given by equation~(\ref{eq:nulling-condition}) can be solved \citep{Nulling} leading to
\begin{equation}
\left\{
\begin{aligned}
  &  p_2/p_1 = \frac{(R_{s,2})(g(R_{s,3})-g(R_{s,1}))}{(R_{s,1})(g(R_{s,2})-g(R_{s,3}))}, \\ 
 &  p_3/p_1 = \frac{(R_{s,3})(g(R_{s,1})-g(R_{s,2}))}{(R_{s,1})(g(R_{s,2})-g(R_{s,3}))},
   \end{aligned}
    \right.
        \label{weights}
\end{equation}
$p_1$ thus being chosen as an arbitrary normalisation.

\subsection{Cumulants of the nulled convergence}
Let us now apply this formalism to our simulated set of convergence maps from different source redshifts. 
We select 8 redshifts from $z_s = 0.57$ to $1.6$ so as to roughly mimic the tomographic strategy of the Euclid photometric survey \citep{Euclid,matteo}. 
We then apply the three-plane solution to every set of three neighbouring planes (our choice of normalisation is always $p_1 =1$) and obtain the effective lensing kernels shown in figure~\ref{kernel} with weights given by
\begin{equation}
    \kappa_{\rm null, a} = \sum_i p_a^i \, \kappa_i,
\end{equation}
the subscript $a$ denoting the $a$th nulled map we construct and where the matrix of weights $p$ is 
\begin{equation}
\label{matrix}
\begin{pmatrix}
1&0&0&0&0&0&0&0\\0&1&0&0&0&0&0&0\\1&-3.6&2.6&0&0&0&0&0\\0&1&-1.6&0.6&0&0&0&0\\0&0&1&-3.4&2.4&0&0&0\\0&0&0&1&-1.6&0.6&0&0\\0&0&0&0&1&-2.2&1.2&0\\0&0&0&0&0&1&-2.2&1.2\\
\end{pmatrix} \!\!\!.
\end{equation}
Obviously, so as not to loose any information, one shall also consider the first two maps (the two lowest redshifts) without nulling applied to keep the same information content before and after nulling.

\begin{figure}
    \centering
    \includegraphics[width=\columnwidth]{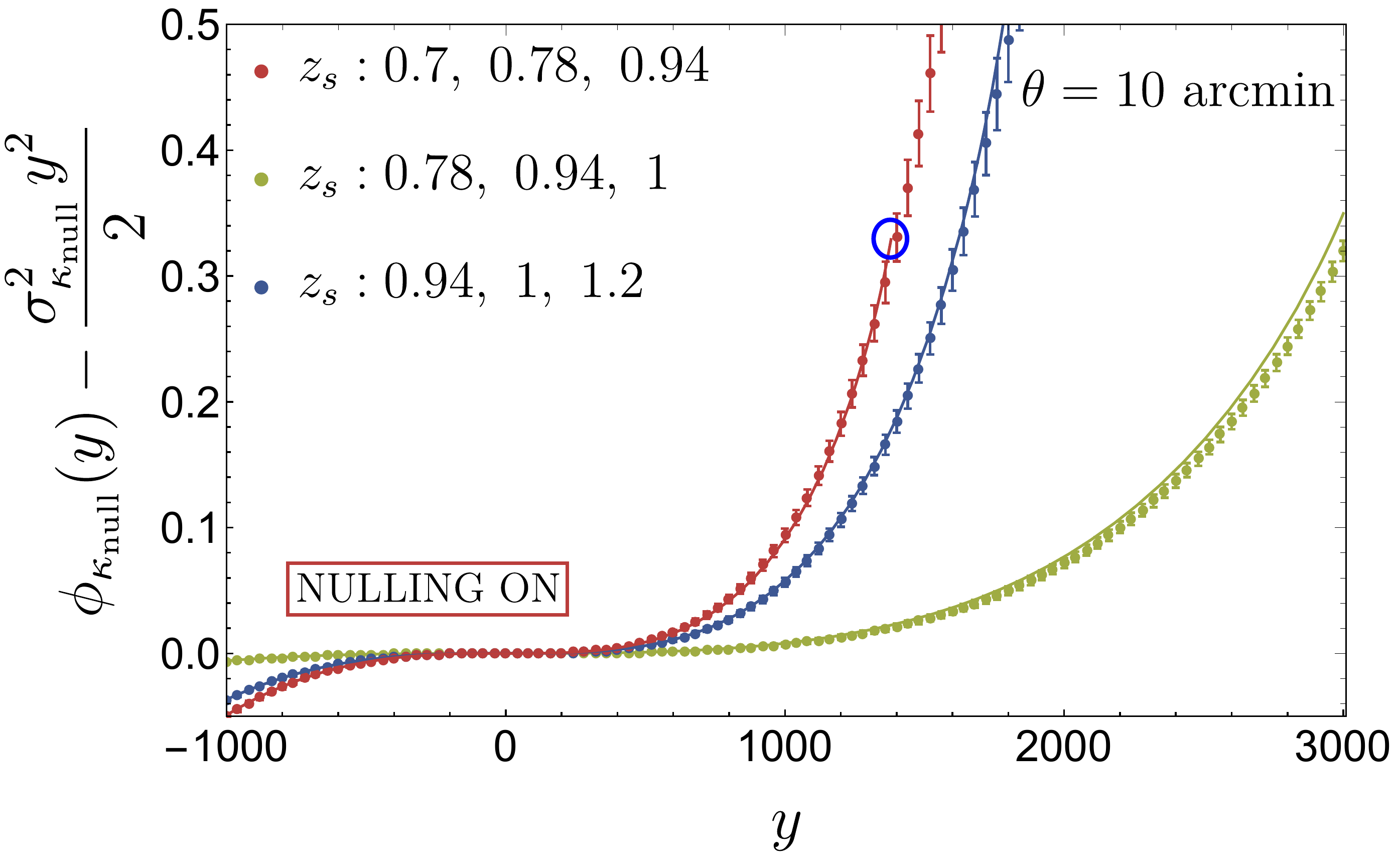}
    \caption{CGF of the nulled convergence for different source redshifts and opening angle $\theta = 10$~arcmin. Data points are taken from the simulation after implementation of the nulling procedure described in section \ref{nulling}. Blue circles indicate critical points as discussed for figure~\ref{CGFplot} in section \ref{CGFvalid}.}
    \label{CGF_nulling}
\end{figure}

For every set of 3 simulated $\kappa$-maps at the redshifts chosen, we take their linear combination and filter the obtained nulled convergence maps with a top-hat window function of angular radius $\theta = 10$~arcmin. Similarly to section~\ref{CGFproj}, we compute the theoretical CGFs for the nulled convergences as well as measure it in the maps. 
Once again the Gaussian contribution is removed to focus on the higher-order contributions. An excellent agreement is found between the simulated and theoretical CGFs in figure~\ref{CGF_nulling}, that is to be contrasted with the case before nulling shown in figure~\ref{CGFplot}. 
As expected, the nulling strategy has allowed us to remove the non-linear effect of the smaller scales and recover a situation where our approach based on cylindrical collapse, and therefore tree-order perturbation theory, is accurate enough. 
This agreement is discussed in more details in the next section when comparing PDFs obtained making use of the nulling procedure. 

\subsection{Nulled convergence PDFs}

\begin{figure}
    \centering
    \includegraphics[width=\columnwidth]{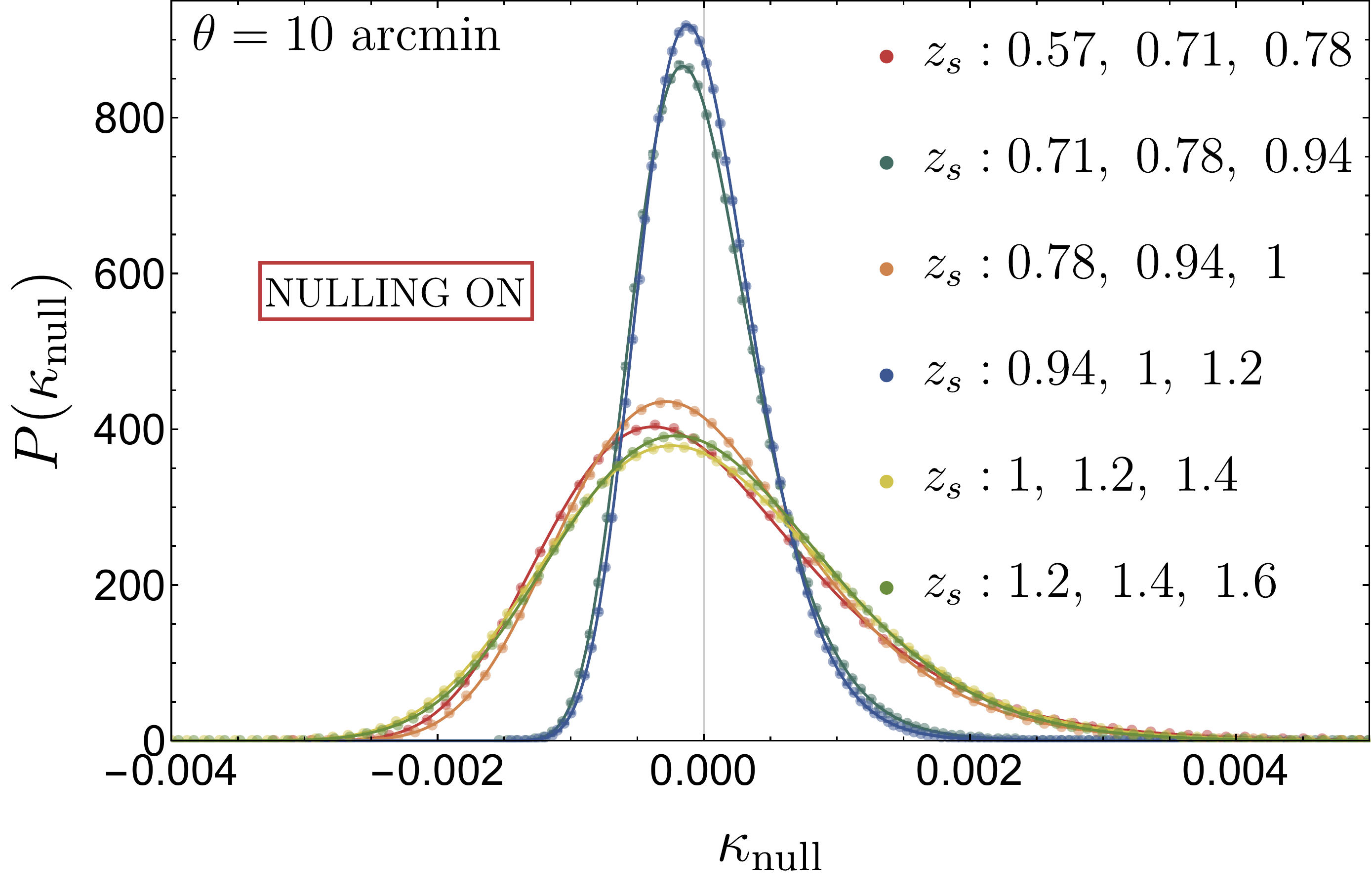}\\
    \includegraphics[width=\columnwidth]{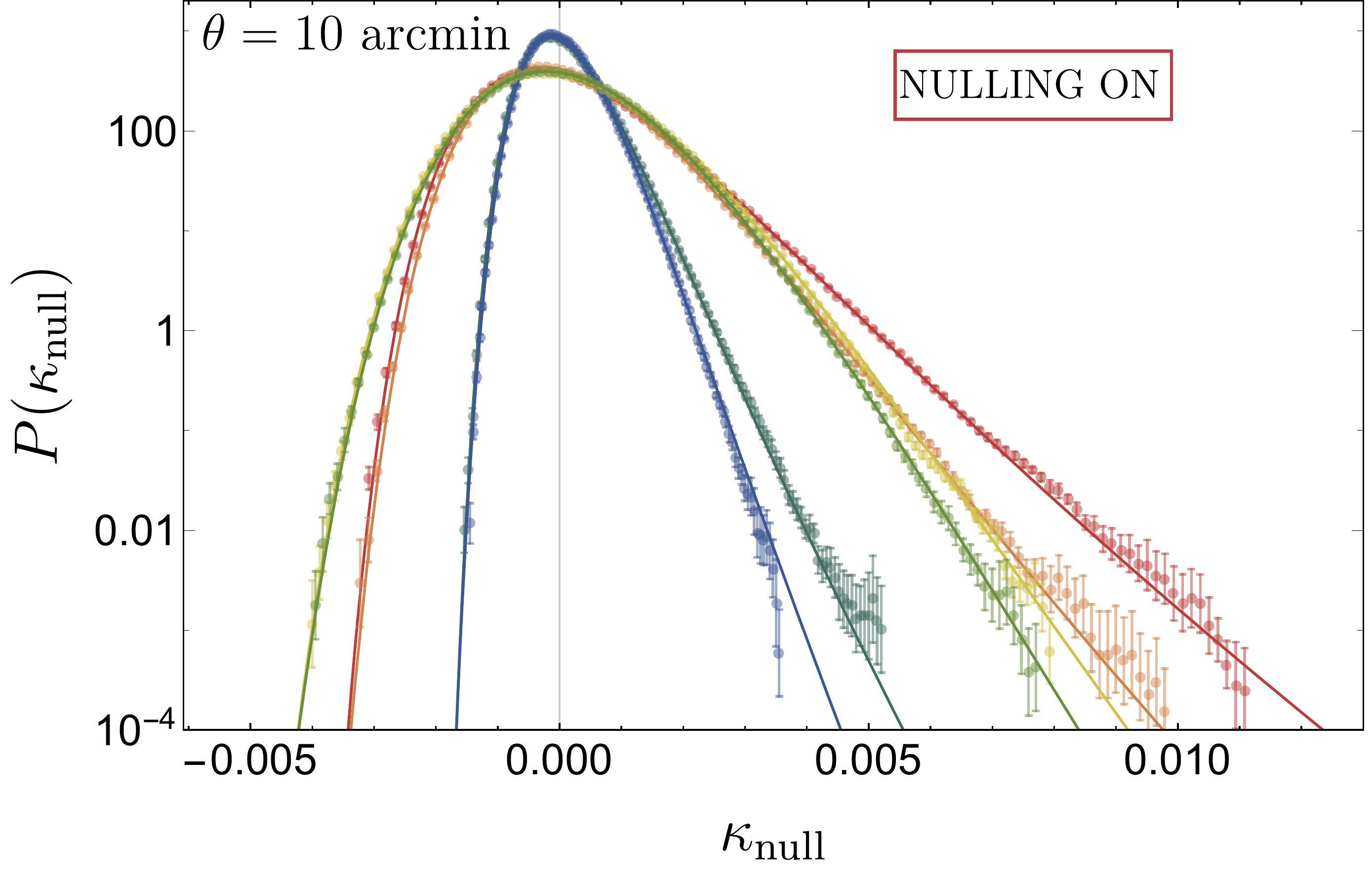}\\
    \includegraphics[width=\columnwidth]{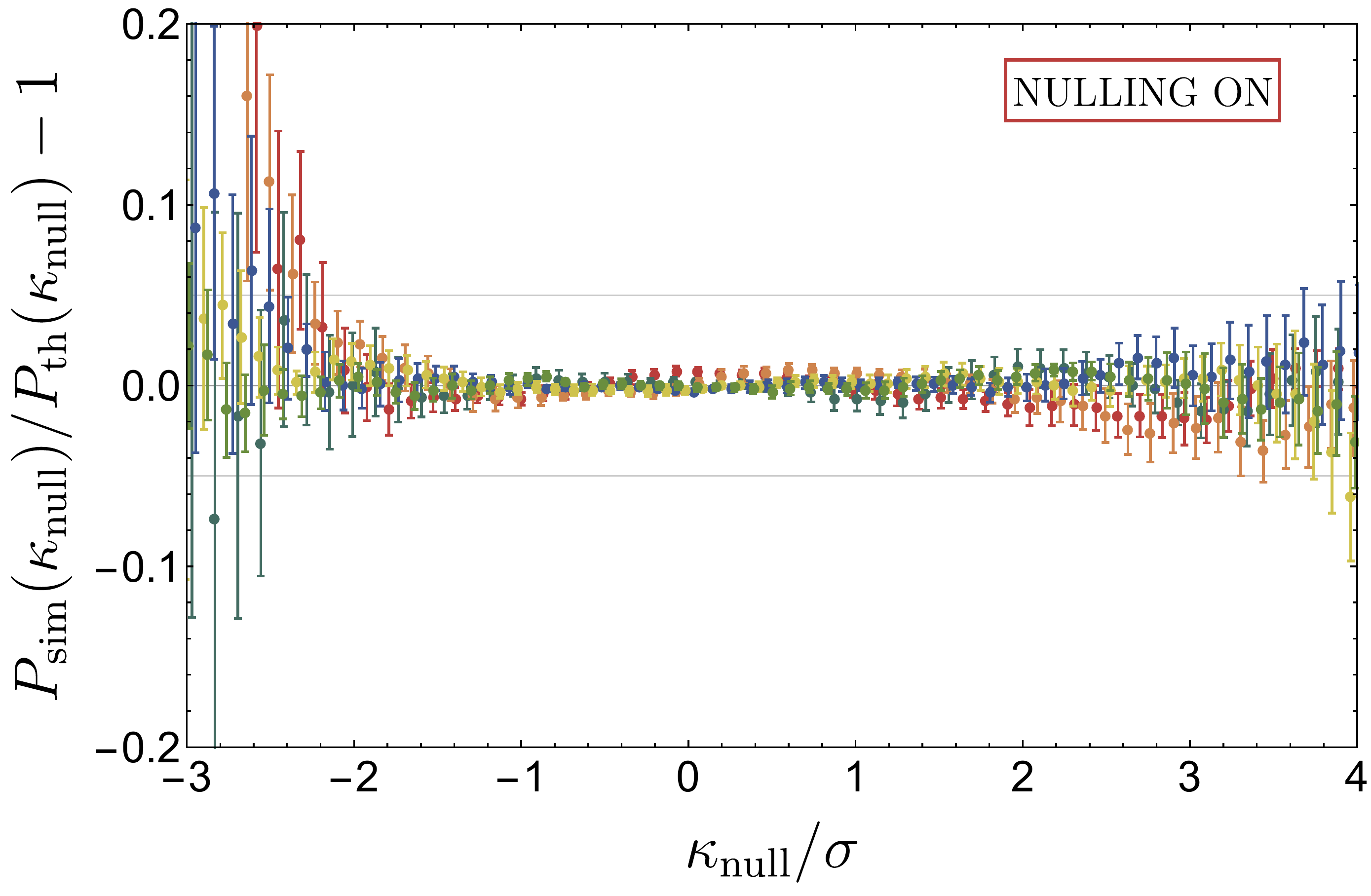}
    \caption{One-point PDF of the nulled weak-lensing convergence maps for different sets of redshift planes ordered as labelled from high variance (red) to lower variance (blue). The associated sets of planes leading to the nulled lensing kernel are displayed in figure~\ref{kernel}. The opening angle is fixed here to $\theta = 10$~arcmin. Solid lines display the LDT predictions given by equation~(\ref{laplacePDF}) while the measurements on the simulated sky are shown with error bars. Top panel: PDF in linear scale. Middle panel: same as top panel in log scale to better display the tails. Bottom panel: residuals of the simulated data compared to the prediction. These plots have to be contrasted with figure~\ref{redshift_pdf} obtained before nulling.}
    \label{nulling_pdf}
\end{figure}

\begin{table}
\resizebox{\columnwidth}{!}{
    \centering
    \bgroup
    \def\arraystretch{1.5}
    \begin{tabular}{|c|c|c|c|c|c|c|}
    \hline
      $z_{s,1}$ & 0.57 & 0.71 & 0.78 & 0.94 & 1 & 1.2 \\
      \hline
      $z_{s,2}$ & 0.71 & 0.78 & 0.94 & 1 & 1.2 & 1.4 \\
      \hline
     $z_{s,3}$ & 0.78 & 0.94 & 1 & 1.2 & 1.4 & 1.6 \\
    \hline
$\sigma^2_{\kappa} \ (10^{-6})$  & 1.2 & 0.25 & 0.95 & 0.21 &  1.2 & 1.1\\
    \hline
    \end{tabular}
    \egroup}
    \caption{\red{Variance of the nulled convergence field fitted from the simulated data for various sets of source redshifts, 3 for each map, as labelled. The opening angle is $\theta = 10$ arcmin. These values are notably used in figure~\ref{nulling_pdf}.}}
    \label{table4}
\end{table}

\red{Table~\ref{table4} shows the fitted variances and} figure~\ref{nulling_pdf} displays the resulting PDFs of the 6 nulled convergence maps for a fixed opening angle of $\theta=10$~arcmin.
They are found to match remarkably well with the theoretical predictions: peaks are described within percent accuracy and tails are accurately captured within a 2-3 $\sigma$ range depending on the redshifts considered.
Also note that the inaccuracy that seems to appear at 2 $\sigma$ in the low-$\kappa$ regime for the red and orange curves happens at a very low probability not shown in the middle panel since the PDF is not symmetric.
By essence, the nulling procedure 
considerably reduces scale mixing and
will thus be more likely to significantly enhance the accuracy of the theoretical predictions. 
Note that at these scales and redshifts, the PDFs are all highly non-Gaussian even though this may not be that clear on figure~\ref{nulling_pdf}. Indeed, the convergence variance, nulled or not, is not a good probe for the level of non-linearity as opposed to the variance of the projected density, nulled or not. 
It is even more the case in a scenario where nulling is applied since the variance depends both on the choice of source planes and the normalisation constant (which is totally arbitrary and could therefore be arbitrarily large!) which renders even less meaningful a direct interpretation of the width of the PDF in terms of level of non-Gaussianity. To do so, one would need to go back to projected densities (that is to say divide by $\kappa_{\rm min}^{\rm null}$).

\red{
Let us highlight that the LDT prediction presented here performs significantly better than a Gaussian as it allows us to model the non-zero higher order cumulants and therefore captures the tails of the PDF. This is clearly seen in Appendix~\ref{app:approx} where a comparison with a Gaussian is displayed.
Because our observable is non-Gaussian, the peak of the distribution (i.e the most likely value) and the mean differ significantly so that a Gaussian is always a bad fit to the PDF, not only in the tails but also around the peak (0 for a Gaussian while negative in the non-Gaussian case as the PDF is skewed towards underdensities).
}
In addition, note that the LDT approach significantly outperforms a standard log-normal approximation for $\kappa + \kappa_{\text{min}}$ as shown quantitatively in appendix~\ref{app:approx}. Additionally, our formalism can potentially be applied to jointly model the statistics of lensing convergence considered here and tracer densities in thick redshift slices described in \cite{cylindres}, a situation where the log-normal approximation has been shown to fail. 

\red{Finally, in appendix~\ref{Halofitsection}, we test the robustness of our approach in particular with regards to the model chosen for the variance. It is shown that the linear approximation used to describe the cross-correlations between scales (see equation~(\ref{variance}) is sufficient as taking the non-linear scale-dependence of the variance given by Halofit does not change the resulting PDF prediction. Note also that using the projected non-linear variance in equation~(\ref{NLPhi}) as taken from Halofit or as measured directly in the simulation does not make any major difference.
}

\section{Discussion} \label{discussion}

\subsection{Shape noise \& source distribution}

\begin{figure}
    \centering
    \includegraphics[width = \columnwidth]{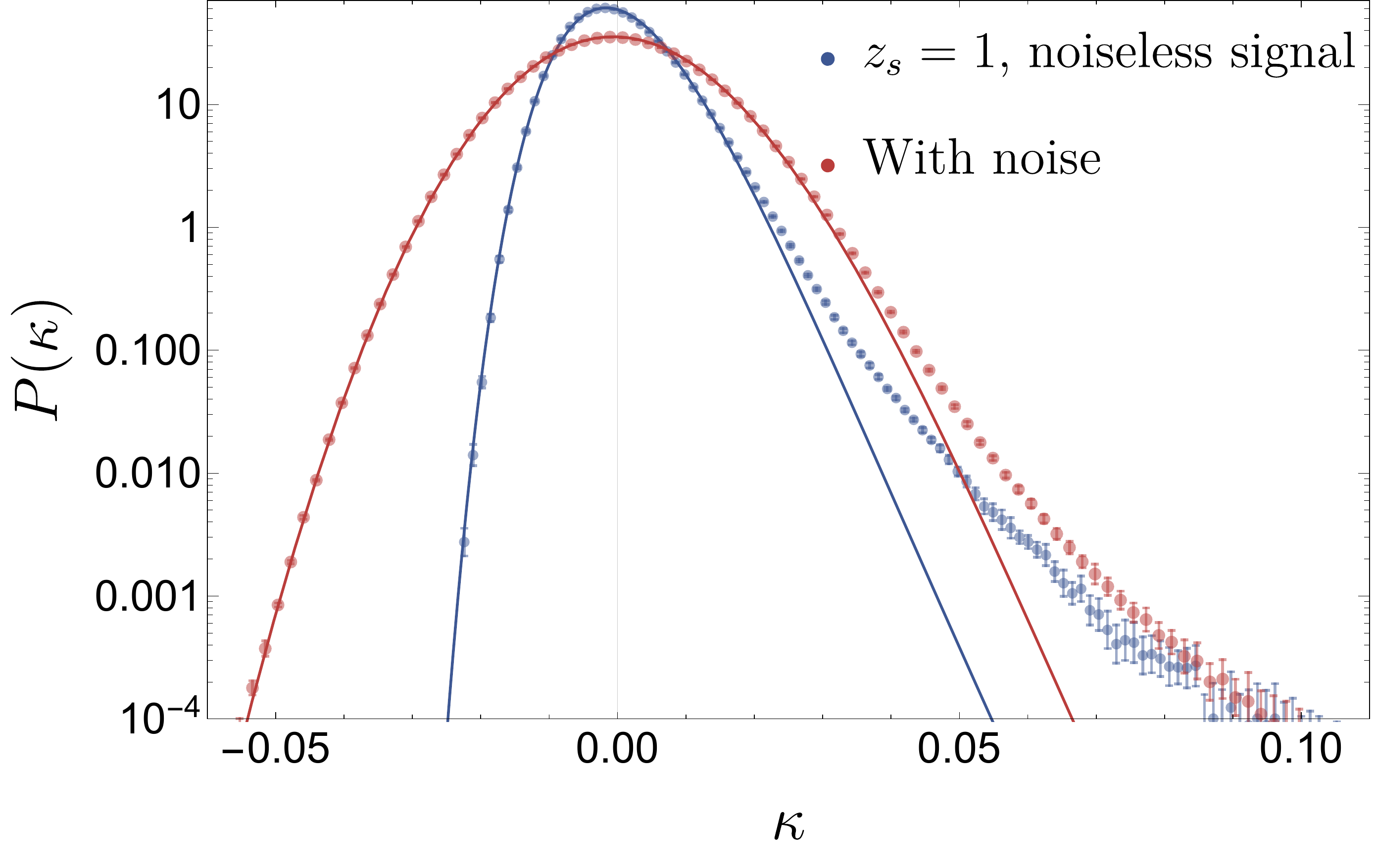}
    \caption{Effect of shape noise on the convergence PDF for opening angle of 10~arcmin at source redshift $z=1$. Points with error bars are measured in the simulation while solid lines are LDT predictions convolved by a Gaussian as described by equation~(\ref{eq:PDFnoise}) and taking $\sigma_{SN} = 0.009$.}
    \label{noise}
\end{figure}

Since the weak-lensing convergence map is obtained from cosmic shear measurements and galaxies themselves are intrinsically elliptical, the observed shear has a contribution from weak-lensing and the intrinsic signal. Shape noise is caused by the variance of the  intrinsic ellipticity, which is the dominant source of noise in shear measurements and impacts the convergence PDF as if it was convolved with a Gaussian centred at zero with variance $\sigma_{SN}^2$ \citep{Clerkin17}

\begin{align}
\label{eq:PDFnoise}
    \mathcal P_{SN}(\kappa) = \frac{1}{\sqrt{2\pi}\sigma_{SN}}\int_{\kappa_{\rm min}}^\infty \!\!\!\! d\kappa' \exp\left(-\frac{(\kappa-\kappa')^2}{2\sigma_{SN}^2}\right) \mathcal P(\kappa')\,.
\end{align}
To estimate the variance of shape noise distribution, we assume
$\sigma_{SN}^2=\sigma_\epsilon^2/(n_{g_s}\Omega_{\theta})$, where  $\sigma_\epsilon^2 = 0.26$ for the ellipticity and $\Omega_{\theta}$ is the solid angle in units of~arcmin$^2$. This is similar to the addition of shape noise performed in \cite{Liu19}. We assume a source galaxy redshift distribution
\begin{align}
    n_s(z) \propto z^{\alpha} \exp\left[-\left(\frac{z}{z_0}\right)^\beta\right]\,,
\end{align}
normalised such that the total source galaxy number density is $n_{g_s}$.
For Euclid (LSST) specifications one has $\alpha =$ 1.3 (1.27), $\beta =$ 1.5 (1.02), $z_0 =$ 0.65 (0.5) and source galaxy number density $n_{g_s}=$ 30 (26)~$\text{arcmin}^{-2}$.\footnote{Taken from figs. 2 and 10 in \cite{Schaan17}.} 
In practice, we add a random noise to each pixel in our simulated maps following a Gaussian with zero mean and variance $\sigma_{SN}^2$.
figure~\ref{noise} shows an example of a measured noisy PDF for an opening angle of 10~arcmin. In this Euclid-like configuration, non-Gaussianities are clearly detectable. It is beyond the scope of this paper to perform realistic forecasts but this result is encouraging and seems promising for application to future weak-lensing experiments.

In addition, note that one would need to account for the fact that convergence is measured not from a single source redshift but from a given source galaxy distribution $n_s(z_s)$. This can be readily done in our formalism. Indeed,
when aiming to predict the weak-lensing convergence measured from $n(z_s)$, one can simply replace the weight function from equation~\eqref{eq:weight} by
\begin{equation}
\label{eq:weightsource}
    \omega_{n_s}(R) = \frac{3\,\Omega_m\,H_0^2}{2\,c^2}\!\! \int\!\! dR_s \frac{D(R)\,D(R_s-R)}{D(R_s)a(R)}\, n_s(z_s)\frac{dz_s}{dR_s}.
\end{equation}

\subsection{Tomographic analysis and nulling}
\label{sec:tomography}
Once nulling is implemented, the volume that is probed necessarily diminishes 
(as the lensing kernel is now confined to a smaller subregion). Hence, the noise for one bin increases. However, the total information is restored once a multi-plane approach is considered.
In a tomographic analysis of weak-lensing data, one would need to consider a multidimensional data vector that contains the nulled convergence PDF measured in each redshift bin considered. 
A consistent analysis of these measurements therefore requires the knowledge of the joint PDF of the nulled convergence field in each bin. 
For two bins, the joint CGF of nulled (or standard convergences) is now given by
\begin{equation}
\label{eq:phi2}
    \phi_{\text{proj},\theta}(y_1,y_2)\! =\! \int_0^{R_{s,\text{max}}} \!\!\! {\rm d}R \phi_{\rm cyl}(\omega_1(R) y_1 + \omega_2(R) y_2,R),
\end{equation}
where $\omega_{1,2}$ are the lensing kernels for the convergence be it nulled or standard. This is obtained from a generalisation of equation~(\ref{cumulant}) for the joint cumulants and following the same path as for equations~(\ref{laplace}-\ref{log}) to define the joint generating functions.
This expression can straightforwardly be generalised to any number of bins. 
By definition, a nulled convergence in a given bin only correlates with its two adjacent neighbours. This property allows us to write the full joint CGF $\phi_{1,\cdots ,N}$ of the {\it nulled} convergence in $N$ bins of redshift as a function of the CGFs of two adjacent (and individual) bins only following (see appendix~\ref{app:phiN} for the derivation)
\begin{equation}
\label{eq:phiN}
\phi_{1,\cdots ,N}(y_1,\!\cdots\!,y_N)\!=\!\sum_{i=1}^{N-1}\phi_{i,i+1}(y_i,y_{i+1})
\!-\!\!\sum_{i=2}^{N-1}\!\phi_{i}(y_i).
\end{equation}
The simplification of the N-bin CGF in equation~(\ref{eq:phiN}) has the advantage of reducing the correlation between all variables down to couples of adjacent ones in terms of the two-bin CGF given by equation~(\ref{eq:phi2}).
This expression is of crucial practical importance as it can be rewritten efficiently in terms of the corresponding PDFs although this lengthy formula is not displayed here for the sake of simplicity. This yet complicated integral has the advantage of reducing the correlation between all variables down to couples of adjacent ones in terms of the two-bin PDFs, and opens the way to reducing the computational complexity of the numerical derivation of multi-dimensional convergence PDF. 
This is however beyond the scope of this paper to investigate this further.

\subsection{Correlated shape noise after nulling}
In a tomographic analysis, let us emphasize that nulling will combine with different weights the redshift bins. 
Hence initially uncorrelated shape noises between bins will become correlated after nulling.

Indeed,
let us decompose the estimator of the convergence in bin $i$ into the "true" convergence and a noise term
\begin{equation}
    \hat \kappa_i=\kappa_i+\epsilon_i,
\end{equation}
where the noise is uncorrelated $\left\langle \epsilon_i\epsilon_j\right\rangle=\delta_{i,j}\sigma_\epsilon^2/(n_{g_s}\Omega_{\theta})$ with $\delta_{i,j}$ the Kronecker delta.
The estimator of the nulled convergence can now be written
\begin{equation}
    \hat \kappa^{\rm null}_i=\kappa^{\rm null}_i+\epsilon^{\rm null}_i,
\end{equation}
with
\begin{equation}
\kappa^{\rm null}_i=\sum_{j=i-2}^{i}p_i^j \kappa_j\quad \text{for}\quad i>2
,
\end{equation}
and
\begin{equation}
\epsilon^{\rm null}_i=\sum_{j=i-2}^{i}p_i^j \epsilon_j,
\end{equation}
with 
$p_i^j$ the weights used to computed the $i$th nulled convergence maps.

From there, it appears that the shape noise of the nulled convergence maps are correlated such that
\begin{equation}
\left\langle \epsilon_i\epsilon_j\right\rangle=\delta_{i,j}\sigma_{i,i}+\delta_{i,j-1}\sigma_{i,i+1}+\delta_{i,j-2}\sigma_{i,i+2}\,,
\end{equation}
with
\begin{equation}
\sigma_{i,i}=\sum_{j=i-2}^{j}\left(p_i^j\right)^2\sigma_\epsilon^2/(n_{g_s}\Omega_{\theta})\,,
\end{equation}
\begin{equation}
\sigma_{i,i+1}=\left(p_i^{i-1}p_{i+1}^{i-1}+p_i^{i}p_{i+1}^{i}\right)\sigma_\epsilon^2/(n_{g_s}\Omega_{\theta})\,,
\end{equation}
and
\begin{equation}
\sigma_{i,i+2}=p_i^{i}p_{i+2}^{i}\sigma_\epsilon^2/(n_{g_s}\Omega_{\theta})\,.
\end{equation}

We now have all the theoretical tools at hand to perform tomographic analysis of weak-lensing data and implement nulling, including the expected correlations in the intrinsic shape noise and arbitrary galaxy distributions. 

\subsection{Prospects}
As was shown in section~\ref{sec:tomography}, the next step to build a consistent pipeline for the tomographic analysis of convergence PDFs will be to investigate the joint statistics between two adjacent bins of nulled convergence. \red{Indeed, because nulled lensing kernels only overlap two by two, some single structures (lenses) along the line-of-sight are counted twice and therefore create correlations between adjacent bins. The joint analysis (which contains the correlation matrix) thus only requires the explicit computation of the joint PDF between adjacent bins. Note that for the original convergence maps (before nulling), one would have to compute the joint PDF of all bins since single structures can be counted many times, that is to say all bins are correlated.}
In addition, one would need to also properly model the shot noise due to the finite number of tracers and cosmic variance that should be within reach of the LDT formalism as was shown by \cite{2016MNRAS.460.1598C} in the 3D case where a theory of the errors was developed, including in particular the effect of the finite volume of the survey.
\red{However, a general treatment beyond two-point statistics is not known using LDT. In other words, the likelihood function for such a non-Gaussian quantity is still an unresolved problem. 
One could try and use large sets of simulations (including mocks with added shape noise) to investigate this issue but this is beyond the scope of this paper.
In addition, post-born corrections, although small for the scales treated in this paper, might become important in the context of a high precision analysis, and certainly are for small opening angles (order of arcmin) and very high redshifts. One could then want to take them into account. A solution could be to compute the dominant corrections, for example on the skewness and kurtosis and find a way to implement those corrections in the cumulant generating function. This idea will be further studied elsewhere.}

Once the full theoretical framework is in place, one might worry about potential observational systematics. In particular, it would be interesting to investigate the impact of photometric errors on the nulling procedure and the resulting PDF, together with testing for the effect of baryons on small scales (which could still pervade if the nulling is not exact) or the potential impact of intrinsic alignments of galaxies.

Another promising avenue will be to combine weak-lensing and galaxy clustering in the spirit of the so-called density-split statistics \citep{FriedrichDES17,GruenDES17} that is able to constrain cosmological parameters even if some degrees of freedom are left such as the galaxy bias and galaxy-matter correlation coefficients. Indeed, one can deduce the tangential shear profile around a line-of-sight with given tracer density from the convergence profile. This analysis can avoid systematics such as additive shear biases or intrinsic alignments if the redshift distributions of the tracer and source sample do not overlap.

Beyond the lensing of galaxy shapes, our formalism, with or without nulling, could potentially be applied to extract information -- complementary to the power spectrum -- from the lensing effect measured with intensity mapping: 
the CMB at high redshift \citep{Jia2016}, 
the CIB \citep{Schaan18}, 
21cm emission probing neutral hydrogen in galaxies \citep{Pourtsidou15}
or the Lyman-alpha forest \citep{Croft18} at intermediate redshifts $2<z<5$, where our predictions are accurate, even without nulling (as shown on figure~\ref{redshift_pdf}). 
In particular, one could apply density-split statistics to 21cm intensity mapping, combining our lensing results here with tracer densities of neutral hydrogen \citep{Leicht19}.
Note however that although we here provide a viable theoretical framework for the modelling of these observables from weak-lensing of intensity maps, some observational challenges are still to be tackled. Noise indeed makes the measurement of a cosmological signal in the PDF even harder than in the power spectrum \citep{Jia2016} that is already challenging. \cite{foreman} for example found that the signal-to-noise ratio for the convergence power spectrum was of order unity for SKA, CHIME and HIRAX which renders its detection only possible through cross-correlations. The hope might come from futuristic experiments \citep{StageII21cm} where the signal-to-noise ratio promises to be better although the modelling of noise will still remain highly challenging, especially for higher order statistics like PDFs.

\section{Conclusions}

In this article, the one-point distribution of the convergence field was derived from first principles using LDT.
The geometry and time-evolution within the cone was taken into account by slicing it up and summing the resulting (supposedly independent) random variables and their individual CGF.
In the limit of small variance, 2D spherical collapse was used to compute the individual CGF by Legendre transform of the so-called rate function which drives the exponential decay of the distribution of the field values.
From there, the PDF can be calculated using an inverse Laplace transform of the full CGF.
Even if the exact calculation has to rely on a numerical integration in the complex plane, we also provided a simple accurate approximation of the PDF predicted by LDT that outperforms the lognormal model. 

We also implemented a nulling procedure -- which boils down to linearly combining  the various redshift bins with coefficients that only depend on the background (scale independent) -- that formally allows us to choose where the lensing kernel along the line-of-sight is effectively not zero, thus avoiding the mixing of scales which is particularly important when one wants to leverage the influence of the very small scales where theoretical models break down.

This formalism was tested against numerical simulations with ray-tracing for a wide range of redshifts and opening angles.
On mildly non-linear scales, as expected, in the absence of nulling, the broad lensing kernel tends to pick up contributions for relatively small scales therefore reducing the range of applicability of theoretical approaches compared to the case of the three dimensional matter density field studied previously in the literature. 
However, in the context of tomographic weak-lensing experiments, once we implemented a nulling strategy, we recovered very accurate predictions for the one-point distribution of the nulled convergence maps.
In practice for an opening angle of 10~arcmin, all the redshift bins we tested between 0.5 and 1.5 were correctly modelled by LDT (after nulling) with no deviation from the simulation given the estimated error bars, see figure~\ref{nulling_pdf}.

Given the potential huge information content of these observables advocated by some recent works \citep{Patton17}, the predictions from first principles developed in this article could be successfully applied to forthcoming data along with the standard power spectrum based analysis and could bring additional information beyond $\Lambda$CDM parameters like massive neutrinos \citep{Liu19} or dark energy \citep{encircling}. 
Let us stress that implementing nulling in weak-lensing analysis is central in order to avoid extracting biased information from the small scales that lack a full theoretical understanding (including due to the effect of baryon physics that needs to be modelled in weak-lensing surveys \citep{Kidsviking,DLS}). Not only this general nulling technique should be used for one-point statistics but could also be applied to standard power spectrum analysis 
(and more generally to the full two-point PDF) in order to disentangle the effects of the different physical scales. 
However, more realistic effects have to be accounted for before the here mentioned formalism could be directly applied to real data. In particular, we have not investigated the precise impact of the galaxy redshift distribution $n_s(z)$ (for which one needs to go from a set of discrete source planes to a source distribution), photometric redshift errors or shape noise, which are left for future works.
Promising extensions include an application of the formalism: i) to compensated filters such as for aperture mass which require the joint modelling of the field at two different scales, ii) to two-point statistics in order to model cosmic variance and iii) to the joint analysis of multiple redshift bins. All of these ideas are within reach of LDT as was shown in the case of the three-dimensional matter density in \cite{2015MNRAS.449L.105B,2016MNRAS.460.1598C} for respectively the multi-scale and two-point statistics.

\section*{Acknowledgements}
This work is partially supported by the SPHERES grant ANR-18-CE31-0009 of the French {\sl Agence Nationale de la Recherche} and by Fondation MERAC.
AB's work is supported by a fellowship from CNES. CU kindly acknowledges funding by the STFC grant RG84196 `Revealing the Structure of the Universe'.
We thank Ken Osato for pointing us to the simulation we used in this paper and Eric Hivon for his help with {\sc healpix}. 
We also warmly thank Emmanuel Schaan, Karim Benabed and Christophe Pichon for fruitful discussions.
This work has made use of the Horizon Cluster hosted by Institut d'Astrophysique de Paris. We thank Stephane Rouberol for running smoothly this cluster for us. 
CU would like to thank Daniel Gruen, Oliver Friedrich and Colin Hill for useful discussions. 

\bibliographystyle{mnras}
\bibliography{biblio} 

\begin{thebibliography}{}
\makeatletter
\relax
\def\mn@urlcharsother{\let\do\@makeother \do\$\do\&\do\#\do\^\do\_\do\%\do\~}
\def\mn@doi{\begingroup\mn@urlcharsother \@ifnextchar [ {\mn@doi@}
  {\mn@doi@[]}}
\def\mn@doi@[#1]#2{\def\@tempa{#1}\ifx\@tempa\@empty \href
  {http://dx.doi.org/#2} {doi:#2}\else \href {http://dx.doi.org/#2} {#1}\fi
  \endgroup}
\def\mn@eprint#1#2{\mn@eprint@#1:#2::\@nil}
\def\mn@eprint@arXiv#1{\href {http://arxiv.org/abs/#1} {{\tt arXiv:#1}}}
\def\mn@eprint@dblp#1{\href {http://dblp.uni-trier.de/rec/bibtex/#1.xml}
  {dblp:#1}}
\def\mn@eprint@#1:#2:#3:#4\@nil{\def\@tempa {#1}\def\@tempb {#2}\def\@tempc
  {#3}\ifx \@tempc \@empty \let \@tempc \@tempb \let \@tempb \@tempa \fi \ifx
  \@tempb \@empty \def\@tempb {arXiv}\fi \@ifundefined
  {mn@eprint@\@tempb}{\@tempb:\@tempc}{\expandafter \expandafter \csname
  mn@eprint@\@tempb\endcsname \expandafter{\@tempc}}}

\bibitem[\protect\citeauthoryear{{Baker}, {Clampitt}, {Jain}  \&
  {Trodden}}{{Baker} et~al.}{2018}]{Baker18}
{Baker} T.,  {Clampitt} J.,  {Jain} B.,   {Trodden} M.,  2018, \mn@doi [\prd]
  {10.1103/PhysRevD.98.023511}, \href
  {https://ui.adsabs.harvard.edu/abs/2018PhRvD..98b3511B} {98, 023511}

\bibitem[\protect\citeauthoryear{{Baratta}, {Bel}, {Plaszczynski}  \&
  {Ealet}}{{Baratta} et~al.}{2019}]{Baratta19}
{Baratta} P.,  {Bel} J.,  {Plaszczynski} S.,   {Ealet} A.,  2019, arXiv
  e-prints, \href {https://ui.adsabs.harvard.edu/abs/2019arXiv190609042B} {p.
  arXiv:1906.09042}

\bibitem[\protect\citeauthoryear{{Barber}, {Munshi}  \& {Valageas}}{{Barber}
  et~al.}{2004}]{Valageas2}
{Barber} A.~J.,  {Munshi} D.,   {Valageas} P.,  2004, \mn@doi [\mnras]
  {10.1111/j.1365-2966.2004.07249.x}, \href
  {https://ui.adsabs.harvard.edu/abs/2004MNRAS.347..667B} {347, 667}

\bibitem[\protect\citeauthoryear{{Baugh}, {Gaztanaga}  \& {Efstathiou}}{{Baugh}
  et~al.}{1995}]{1995MNRAS.274.1049B}
{Baugh} C.~M.,  {Gaztanaga} E.,   {Efstathiou} G.,  1995, \mn@doi [\mnras]
  {10.1093/mnras/274.4.1049}, \href
  {https://ui.adsabs.harvard.edu/abs/1995MNRAS.274.1049B} {274, 1049}

\bibitem[\protect\citeauthoryear{{Bel} et~al.,}{{Bel} et~al.}{2016}]{Bel16}
{Bel} J.,  et~al., 2016, \mn@doi [\aap] {10.1051/0004-6361/201526455}, \href
  {http://adsabs.harvard.edu/abs/2016A%26A...588A..51B} {588, A51}

\bibitem[\protect\citeauthoryear{{Bernardeau}}{{Bernardeau}}{1992}]{Bernardeau1992}
{Bernardeau} F.,  1992, \mn@doi [\apj] {10.1086/171398}, \href
  {https://ui.adsabs.harvard.edu/abs/1992ApJ...392....1B} {392, 1}

\bibitem[\protect\citeauthoryear{{Bernardeau}}{{Bernardeau}}{1994}]{Bernardeau1994}
{Bernardeau} F.,  1994, \aap, \href
  {https://ui.adsabs.harvard.edu/abs/1994A&A...291..697B} {291, 697}

\bibitem[\protect\citeauthoryear{{Bernardeau}}{{Bernardeau}}{1995}]{Bernardeau1995}
{Bernardeau} F.,  1995, \aap, \href
  {https://ui.adsabs.harvard.edu/abs/1995A&A...301..309B} {301, 309}

\bibitem[\protect\citeauthoryear{{Bernardeau} \& {Kofman}}{{Bernardeau} \&
  {Kofman}}{1995}]{BernardeauKofman}
{Bernardeau} F.,  {Kofman} L.,  1995, \mn@doi [\apj] {10.1086/175542}, \href
  {https://ui.adsabs.harvard.edu/abs/1995ApJ...443..479B} {443, 479}

\bibitem[\protect\citeauthoryear{{Bernardeau} \& {Reimberg}}{{Bernardeau} \&
  {Reimberg}}{2016}]{seminalLDT}
{Bernardeau} F.,  {Reimberg} P.,  2016, \mn@doi [\prd]
  {10.1103/PhysRevD.94.063520}, \href
  {https://ui.adsabs.harvard.edu/abs/2016PhRvD..94f3520B} {94, 063520}

\bibitem[\protect\citeauthoryear{{Bernardeau} \& {Valageas}}{{Bernardeau} \&
  {Valageas}}{2000}]{BernardeauValageas}
{Bernardeau} F.,  {Valageas} P.,  2000, \aap, \href
  {https://ui.adsabs.harvard.edu/abs/2000A&A...364....1B} {364, 1}

\bibitem[\protect\citeauthoryear{{Bernardeau}, {van Waerbeke}  \&
  {Mellier}}{{Bernardeau} et~al.}{1997}]{Bernardeau1997}
{Bernardeau} F.,  {van Waerbeke} L.,   {Mellier} Y.,  1997, \aap, \href
  {https://ui.adsabs.harvard.edu/abs/1997A&A...322....1B} {322, 1}

\bibitem[\protect\citeauthoryear{{Bernardeau}, {Colombi}, {Gazta{\~n}aga}  \&
  {Scoccimarro}}{{Bernardeau} et~al.}{2002}]{BernardeauReview}
{Bernardeau} F.,  {Colombi} S.,  {Gazta{\~n}aga} E.,   {Scoccimarro} R.,  2002,
  \mn@doi [\physrep] {10.1016/S0370-1573(02)00135-7}, \href
  {https://ui.adsabs.harvard.edu/abs/2002PhR...367....1B} {367, 1}

\bibitem[\protect\citeauthoryear{{Bernardeau}, {Pichon}  \&
  {Codis}}{{Bernardeau} et~al.}{2014a}]{2014PhRvD..90j3519B}
{Bernardeau} F.,  {Pichon} C.,   {Codis} S.,  2014a, \mn@doi [\prd]
  {10.1103/PhysRevD.90.103519}, \href
  {https://ui.adsabs.harvard.edu/abs/2014PhRvD..90j3519B} {90, 103519}

\bibitem[\protect\citeauthoryear{{Bernardeau}, {Nishimichi}  \&
  {Taruya}}{{Bernardeau} et~al.}{2014b}]{Nulling}
{Bernardeau} F.,  {Nishimichi} T.,   {Taruya} A.,  2014b, \mn@doi [\mnras]
  {10.1093/mnras/stu1861}, \href
  {https://ui.adsabs.harvard.edu/abs/2014MNRAS.445.1526B} {445, 1526}

\bibitem[\protect\citeauthoryear{{Bernardeau}, {Codis}  \&
  {Pichon}}{{Bernardeau} et~al.}{2015}]{2015MNRAS.449L.105B}
{Bernardeau} F.,  {Codis} S.,   {Pichon} C.,  2015, \mn@doi [\mnras]
  {10.1093/mnrasl/slv028}, \href
  {https://ui.adsabs.harvard.edu/abs/2015MNRAS.449L.105B} {449, L105}

\bibitem[\protect\citeauthoryear{{Brouwer} et~al.,}{{Brouwer}
  et~al.}{2018}]{Brouwer18troughs}
{Brouwer} M.~M.,  et~al., 2018, \mn@doi [\mnras] {10.1093/mnras/sty2589}, \href
  {https://ui.adsabs.harvard.edu/abs/2018MNRAS.481.5189B} {481, 5189}

\bibitem[\protect\citeauthoryear{{Clerkin} et~al.,}{{Clerkin}
  et~al.}{2017}]{Clerkin17}
{Clerkin} L.,  et~al., 2017, \mn@doi [\mnras] {10.1093/mnras/stw2106}, \href
  {https://ui.adsabs.harvard.edu/abs/2017MNRAS.466.1444C} {466, 1444}

\bibitem[\protect\citeauthoryear{{Codis}, {Pichon}, {Bernardeau}, {Uhlemann}
  \& {Prunet}}{{Codis} et~al.}{2016a}]{encircling}
{Codis} S.,  {Pichon} C.,  {Bernardeau} F.,  {Uhlemann} C.,   {Prunet} S.,
  2016a, \mn@doi [\mnras] {10.1093/mnras/stw1084}, \href
  {https://ui.adsabs.harvard.edu/abs/2016MNRAS.460.1549C} {460, 1549}

\bibitem[\protect\citeauthoryear{{Codis}, {Bernardeau}  \& {Pichon}}{{Codis}
  et~al.}{2016b}]{2016MNRAS.460.1598C}
{Codis} S.,  {Bernardeau} F.,   {Pichon} C.,  2016b, \mn@doi [\mnras]
  {10.1093/mnras/stw1103}, \href
  {https://ui.adsabs.harvard.edu/abs/2016MNRAS.460.1598C} {460, 1598}

\bibitem[\protect\citeauthoryear{{Coles} \& {Jones}}{{Coles} \&
  {Jones}}{1991}]{ColesJones91}
{Coles} P.,  {Jones} B.,  1991, \mn@doi [\mnras] {10.1093/mnras/248.1.1}, \href
  {http://adsabs.harvard.edu/abs/1991MNRAS.248....1C} {248, 1}

\bibitem[\protect\citeauthoryear{{Colombi}}{{Colombi}}{1994}]{Colombi94}
{Colombi} S.,  1994, \mn@doi [\apj] {10.1086/174834}, \href
  {https://ui.adsabs.harvard.edu/abs/1994ApJ...435..536C} {435, 536}

\bibitem[\protect\citeauthoryear{{Cosmic Visions 21 cm Collaboration}
  et~al.,}{{Cosmic Visions 21 cm Collaboration} et~al.}{2018}]{StageII21cm}
{Cosmic Visions 21 cm Collaboration} et~al., 2018, arXiv e-prints, \href
  {https://ui.adsabs.harvard.edu/abs/2018arXiv181009572C} {p. arXiv:1810.09572}

\bibitem[\protect\citeauthoryear{{Croft}, {Romeo}  \& {Metcalf}}{{Croft}
  et~al.}{2018}]{Croft18}
{Croft} R. A.~C.,  {Romeo} A.,   {Metcalf} R.~B.,  2018, \mn@doi [\mnras]
  {10.1093/mnras/sty650}, \href
  {https://ui.adsabs.harvard.edu/abs/2018MNRAS.477.1814C} {477, 1814}

\bibitem[\protect\citeauthoryear{{Dolag}, {Komatsu}  \& {Sunyaev}}{{Dolag}
  et~al.}{2016}]{Dolag16}
{Dolag} K.,  {Komatsu} E.,   {Sunyaev} R.,  2016, \mn@doi [\mnras]
  {10.1093/mnras/stw2035}, \href
  {https://ui.adsabs.harvard.edu/abs/2016MNRAS.463.1797D} {463, 1797}

\bibitem[\protect\citeauthoryear{{Foreman}, {Meerburg}, {van Engelen}  \&
  {Meyers}}{{Foreman} et~al.}{2018}]{foreman}
{Foreman} S.,  {Meerburg} P.~D.,  {van Engelen} A.,   {Meyers} J.,  2018,
  \mn@doi [\jcap] {10.1088/1475-7516/2018/07/046}, \href
  {https://ui.adsabs.harvard.edu/abs/2018JCAP...07..046F} {2018, 046}

\bibitem[\protect\citeauthoryear{{Friedrich} et~al.,}{{Friedrich}
  et~al.}{2018}]{FriedrichDES17}
{Friedrich} O.,  et~al., 2018, \mn@doi [\prd] {10.1103/PhysRevD.98.023508},
  \href {https://ui.adsabs.harvard.edu/abs/2018PhRvD..98b3508F} {98, 023508}

\bibitem[\protect\citeauthoryear{{Gazta{\~n}aga}, {Fosalba}  \&
  {Elizalde}}{{Gazta{\~n}aga} et~al.}{2000}]{Gaztanaga00}
{Gazta{\~n}aga} E.,  {Fosalba} P.,   {Elizalde} E.,  2000, \mn@doi [\apj]
  {10.1086/309249}, \href
  {https://ui.adsabs.harvard.edu/abs/2000ApJ...539..522G} {539, 522}

\bibitem[\protect\citeauthoryear{{Gouin} et~al.,}{{Gouin}
  et~al.}{2019}]{Celine}
{Gouin} C.,  et~al., 2019, \mn@doi [\aap] {10.1051/0004-6361/201834199}, \href
  {https://ui.adsabs.harvard.edu/abs/2019A&A...626A..72G} {626, A72}

\bibitem[\protect\citeauthoryear{{Gruen} et~al.,}{{Gruen}
  et~al.}{2016}]{Gruen16troughs}
{Gruen} D.,  et~al., 2016, \mn@doi [\mnras] {10.1093/mnras/stv2506}, \href
  {http://adsabs.harvard.edu/abs/2016MNRAS.455.3367G} {455, 3367}

\bibitem[\protect\citeauthoryear{{Gruen} et~al.,}{{Gruen}
  et~al.}{2018}]{GruenDES17}
{Gruen} D.,  et~al., 2018, \mn@doi [\prd] {10.1103/PhysRevD.98.023507}, \href
  {https://ui.adsabs.harvard.edu/abs/2018PhRvD..98b3507G} {98, 023507}

\bibitem[\protect\citeauthoryear{{Hilbert}, {Hartlap}  \&
  {Schneider}}{{Hilbert} et~al.}{2011}]{Hilbert11}
{Hilbert} S.,  {Hartlap} J.,   {Schneider} P.,  2011, \mn@doi [\aap]
  {10.1051/0004-6361/201117294}, \href
  {http://adsabs.harvard.edu/abs/2011A%26A...536A..85H} {536, A85}

\bibitem[\protect\citeauthoryear{{Hilbert} et~al.,}{{Hilbert}
  et~al.}{2019}]{HilbertRayTracing}
{Hilbert} S.,  et~al., 2019, arXiv e-prints, \href
  {https://ui.adsabs.harvard.edu/abs/2019arXiv191010625H} {p. arXiv:1910.10625}

\bibitem[\protect\citeauthoryear{{Hildebrandt} et~al.,}{{Hildebrandt}
  et~al.}{2018}]{Kidsviking}
{Hildebrandt} H.,  et~al., 2018, arXiv e-prints, \href
  {https://ui.adsabs.harvard.edu/abs/2018arXiv181206076H} {p. arXiv:1812.06076}

\bibitem[\protect\citeauthoryear{{Hurtado-Gil}, {Mart{\'{\i}}nez},
  {Arnalte-Mur}, {Pons-Border{\'{\i}}a}, {Pareja-Flores}  \&
  {Paredes}}{{Hurtado-Gil} et~al.}{2017}]{Hurtado-Gil17}
{Hurtado-Gil} L.,  {Mart{\'{\i}}nez} V.~J.,  {Arnalte-Mur} P.,
  {Pons-Border{\'{\i}}a} M.-J.,  {Pareja-Flores} C.,   {Paredes} S.,  2017,
  \mn@doi [\aap] {10.1051/0004-6361/201629097}, \href
  {http://adsabs.harvard.edu/abs/2017A%26A...601A..40H} {601, A40}

\bibitem[\protect\citeauthoryear{{Ivanov}, {Kaurov}  \& {Sibiryakov}}{{Ivanov}
  et~al.}{2019}]{Ivanov19}
{Ivanov} M.~M.,  {Kaurov} A.~A.,   {Sibiryakov} S.,  2019, \mn@doi [\jcap]
  {10.1088/1475-7516/2019/03/009}, \href
  {https://ui.adsabs.harvard.edu/abs/2019JCAP...03..009I} {2019, 009}

\bibitem[\protect\citeauthoryear{{Ivezi{\'c}} et~al.,}{{Ivezi{\'c}}
  et~al.}{2019}]{LSST}
{Ivezi{\'c}} {\v{Z}}.,  et~al., 2019, \mn@doi [\apj]
  {10.3847/1538-4357/ab042c}, \href
  {https://ui.adsabs.harvard.edu/abs/2019ApJ...873..111I} {873, 111}

\bibitem[\protect\citeauthoryear{Jones \& Williams}{Jones \&
  Williams}{2017}]{sphere}
Jones B.~D.,  Williams J.~R.,  2017, \mn@doi [Engineering Computations]
  {10.1108/EC-02-2016-0052}, 34, 1204

\bibitem[\protect\citeauthoryear{{Kacprzak} et~al.,}{{Kacprzak}
  et~al.}{2016}]{KacprzakDES16}
{Kacprzak} T.,  et~al., 2016, \mn@doi [\mnras] {10.1093/mnras/stw2070}, \href
  {https://ui.adsabs.harvard.edu/abs/2016MNRAS.463.3653K} {463, 3653}

\bibitem[\protect\citeauthoryear{{Kaiser} \& {Squires}}{{Kaiser} \&
  {Squires}}{1993}]{KaiserSquires}
{Kaiser} N.,  {Squires} G.,  1993, \mn@doi [\apj] {10.1086/172297}, \href
  {https://ui.adsabs.harvard.edu/abs/1993ApJ...404..441K} {404, 441}

\bibitem[\protect\citeauthoryear{{Kayo}, {Taruya}  \& {Suto}}{{Kayo}
  et~al.}{2001}]{Kayo01}
{Kayo} I.,  {Taruya} A.,   {Suto} Y.,  2001, \mn@doi [\apj] {10.1086/323227},
  \href {http://adsabs.harvard.edu/abs/2001ApJ...561...22K} {561, 22}

\bibitem[\protect\citeauthoryear{{Kilbinger}}{{Kilbinger}}{2015}]{kilbinger15}
{Kilbinger} M.,  2015, \mn@doi [Reports on Progress in Physics]
  {10.1088/0034-4885/78/8/086901}, \href
  {https://ui.adsabs.harvard.edu/abs/2015RPPh...78h6901K} {78, 086901}

\bibitem[\protect\citeauthoryear{{Klypin}, {Prada}, {Betancort-Rijo}  \&
  {Albareti}}{{Klypin} et~al.}{2018}]{2018MNRAS.481.4588K}
{Klypin} A.,  {Prada} F.,  {Betancort-Rijo} J.,   {Albareti} F.~D.,  2018,
  \mn@doi [\mnras] {10.1093/mnras/sty2613}, \href
  {https://ui.adsabs.harvard.edu/abs/2018MNRAS.481.4588K} {481, 4588}

\bibitem[\protect\citeauthoryear{{Krause}, {Chang}, {Dor{\'e}}  \&
  {Umetsu}}{{Krause} et~al.}{2013}]{voids}
{Krause} E.,  {Chang} T.-C.,  {Dor{\'e}} O.,   {Umetsu} K.,  2013, \mn@doi
  [\apjl] {10.1088/2041-8205/762/2/L20}, \href
  {https://ui.adsabs.harvard.edu/abs/2013ApJ...762L..20K} {762, L20}

\bibitem[\protect\citeauthoryear{{Laureijs} et~al.,}{{Laureijs}
  et~al.}{2011}]{Euclid}
{Laureijs} R.,  et~al., 2011, arXiv e-prints, \href
  {https://ui.adsabs.harvard.edu/abs/2011arXiv1110.3193L} {p. arXiv:1110.3193}

\bibitem[\protect\citeauthoryear{{Leicht}, {Uhlemann}, {Villaescusa-Navarro},
  {Codis}, {Hernquist}  \& {Genel}}{{Leicht} et~al.}{2019}]{Leicht19}
{Leicht} O.,  {Uhlemann} C.,  {Villaescusa-Navarro} F.,  {Codis} S.,
  {Hernquist} L.,   {Genel} S.,  2019, \mn@doi [\mnras]
  {10.1093/mnras/sty3469}, \href
  {https://ui.adsabs.harvard.edu/abs/2019MNRAS.484..269L} {484, 269}

\bibitem[\protect\citeauthoryear{Lewis \& Bridle}{Lewis \& Bridle}{2002}]{camb}
Lewis A.,  Bridle S.,  2002, \mn@doi [\prd] {10.1103/PhysRevD.66.103511}, 66,
  103511

\bibitem[\protect\citeauthoryear{{Liu} \& {Madhavacheril}}{{Liu} \&
  {Madhavacheril}}{2019}]{Liu19}
{Liu} J.,  {Madhavacheril} M.~S.,  2019, \mn@doi [\prd]
  {10.1103/PhysRevD.99.083508}, \href
  {https://ui.adsabs.harvard.edu/abs/2019PhRvD..99h3508L} {99, 083508}

\bibitem[\protect\citeauthoryear{{Liu}, {Hill}, {Sherwin}, {Petri}, {B{\"o}hm}
  \& {Haiman}}{{Liu} et~al.}{2016}]{Jia2016}
{Liu} J.,  {Hill} J.~C.,  {Sherwin} B.~D.,  {Petri} A.,  {B{\"o}hm} V.,
  {Haiman} Z.,  2016, \mn@doi [\prd] {10.1103/PhysRevD.94.103501}, \href
  {https://ui.adsabs.harvard.edu/abs/2016PhRvD..94j3501L} {94, 103501}

\bibitem[\protect\citeauthoryear{Mellier}{Mellier}{1999}]{kappadef}
Mellier Y.,  1999, \mn@doi [Annual Review of Astronomy and Astrophysics]
  {10.1146/annurev.astro.37.1.127}, 37, 127

\bibitem[\protect\citeauthoryear{{Miyazaki} et~al.,}{{Miyazaki}
  et~al.}{2012}]{HSC}
{Miyazaki} S.,  et~al., 2012, in \procspie. p. 84460Z,
  \mn@doi{10.1117/12.926844}

\bibitem[\protect\citeauthoryear{{Munshi} \& {Jain}}{{Munshi} \&
  {Jain}}{2000}]{Munshi00}
{Munshi} D.,  {Jain} B.,  2000, \mn@doi [\mnras]
  {10.1046/j.1365-8711.2000.03694.x}, \href
  {http://adsabs.harvard.edu/abs/2000MNRAS.318..109M} {318, 109}

\bibitem[\protect\citeauthoryear{{Munshi}, {Coles}  \& {Kilbinger}}{{Munshi}
  et~al.}{2014a}]{Munshi14}
{Munshi} D.,  {Coles} P.,   {Kilbinger} M.,  2014a, \mn@doi [\jcap]
  {10.1088/1475-7516/2014/04/004}, \href
  {http://adsabs.harvard.edu/abs/2014JCAP...04..004M} {4, 004}

\bibitem[\protect\citeauthoryear{{Munshi}, {Joudaki}, {Coles}, {Smidt}  \&
  {Kay}}{{Munshi} et~al.}{2014b}]{Munshi14SZ+WL}
{Munshi} D.,  {Joudaki} S.,  {Coles} P.,  {Smidt} J.,   {Kay} S.~T.,  2014b,
  \mn@doi [\mnras] {10.1093/mnras/stu794}, \href
  {https://ui.adsabs.harvard.edu/abs/2014MNRAS.442...69M} {442, 69}

\bibitem[\protect\citeauthoryear{{Pajer} \& {van der Woude}}{{Pajer} \& {van
  der Woude}}{2018}]{vanderWoude017}
{Pajer} E.,  {van der Woude} D.,  2018, \mn@doi [\jcap]
  {10.1088/1475-7516/2018/05/039}, \href
  {https://ui.adsabs.harvard.edu/abs/2018JCAP...05..039P} {2018, 039}

\bibitem[\protect\citeauthoryear{Patton, Blazek, Honscheid, Huff, Melchior,
  Ross  \& Suchyta}{Patton et~al.}{2017}]{Patton17}
Patton K.,  Blazek J.,  Honscheid K.,  Huff E.,  Melchior P.,  Ross A.~J.,
  Suchyta E.,  2017, \mn@doi [Monthly Notices of the Royal Astronomical
  Society] {10.1093/mnras/stx1626}, 472, 439

\bibitem[\protect\citeauthoryear{{Peebles}}{{Peebles}}{1980}]{Peebles}
{Peebles} P.~J.~E.,  1980, {The large-scale structure of the universe}

\bibitem[\protect\citeauthoryear{{Peel}, {Pettorino}, {Giocoli}, {Starck}  \&
  {Baldi}}{{Peel} et~al.}{2018}]{Peel18}
{Peel} A.,  {Pettorino} V.,  {Giocoli} C.,  {Starck} J.-L.,   {Baldi} M.,
  2018, \mn@doi [\aap] {10.1051/0004-6361/201833481}, \href
  {https://ui.adsabs.harvard.edu/abs/2018A&A...619A..38P} {619, A38}

\bibitem[\protect\citeauthoryear{{Petri}, {Haiman}, {Hui}, {May}  \&
  {Kratochvil}}{{Petri} et~al.}{2013}]{Petri13}
{Petri} A.,  {Haiman} Z.,  {Hui} L.,  {May} M.,   {Kratochvil} J.~M.,  2013,
  \mn@doi [\prd] {10.1103/PhysRevD.88.123002}, \href
  {https://ui.adsabs.harvard.edu/abs/2013PhRvD..88l3002P} {88, 123002}

\bibitem[\protect\citeauthoryear{{Petri}, {May}  \& {Haiman}}{{Petri}
  et~al.}{2016}]{Petri16cos}
{Petri} A.,  {May} M.,   {Haiman} Z.,  2016, \mn@doi [\prd]
  {10.1103/PhysRevD.94.063534}, \href
  {https://ui.adsabs.harvard.edu/abs/2016PhRvD..94f3534P} {94, 063534}

\bibitem[\protect\citeauthoryear{{Petri}, {Haiman}  \& {May}}{{Petri}
  et~al.}{2017}]{Born}
{Petri} A.,  {Haiman} Z.,   {May} M.,  2017, \mn@doi [\prd]
  {10.1103/PhysRevD.95.123503}, \href
  {https://ui.adsabs.harvard.edu/abs/2017PhRvD..95l3503P} {95, 123503}

\bibitem[\protect\citeauthoryear{{Pichon}, {Thi{\'e}baut}, {Prunet}, {Benabed},
  {Colombi}, {Sousbie}  \& {Teyssier}}{{Pichon} et~al.}{2010}]{aski}
{Pichon} C.,  {Thi{\'e}baut} E.,  {Prunet} S.,  {Benabed} K.,  {Colombi} S.,
  {Sousbie} T.,   {Teyssier} R.,  2010, \mn@doi [\mnras]
  {10.1111/j.1365-2966.2009.15609.x}, \href
  {https://ui.adsabs.harvard.edu/abs/2010MNRAS.401..705P} {401, 705}

\bibitem[\protect\citeauthoryear{{Planck Collaboration} et~al.,}{{Planck
  Collaboration} et~al.}{2016}]{planck}
{Planck Collaboration} et~al., 2016, \mn@doi [\aap]
  {10.1051/0004-6361/201527101}, \href
  {https://ui.adsabs.harvard.edu/abs/2016A&A...594A...1P} {594, A1}

\bibitem[\protect\citeauthoryear{{Pourtsidou} \& {Metcalf}}{{Pourtsidou} \&
  {Metcalf}}{2015}]{Pourtsidou15}
{Pourtsidou} A.,  {Metcalf} R.~B.,  2015, \mn@doi [\mnras]
  {10.1093/mnras/stv102}, \href
  {https://ui.adsabs.harvard.edu/abs/2015MNRAS.448.2368P} {448, 2368}

\bibitem[\protect\citeauthoryear{{Reimberg} \& {Bernardeau}}{{Reimberg} \&
  {Bernardeau}}{2018}]{paolo}
{Reimberg} P.,  {Bernardeau} F.,  2018, \mn@doi [\prd]
  {10.1103/PhysRevD.97.023524}, \href
  {https://ui.adsabs.harvard.edu/abs/2018PhRvD..97b3524R} {97, 023524}

\bibitem[\protect\citeauthoryear{{Repp} \& {Szapudi}}{{Repp} \&
  {Szapudi}}{2018}]{Repp18}
{Repp} A.,  {Szapudi} I.,  2018, \mn@doi [\mnras] {10.1093/mnras/stx2615},
  \href {https://ui.adsabs.harvard.edu/abs/2018MNRAS.473.3598R} {473, 3598}

\bibitem[\protect\citeauthoryear{{Rizzato}, {Benabed}, {Bernardeau}  \&
  {Lacasa}}{{Rizzato} et~al.}{2018}]{matteo}
{Rizzato} M.,  {Benabed} K.,  {Bernardeau} F.,   {Lacasa} F.,  2018, arXiv
  e-prints, \href {https://ui.adsabs.harvard.edu/abs/2018arXiv181207437R} {p.
  arXiv:1812.07437}

\bibitem[\protect\citeauthoryear{{Schaan}, {Krause}, {Eifler}, {Dor{\'e}},
  {Miyatake}, {Rhodes}  \& {Spergel}}{{Schaan} et~al.}{2017}]{Schaan17}
{Schaan} E.,  {Krause} E.,  {Eifler} T.,  {Dor{\'e}} O.,  {Miyatake} H.,
  {Rhodes} J.,   {Spergel} D.~N.,  2017, \mn@doi [\prd]
  {10.1103/PhysRevD.95.123512}, \href
  {https://ui.adsabs.harvard.edu/abs/2017PhRvD..95l3512S} {95, 123512}

\bibitem[\protect\citeauthoryear{{Schaan}, {Ferraro}  \& {Spergel}}{{Schaan}
  et~al.}{2018}]{Schaan18}
{Schaan} E.,  {Ferraro} S.,   {Spergel} D.~N.,  2018, \mn@doi [\prd]
  {10.1103/PhysRevD.97.123539}, \href
  {https://ui.adsabs.harvard.edu/abs/2018PhRvD..97l3539S} {97, 123539}

\bibitem[\protect\citeauthoryear{{Schneider}, {Teyssier}, {Stadel}, {Chisari},
  {Le Brun}, {Amara}  \& {Refregier}}{{Schneider}
  et~al.}{2019}]{2019JCAP...03..020S}
{Schneider} A.,  {Teyssier} R.,  {Stadel} J.,  {Chisari} N.~E.,  {Le Brun} A.
  M.~C.,  {Amara} A.,   {Refregier} A.,  2019, \mn@doi [\jcap]
  {10.1088/1475-7516/2019/03/020}, \href
  {https://ui.adsabs.harvard.edu/abs/2019JCAP...03..020S} {2019, 020}

\bibitem[\protect\citeauthoryear{{Shin}, {Kim}, {Pichon}, {Jeong}  \&
  {Park}}{{Shin} et~al.}{2017}]{Shin17}
{Shin} J.,  {Kim} J.,  {Pichon} C.,  {Jeong} D.,   {Park} C.,  2017, \mn@doi
  [\apj] {10.3847/1538-4357/aa74b9}, \href
  {https://ui.adsabs.harvard.edu/abs/2017ApJ...843...73S} {843, 73}

\bibitem[\protect\citeauthoryear{{Takahashi}, {Sato}, {Nishimichi}, {Taruya}
  \& {Oguri}}{{Takahashi} et~al.}{2012}]{Halofit}
{Takahashi} R.,  {Sato} M.,  {Nishimichi} T.,  {Taruya} A.,   {Oguri} M.,
  2012, \mn@doi [\apj] {10.1088/0004-637X/761/2/152}, \href
  {https://ui.adsabs.harvard.edu/abs/2012ApJ...761..152T} {761, 152}

\bibitem[\protect\citeauthoryear{{Takahashi}, {Hamana}, {Shirasaki},
  {Namikawa}, {Nishimichi}, {Osato}  \& {Shiroyama}}{{Takahashi}
  et~al.}{2017}]{Simulation}
{Takahashi} R.,  {Hamana} T.,  {Shirasaki} M.,  {Namikawa} T.,  {Nishimichi}
  T.,  {Osato} K.,   {Shiroyama} K.,  2017, \mn@doi [\apj]
  {10.3847/1538-4357/aa943d}, 850, 24

\bibitem[\protect\citeauthoryear{{Taruya}, {Takada}, {Hamana}, {Kayo}  \&
  {Futamase}}{{Taruya} et~al.}{2002}]{Taruya02}
{Taruya} A.,  {Takada} M.,  {Hamana} T.,  {Kayo} I.,   {Futamase} T.,  2002,
  \mn@doi [\apj] {10.1086/340048}, \href
  {http://adsabs.harvard.edu/abs/2002ApJ...571..638T} {571, 638}

\bibitem[\protect\citeauthoryear{{Taylor}, {Bernardeau}  \&
  {Kitching}}{{Taylor} et~al.}{2018}]{k-cut}
{Taylor} P.~L.,  {Bernardeau} F.,   {Kitching} T.~D.,  2018, \mn@doi [\prd]
  {10.1103/PhysRevD.98.083514}, \href
  {https://ui.adsabs.harvard.edu/abs/2018PhRvD..98h3514T} {98, 083514}

\bibitem[\protect\citeauthoryear{{The Dark Energy Survey Collaboration}}{{The
  Dark Energy Survey Collaboration}}{2005}]{DES}
{The Dark Energy Survey Collaboration} 2005, arXiv e-prints, \href
  {https://ui.adsabs.harvard.edu/abs/2005astro.ph.10346T} {pp
  astro--ph/0510346}

\bibitem[\protect\citeauthoryear{{Thiele}, {Hill}  \& {Smith}}{{Thiele}
  et~al.}{2018}]{Thiele18}
{Thiele} L.,  {Hill} J.~C.,   {Smith} K.~M.,  2018, arXiv e-prints, \href
  {https://ui.adsabs.harvard.edu/abs/2018arXiv181205584T} {p. arXiv:1812.05584}

\bibitem[\protect\citeauthoryear{{Touchette}}{{Touchette}}{2011}]{touchette}
{Touchette} H.,  2011, arXiv e-prints, \href
  {https://ui.adsabs.harvard.edu/abs/2011arXiv1106.4146T} {p. arXiv:1106.4146}

\bibitem[\protect\citeauthoryear{{Uhlemann}, {Codis}, {Pichon}, {Bernardeau}
  \& {Reimberg}}{{Uhlemann} et~al.}{2016}]{saddle}
{Uhlemann} C.,  {Codis} S.,  {Pichon} C.,  {Bernardeau} F.,   {Reimberg} P.,
  2016, \mn@doi [\mnras] {10.1093/mnras/stw1074}, \href
  {https://ui.adsabs.harvard.edu/abs/2016MNRAS.460.1529U} {460, 1529}

\bibitem[\protect\citeauthoryear{{Uhlemann} et~al.,}{{Uhlemann}
  et~al.}{2018a}]{bias}
{Uhlemann} C.,  et~al., 2018a, \mn@doi [\mnras] {10.1093/mnras/stx2616}, \href
  {https://ui.adsabs.harvard.edu/abs/2018MNRAS.473.5098U} {473, 5098}

\bibitem[\protect\citeauthoryear{{Uhlemann}, {Pajer}, {Pichon}, {Nishimichi},
  {Codis}  \& {Bernardeau}}{{Uhlemann} et~al.}{2018b}]{NonGaussianities}
{Uhlemann} C.,  {Pajer} E.,  {Pichon} C.,  {Nishimichi} T.,  {Codis} S.,
  {Bernardeau} F.,  2018b, \mn@doi [\mnras] {10.1093/mnras/stx2623}, \href
  {https://ui.adsabs.harvard.edu/abs/2018MNRAS.474.2853U} {474, 2853}

\bibitem[\protect\citeauthoryear{{Uhlemann}, {Pichon}, {Codis}, {L'Huillier},
  {Kim}, {Bernardeau}, {Park}  \& {Prunet}}{{Uhlemann}
  et~al.}{2018c}]{cylindres}
{Uhlemann} C.,  {Pichon} C.,  {Codis} S.,  {L'Huillier} B.,  {Kim} J.,
  {Bernardeau} F.,  {Park} C.,   {Prunet} S.,  2018c, \mn@doi [\mnras]
  {10.1093/mnras/sty664}, \href
  {https://ui.adsabs.harvard.edu/abs/2018MNRAS.477.2772U} {477, 2772}

\bibitem[\protect\citeauthoryear{{Valageas}}{{Valageas}}{2000}]{Valageas1}
{Valageas} P.,  2000, \aap, \href
  {https://ui.adsabs.harvard.edu/abs/2000A&A...356..771V} {356, 771}

\bibitem[\protect\citeauthoryear{{Valageas}}{{Valageas}}{2002}]{Valageas}
{Valageas} P.,  2002, \mn@doi [\aap] {10.1051/0004-6361:20011663}, \href
  {https://ui.adsabs.harvard.edu/abs/2002A&A...382..412V} {382, 412}

\bibitem[\protect\citeauthoryear{{Vicinanza}, {Cardone}, {Maoli}, {Scaramella}
  \& {Er}}{{Vicinanza} et~al.}{2018}]{Vicinanza18}
{Vicinanza} M.,  {Cardone} V.~F.,  {Maoli} R.,  {Scaramella} R.,   {Er} X.,
  2018, \mn@doi [\prd] {10.1103/PhysRevD.97.023519}, \href
  {https://ui.adsabs.harvard.edu/abs/2018PhRvD..97b3519V} {97, 023519}

\bibitem[\protect\citeauthoryear{{Weiss}, {Schneider}, {Sgier}, {Kacprzak},
  {Amara}  \& {Refregier}}{{Weiss} et~al.}{2019}]{BaryonEffect}
{Weiss} A.~J.,  {Schneider} A.,  {Sgier} R.,  {Kacprzak} T.,  {Amara} A.,
  {Refregier} A.,  2019, arXiv e-prints, \href
  {https://ui.adsabs.harvard.edu/abs/2019arXiv190511636W} {p. arXiv:1905.11636}

\bibitem[\protect\citeauthoryear{{Xavier}, {Abdalla}  \& {Joachimi}}{{Xavier}
  et~al.}{2016}]{Xavier16}
{Xavier} H.~S.,  {Abdalla} F.~B.,   {Joachimi} B.,  2016, \mn@doi [\mnras]
  {10.1093/mnras/stw874}, \href
  {http://adsabs.harvard.edu/abs/2016MNRAS.459.3693X} {459, 3693}

\bibitem[\protect\citeauthoryear{{Yoon}, {Jee}, {Tyson}, {Schmidt}, {Wittman}
  \& {Choi}}{{Yoon} et~al.}{2019}]{DLS}
{Yoon} M.,  {Jee} M.~J.,  {Tyson} J.~A.,  {Schmidt} S.,  {Wittman} D.,   {Choi}
  A.,  2019, \mn@doi [\apj] {10.3847/1538-4357/aaf3a9}, \href
  {https://ui.adsabs.harvard.edu/abs/2019ApJ...870..111Y} {870, 111}

\bibitem[\protect\citeauthoryear{{de Jong}, {Verdoes Kleijn}, {Kuijken}  \&
  {Valentijn}}{{de Jong} et~al.}{2013}]{kids}
{de Jong} J. T.~A.,  {Verdoes Kleijn} G.~A.,  {Kuijken} K.~H.,   {Valentijn}
  E.~A.,  2013, \mn@doi [Experimental Astronomy] {10.1007/s10686-012-9306-1},
  \href {https://ui.adsabs.harvard.edu/abs/2013ExA....35...25D} {35, 25}

\makeatother
\end{thebibliography}

\appendix

\section{On the top-hat filtering of {\sc healpix} convergence maps} \label{top-hat filter}

To convolve the {\sc healpix} maps with a top-hat window of the desired angular radius, we use the \textit{query\_disc} function of {\sc healpy} to find all pixels whose centres are located within a disk centred at one specific pixel $p$. Then we reassign the value of $p$ as being the mean of all the pixels inside the disk. Though the disk radius is at least 11.6 times bigger than the pixel "size" -- {\sc healpix} pixels all possess the same area but different shapes -- there is still a potential effect of considering the centres of pixels rather than making a weighted mean of all pixels within the disk based on the area actually inside. The goal of this section is to argue that what we did is nonetheless enough to get accurate enough PDFs for the comparison with our theoretical predictions to hold some sense.

\subsection{A formal argument} \label{tophatformal}

Since all pixels have different shapes but the same area, and since we are working with sufficiently small angles so that the small angle approximation applies, we can make the argument that on average everything amounts to considering a regularly-spaced grid on which we are drawing a disk of radius roughly 11.6 times the pixel size. By considering only the centres of pixels, some pixels that are not entirely in the disk are still counted as if it was the case and some that are partially inside are not taken into account since their centres is not inside the disk. The idea is to make an explicit computation to see whether those configurations can compensate each other so that the filtering scheme is still acceptable.

One can easily come up with a formula to compute the number of pixels whose centres lie within a disk of radius $R$ grid unit size,

\begin{equation}
    n_{\rm pix} = 4 \sum_{i=0}^{\lfloor R\rfloor} \lfloor \sqrt{R^2-i^2}\rfloor + 1.
\end{equation}
On the other hand, there is no analytic formula for the weight to actually apply to each pixel but we make use of the simulated result of \cite{sphere} to find the ratio of each pixel inside the disk. Now we compare the sum of all non zero weights to $n_{\rm pix}$, if the two numbers are close then the filtering scheme is appropriate. For the resolution of the maps we use (201,326,592 pixels for a full-sky) and a top-hat filter of radius 10~arcmin, the difference between the sum of all weights and $n_{\rm pix}$ is less that 1 per cent being very acceptable -- the difference with the theoretical predictions is at best of the order of a percent -- with an average of $n_{\rm pix} =$ 421 pixels per disk (sum of all weights is $\simeq$ 425.22).

\subsection{Testing the filtering with higher resolution}

\begin{figure}
    \centering
    \includegraphics[width=\columnwidth]{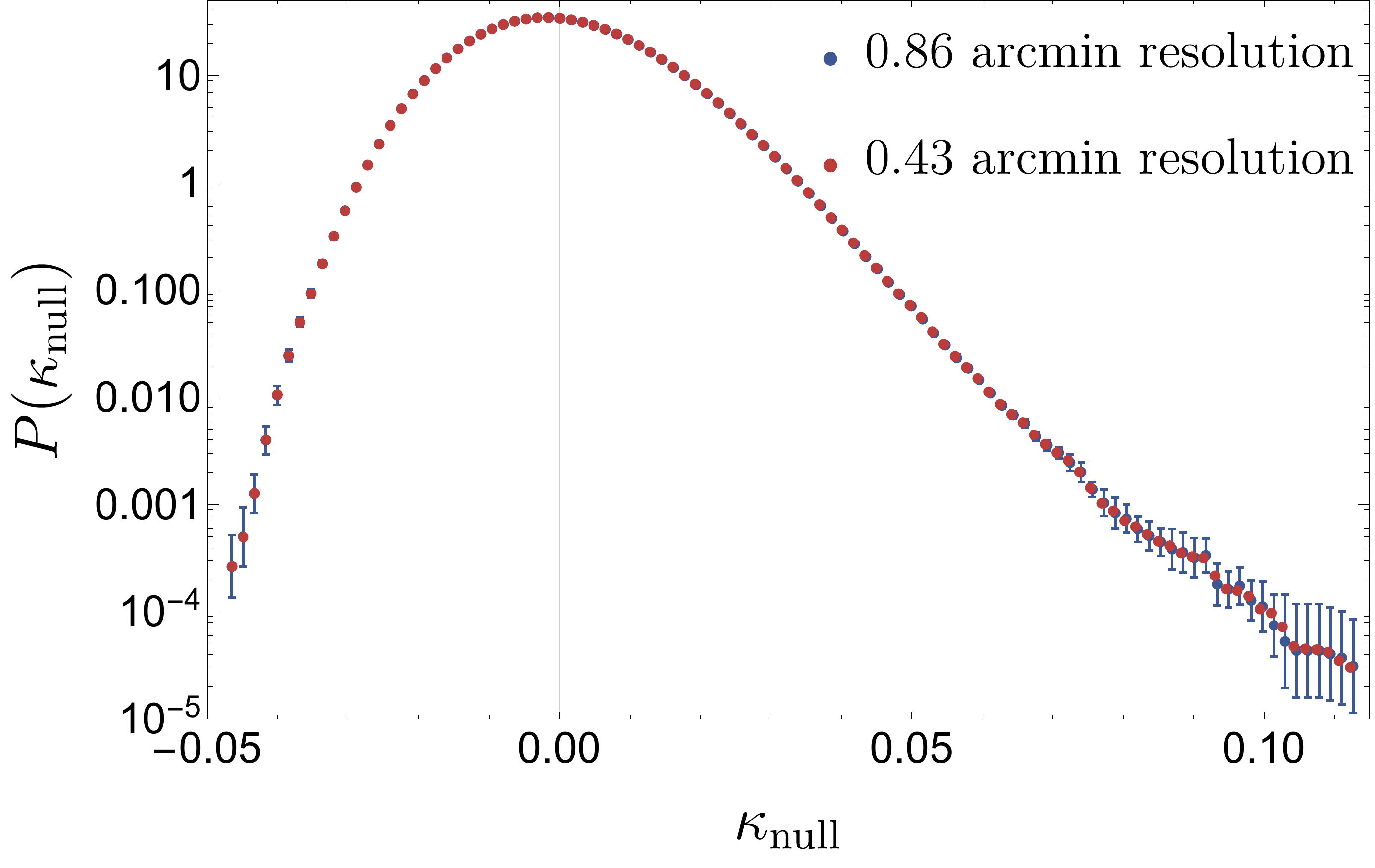}
    \caption{Comparison of our filtering scheme for 2 different resolutions of the same map. The difference completely lies within the measured error bars.}
    \label{top-hat}
\end{figure}

Another way to convince ourselves that the filtering is accurate is to greatly improve the resolution of one map, thus reducing the effect of only considering centres of pixels, and filter it again for comparison. Using the \textit{ud\_grade} function we double the resolution of one map at redshift $z_s = 2.0548$ and filter it again with the \textit{query\_disc} method. Note that for such a resolution, top-hat of radius 10~arcmin and using the previous argument, the difference between $n_p$ and the sum of weights is less than 0.1 per cent. The result is shown in figure~\ref{top-hat} and tends to validate our filtering scheme since no huge differences are displayed.

\section{\red{Error bars' estimate}}
\label{justify}
\red{
In the main text, error bars on the measured PDFs are estimated via the error on the mean amongst eight subvolumes of one simulated full-sky map. As such, for each bin $i$ of convergence in the measured PDF $\hat{\mathcal{P}}_{ i}$, we get an estimate error given by}

\begin{equation}
\red{
    \operatorname{Err}_{i}=\sqrt{\frac{1}{7}\left[\sum_{j=1}^{8}\left(\frac{\hat{\mathcal{P}}_{ j}\left(\kappa_{i}\right)}{N}\right)^{2}-\left(\sum_{j=1}^{8} \frac{\hat{\mathcal{P}}_{ j}\left(\kappa_{i}\right)}{N}\right)^{2}\right]}.}
    \label{error_eq}
\end{equation}

\begin{figure}
    \centering
   \includegraphics[width=\columnwidth]{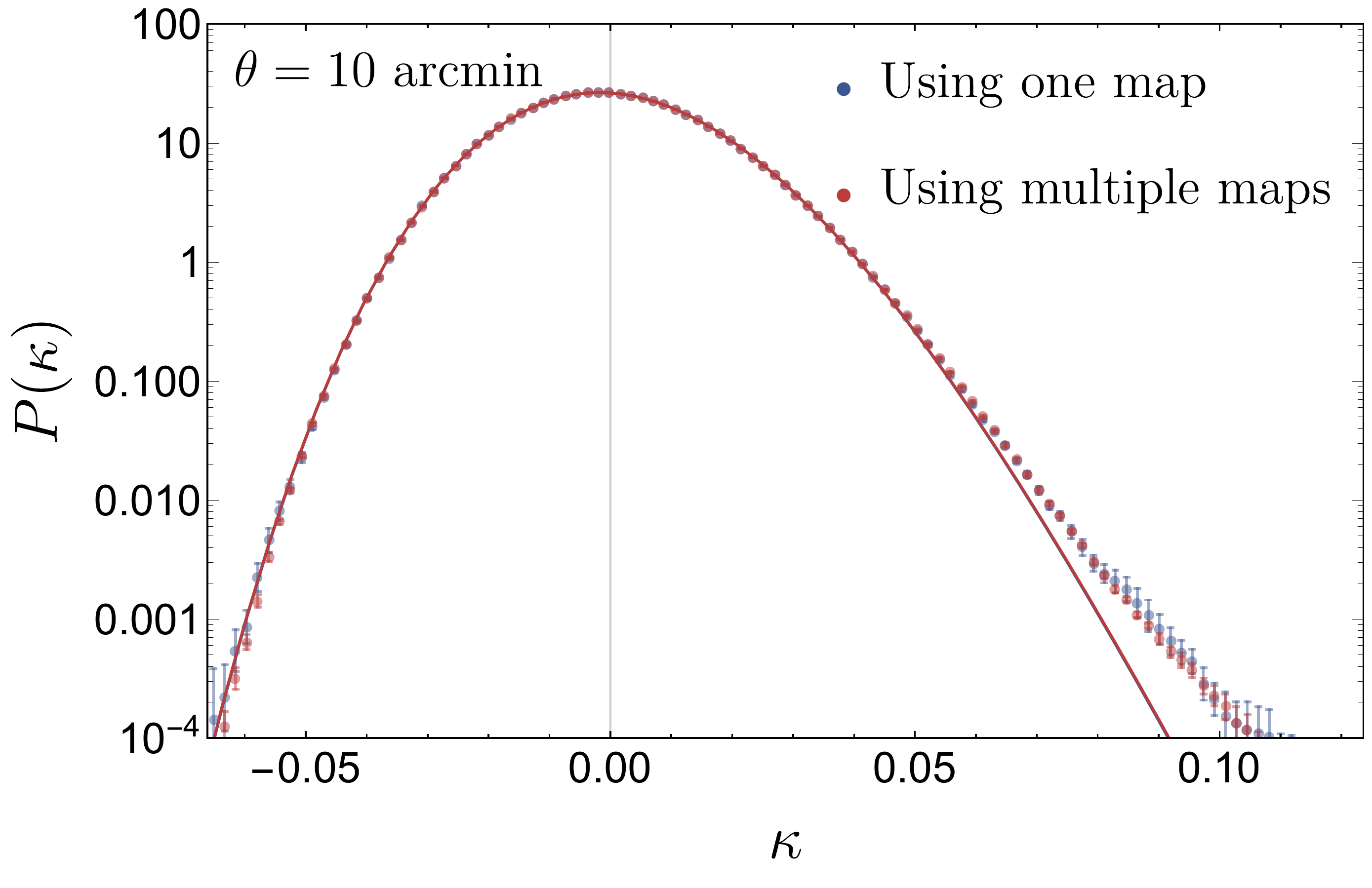}\\
    \includegraphics[width=\columnwidth]{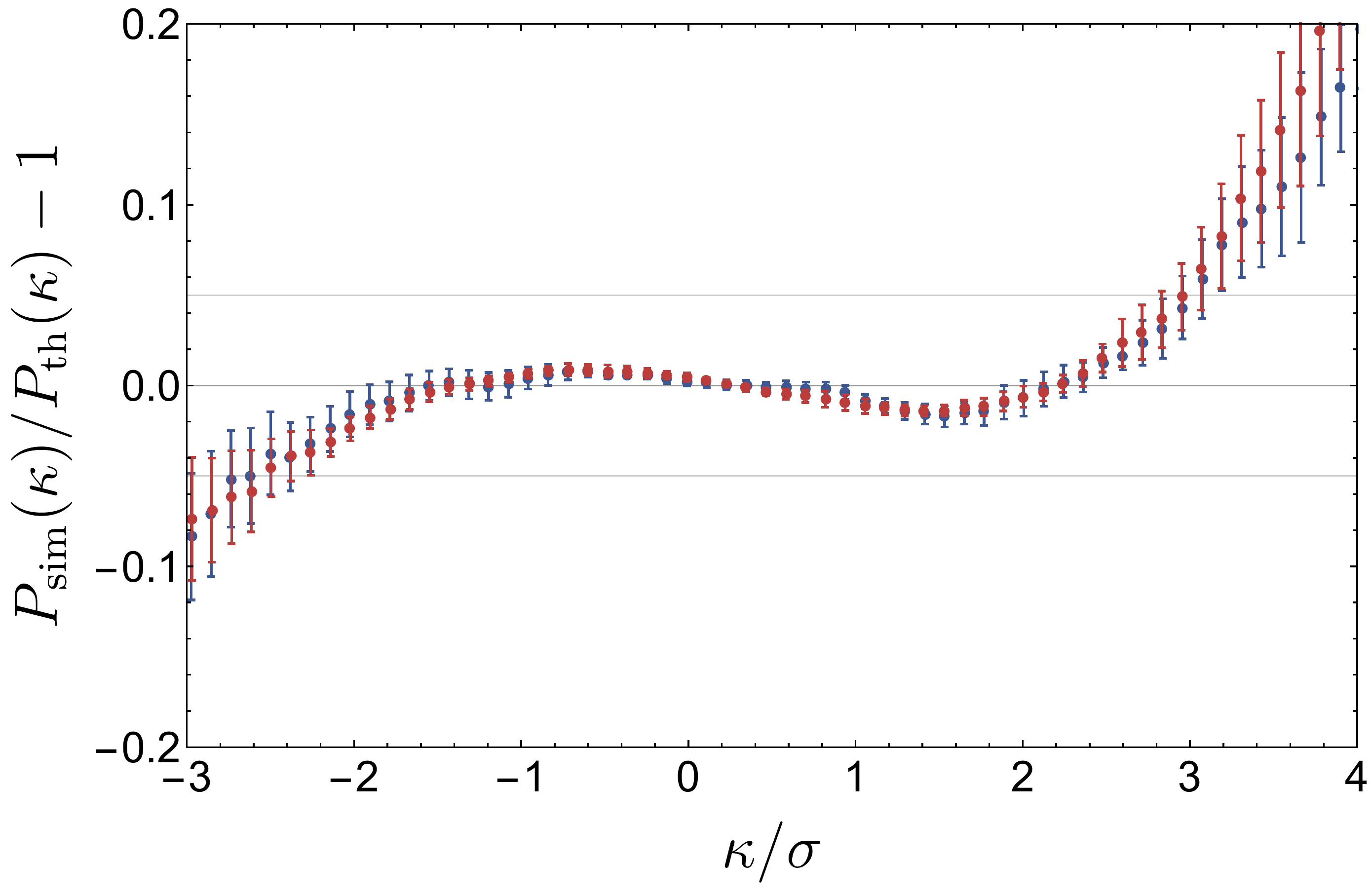}
    \caption{\red{One point PDF of the weak-lensing convergence field for source redshift $z_s = 3.1$ and opening angle $\theta = 10$ arcmin. The measured PDFs are made with two different methods, one using a single simulated map with error bars computed from 8 subvolumes (blue) and one which is computed from the average of 8 realisations of the convergence random field and error bars are computed using the dispersion between each realisation (red). Top panel : PDF in log scale. Bottom panel : residuals of the simulated data compared to the prediction.}}
    \label{errorbars}
\end{figure}

To test the robustness of this estimated error bar, for one source redshift $z_s = 3.1$ and one opening angle $\theta = 10$ arcmin, we measured the PDF both with our method (based on one simulated map and 8 subvolumes) and with 8 simulated maps with error bars computed in a similar way but with the whole set of maps instead of subvolumes. We then fitted the variance in both cases and compared to the theory. The result is depicted in figure~\ref{errorbars} where the top panel shows the resulting PDFs with the method used in this paper (blue) and the method using multiple simulated maps (red). First note that the theory is indistinguishable from one method to the other which is not surprising since using multiple maps only affects the statistical description of the rare events, hence the tails of the PDF which have very little impact on the variance of the overall distribution. It is also noticeable that the error bars themselves are consistent ones with the others, especially when the residuals compared to our theory is plotted. 
Hence, all the conclusions drawn in the main text are robust with regards to how error bars are estimated in the simulations.
Moreover, let us highlight that in order to get a good idea of the cosmic variance PDFs are subject to when one looks at a single realisation of the Universe, the relevant quantity is the dispersion between the various full-sky maps which correspond to different realisations of the Universe. In practice, this multiplies the error bars obtained from (\ref{error_eq}) by a factor $\sqrt{7}$. Moreover, let us emphasise that the question of the numerical convergence of ray tracing simulations is still quite open. This is especially the case for higher order statistics such as PDFs for which the accuracy cannot be trusted below a few percents notably in the tails (see for instance \citep{HilbertRayTracing} for a recent comparison project). Hence, we should not try and reduce artificially the error bars by considering more maps as one would enter a regime where theoretical and numerical errors should be accounted for. This is however not the purpose of this work to investigate the numerical convergence of the simulation. Such a study is deferred to future works.

\section{\red{Gaussian and log-normal approximations}}
\label{app:approx}

\begin{figure}
    \centering
    \includegraphics[width=\columnwidth]{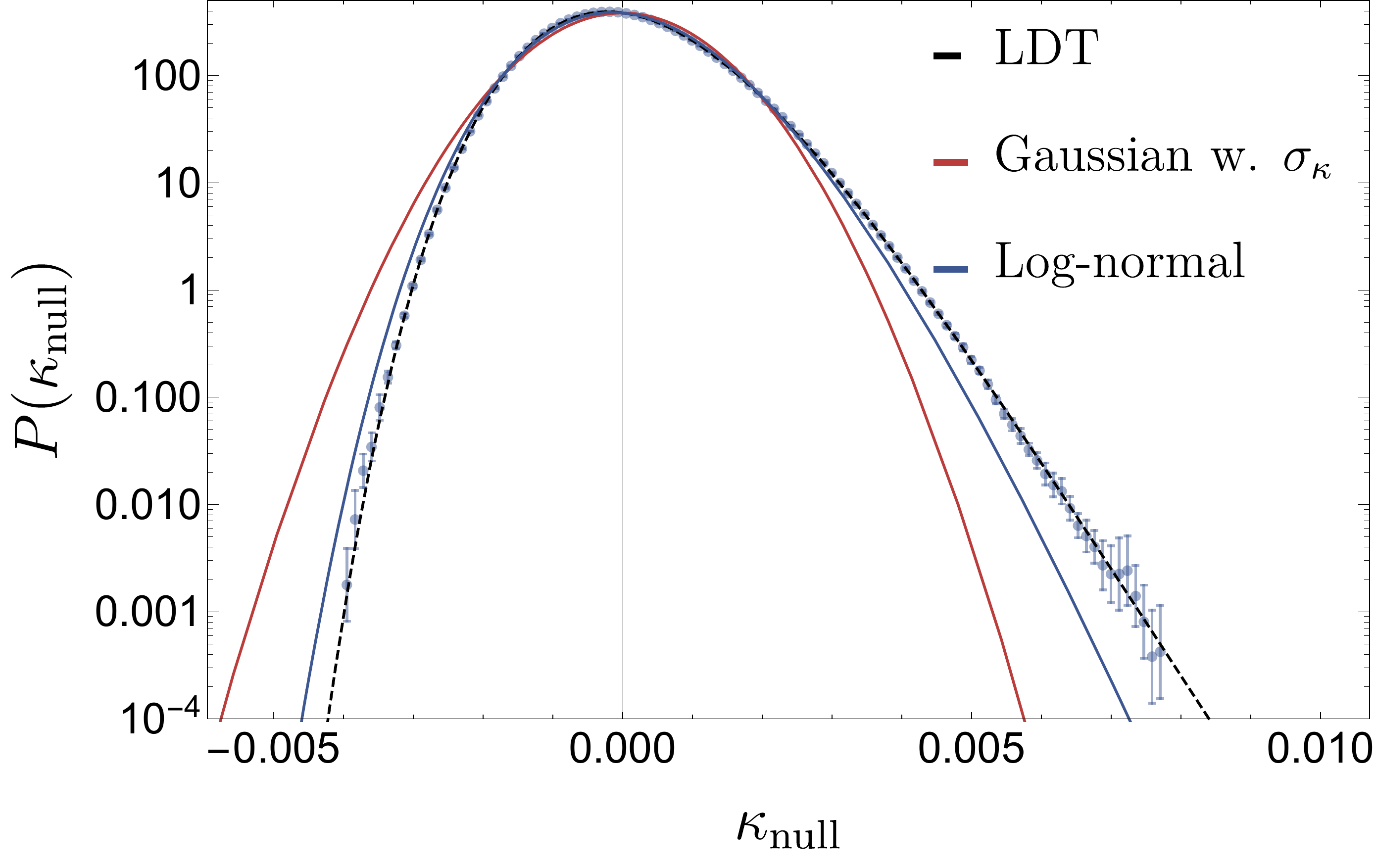}\\
    \includegraphics[width=\columnwidth]{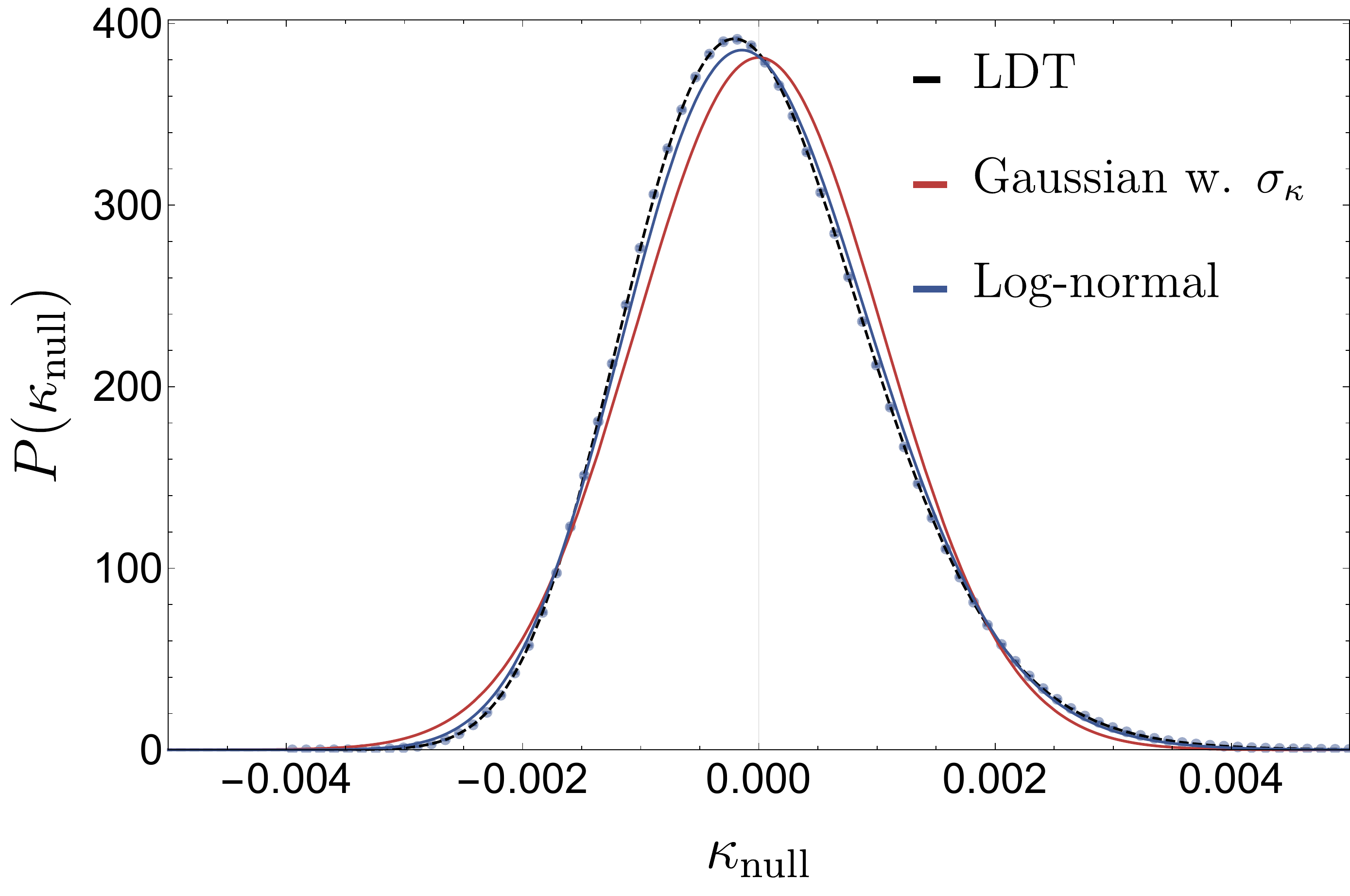}\\
    \includegraphics[width=\columnwidth]{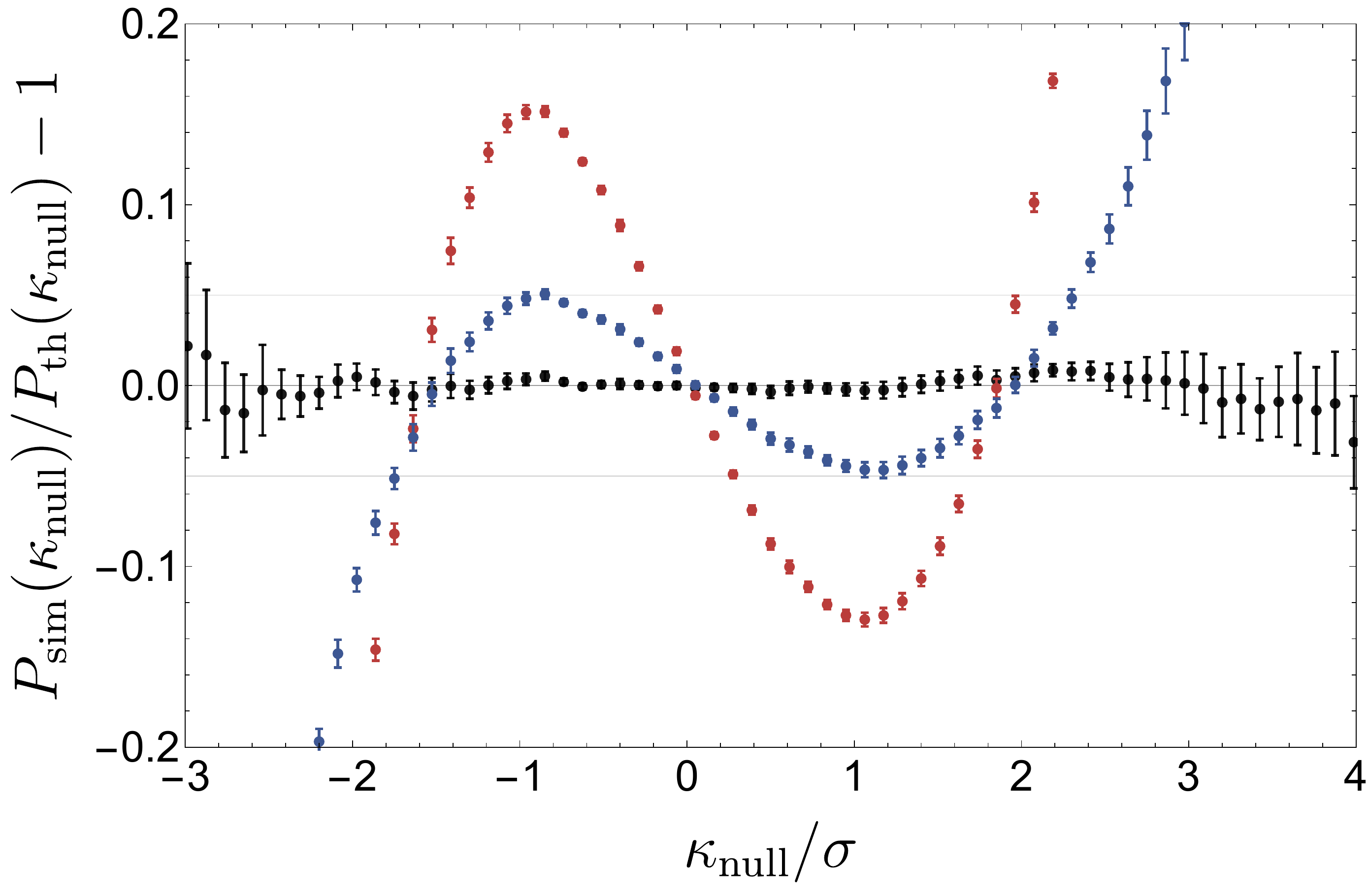}
    \caption{\red{PDF of the nulled convergence $\kappa_{\text{null}}$ for source planes located at redshifts 1.2, 1.4 and 1.6. The opening angle is $\theta$ = 10~arcmin. Different theoretical models are displayed with solid lines and colour-coded as labelled, while measurements are shown with blue error bars. Top panel: PDF in log-scale. Middle: PDF in linear scale. Bottom panel: residuals of the measurements compared to the various theoretical prescriptions.}}
    \label{Gaussplot}
\end{figure}

\red{Let us compare the LDT approach to other commonly used approaches namely a Gaussian with the measured variance and a log-normal approximation. This comparison, although relevant in every cases presented throughout the paper, is exemplified in the case of a nulled convergence map where the LDT approach was shown to make more sense. The conclusions would however be equivalent for any convergence map. We display in figure~\ref{Gaussplot} the comparison between the LDT approach developped in this paper (dashed black line), a Gaussian PDF with the correct variance to explicit the need of high order cumulants (red), the assumption of a lognormal PDF for $\kappa + \kappa_{\rm min}$ (blue), and measurements in the simulation (blue error bars). It is expected that the Gaussian approach does not reproduce the tails of the PDF since it predicts a zero value for the high order cumulants which probe the tails. However, on top of that, because the convergence field is not Gaussian and thus its most probable value is not equal to its mean, it is clear that the Gaussian model povides a poor fit everywhere, not only in the tails but also around the peak. One can moreover note that the theoretical approach based on LDT developed in this article is clearly more accurate than a log-normal approximation (whose skewness is largely off as shown in the residuals).}

\section{Testing the model with Halofit} \label{Halofitsection}

The revisited Halofit model as described in \cite{Halofit} is an accurate fitting model for the non-linear matter power spectrum whose parameters are fitted using state-of-the-art high-resolution N-body simulations. 
This empirical model is useful to our purposes as it enables us to test two things : first the impact of using a non-linear variance in the line-of-sight integration and second, the possibility to reduce -- at least partially -- the free character of the projected non-linear variance, although the Halofit variance is still fitted from a numerical simulation and not a first principle derived parameter. 
We take an example from nulling where the theory is most relevant and test different scenarios.
\begin{figure}
    \centering
    \includegraphics[width=\columnwidth]{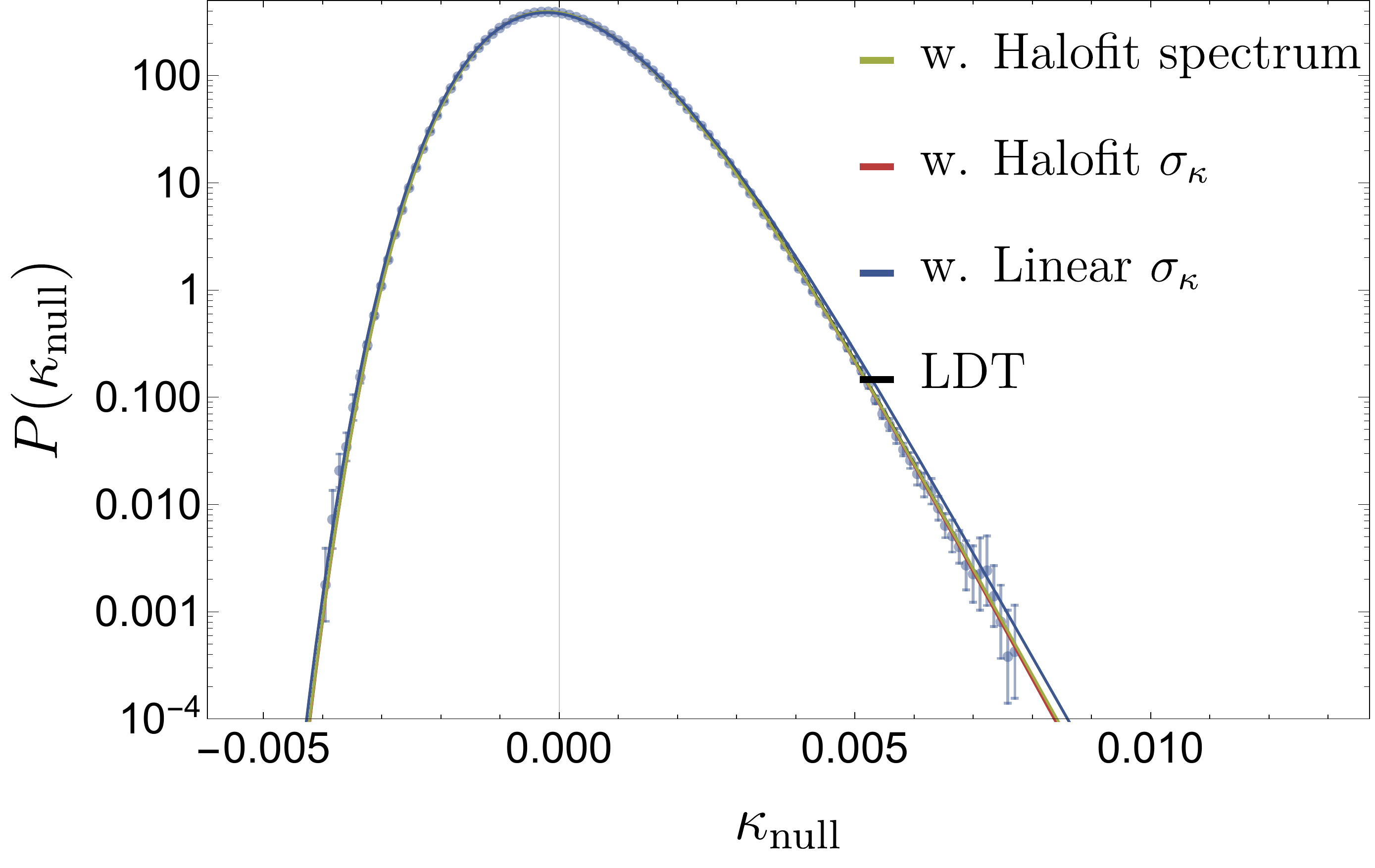}\\
    \includegraphics[width=\columnwidth]{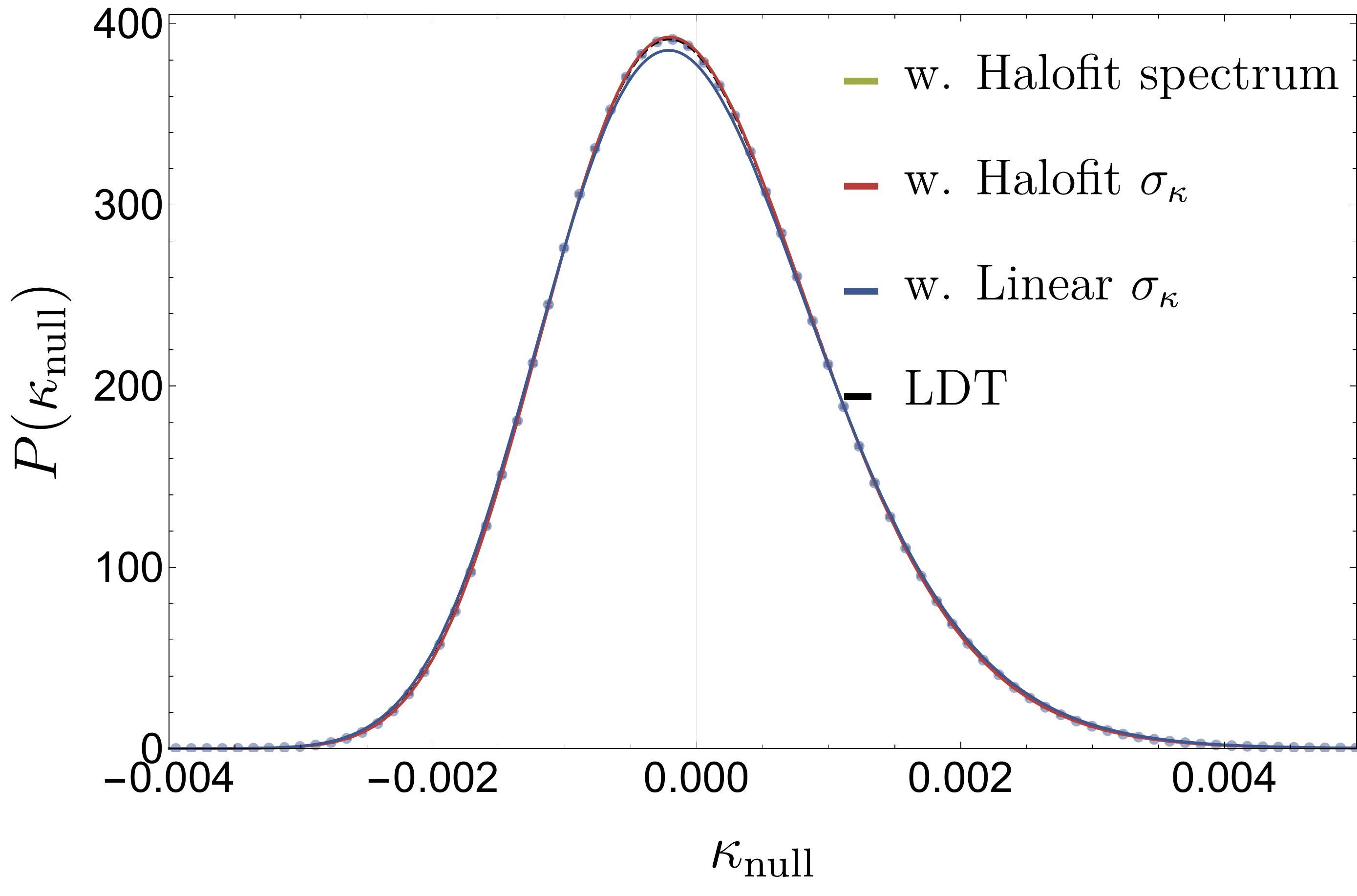}\\
    \includegraphics[width=\columnwidth]{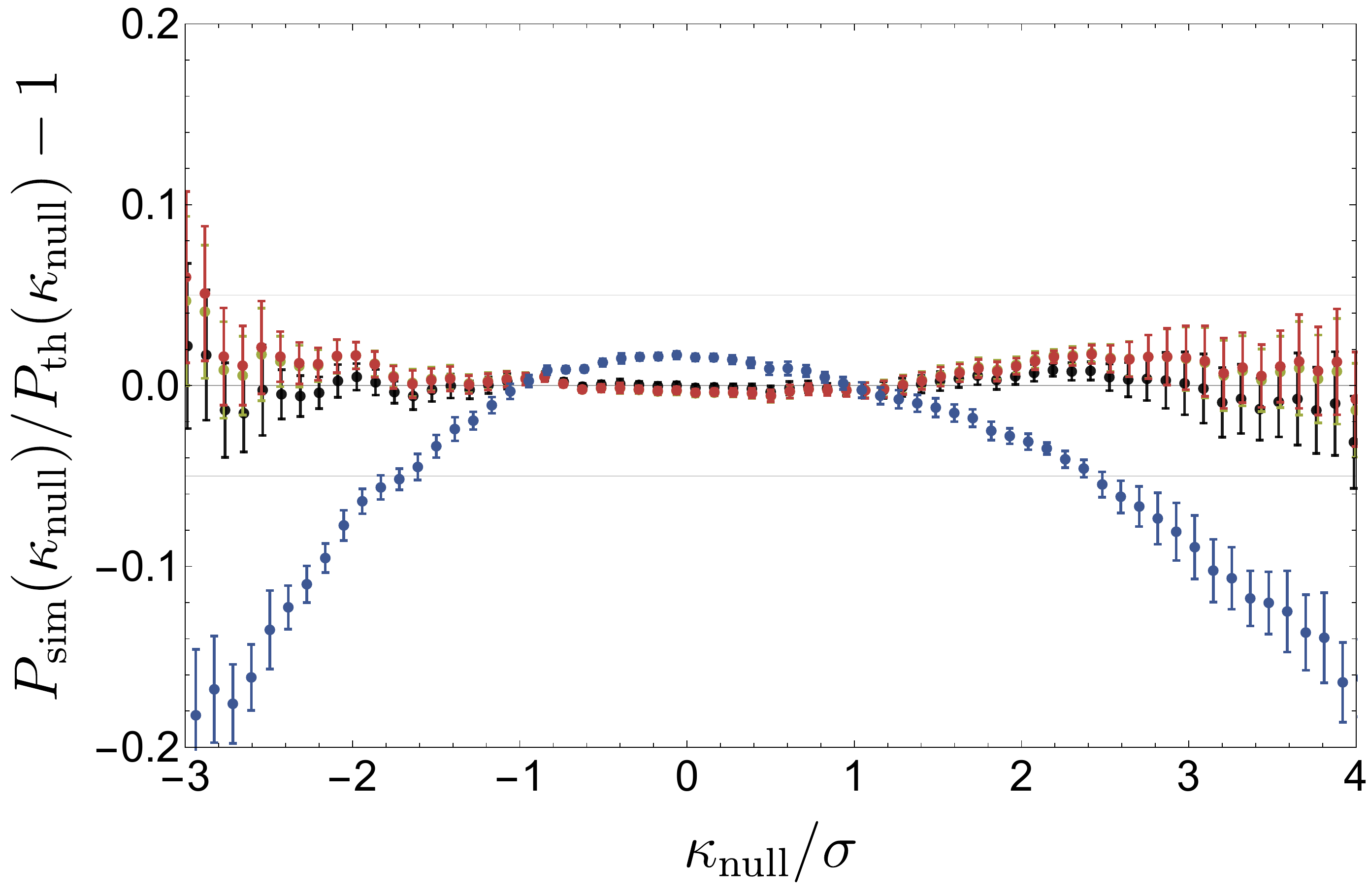}
    \caption{PDF of the nulled convergence $\kappa_{\text{null}}$ for source planes located at redshifts 1.2, 1.4 and 1.6. The opening angle is $\theta$ = 10~arcmin. Different theoretical models are displayed with solid lines and colour-coded as labelled, while measurements are shown with blue error bars. Top panel: PDF in log-scale. Middle: PDF in linear scale. Bottom panel: residuals of the measurements compared to the various theoretical prescriptions.}
    \label{Halofitplot}
\end{figure}

The result is shown in figure~\ref{Halofitplot} where we compare LDT with fitted projected variance which is our approach throughout this work (dashed black), LDT with linear projected variance (blue) i.e when no more free parameter is needed since we use the linear result, LDT with non-linear projected variance computed with Halofit (red), LDT with Halofit power spectrum (green) as opposed to using the linear power spectrum for the integration along the line-of-sight and where the free parameter is thus also the Halofit projected variance.

Note that this is exemplified in the nulling case most relevant for our approach but is still true for the usual line-of-sight integration kernel. 
First note that adopting the Halofit projected variance also reproduces the data to a very good accuracy which is not surprising given that the Halofit model is constructed so as to match the non-linear properties of the matter power spectrum. A more interesting note is that there is a quasi perfect match -- their residuals are hardly distinguishable -- between the Halofit power spectrum approach (green) and the standard theory with the projected Halofit variance taken as a free parameter (red). Again we give a plot in the context of nulling for its relevance but the same thing applies to all plots given in this paper. 
This illustrates the validity of one of the main hypotheses underlying our theoretical approach, which is that one can safely consider the linear power spectrum for each thin slice within the cone as long as the cone itself is re-scaled with the actual non-linear variance (equivalently, cross correlations among different scales are well accounted for by linear theory while the overall amplitude is poorly so and has to be modelled separately   -- e.g with Halofit -- or fitted). 

Lastly, taking the linear variance as our free parameter does not work very well, as expected and in agreement with the previous remark. From the residual plot, it is clear that the main effect is due to a bad modelling of the variance (residual is dominantly quadratic thus pointing towards a bad modelling of the variance).

\section{From perturbations to large deviations} 
\label{app:theory}

Let us recall here some results that lead, from a perturbation theory point of view, to the large deviation theory point of view. As is illustrated in this work, this framework is in our case useful to describe statistics of the cosmic density field as convolved in very symmetric shapes such as spheres or long cylinders which are seen as 2D spheres. 

First let us denote $\rho = \zeta(\tau)$ the non-linear transformation that relates the final normalised density $\rho$ in a $D$-dimensional sphere ($D = 2, 3$) of radius $R$ to the initial density fluctuation $\tau$ in a sphere of radius $r$ determined by mass conservation. 
It has been shown in \cite{Bernardeau1992} and \cite{Bernardeau1995} that this so-called spherical collapse mapping obeys the differential equation
\begin{equation}
    -\zeta \tau^{2} \zeta^{\prime \prime}+c\left(\tau \zeta^{\prime}\right)^{2}-\frac{3}{2} \zeta \tau \zeta^{\prime}+\frac{3}{2} \zeta^{2}\left(\zeta-1\right)=0
    \label{MasterEq}
\end{equation}
where $c = 3/2, 4/3$ respectively for the 2D and 3D case. The particular beauty and generality of this equation is that one has
\begin{equation}
    \zeta(\tau) \simeq 1+\tau+\mathcal{O}\left(\tau^{2}\right)
\end{equation}
where the successive orders allow us to predict the cumulants of the field after spherical collapse, which appeared to correspond exactly to the tree order cumulants of the unsmoothed\footnote{Unsmoothed as in the field is not convolved by any window.} non-linear density field, thus underlying the fact that the dynamics of density fluctuations in our Universe is on average given by the spherical collapse (at least as long as the one-point statistics of the field is concerned).
Strictly speaking, equation~(\ref{MasterEq}) in only valid for an Einstein-de Sitter background but is a very good approximation for a general cosmological background. An explicit fit for $\zeta$ can be found and is given by
\begin{equation}
    \zeta(\tau)=(1-\tau / \nu)^{-\nu}
    \label{collapse2}
\end{equation}
where $\nu$ depends on the dimension and cosmology. In 3D, $\nu_{\text{3D}} = 21/13$ provides for an Einstein-de Sitter Universe a very good description for the dynamics and  reproduces exactly the high-$z$ skewness, for instance in the unsmoothed case: $S_3^{\text{3D}} = 3(1+1/\nu_{\text{3D}}) = 34/7$. 
Similarly in the 2D case, in order to get $S_3^{\text{2D}} = 3(1+1/\nu_{\text{2D}}) = 36/7$ one shall fix $\nu_{\text{2D}} = 1.4$. 
With the help of a Steepest-descent method, \cite{Valageas} showed that the most likely dynamics (amongst all possible mappings between the initial and final density fields) is the one respecting the symmetry. Thus we can make use of the spherical collapse solution \ref{collapse2}.

Finally, we note that the large deviation principle defined in equation~(\ref{LDP}) is generally satisfied by the field variable $\tau$ especially if it is Gaussian distributed. From there the most probable dynamics -- spherical collapse -- to obtain the final densities can be formulated in terms of the contraction principle shown in equation~(\ref{contraction}). Once stated in this language, one can make use of Varadhan's theorem given in equation~(\ref{varadhan}) and thus express the SCGF of our density field convolved in spheres or long cylinders.

For our purposes getting the tree order unsmoothed cumulants is enough since the filtering effects in the large deviation formalism are taken into account via scale-dependent terms expressed in terms of the linear variance at various scales and not via the parametrisation of the spherical collapse. However, building on what we briefly explained, \cite{Bernardeau1994} also showed how to include filtering effects to the cumulants one can get order by order. Let us recap some of those results for a top-hat window of radius $R$ and where $\gamma_p$ stands for ${\rm d^p}\log(\sigma^2(R))/{\rm d}\log(R)^p$. In 3D, the skewness and kurtosis at tree-order are
\begin{equation}
\begin{aligned}
S_3^{3D} &= \frac{34}7 + \gamma_1, \\
S_4^{3D} &= \frac{60712}{1323} + \frac{62}3 \, \gamma_1 + \frac{7}3 \, \gamma_1^2 + \frac{2}3 \, \gamma_2, 
\end{aligned}
\end{equation}
while in 2D, 
\begin{equation}
\begin{aligned}
S_3^{2D} &= \frac{36}7 + 3/2 \, \gamma_1, \\
S_4^{2D} &=\frac{2540}{49} + 33 \, \gamma_1 + \frac{21}{4} \, \gamma_1^2 + \gamma_2 .
\end{aligned}
\end{equation}

\section{Analytical shortcut} \label{analytical_shortcut}
\begin{figure}
    \centering
    \includegraphics[width=\columnwidth]{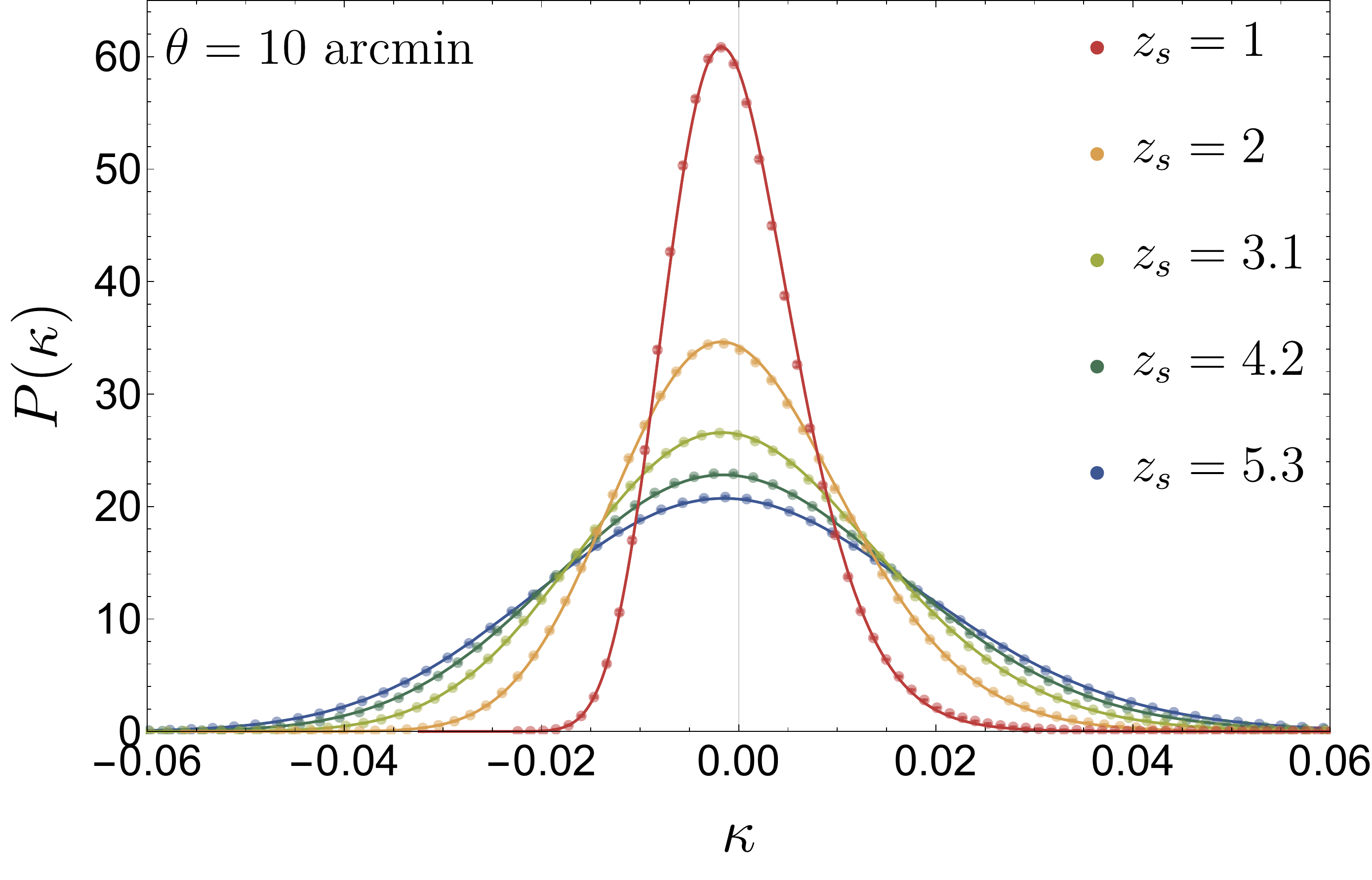}\\
    \includegraphics[width=\columnwidth]{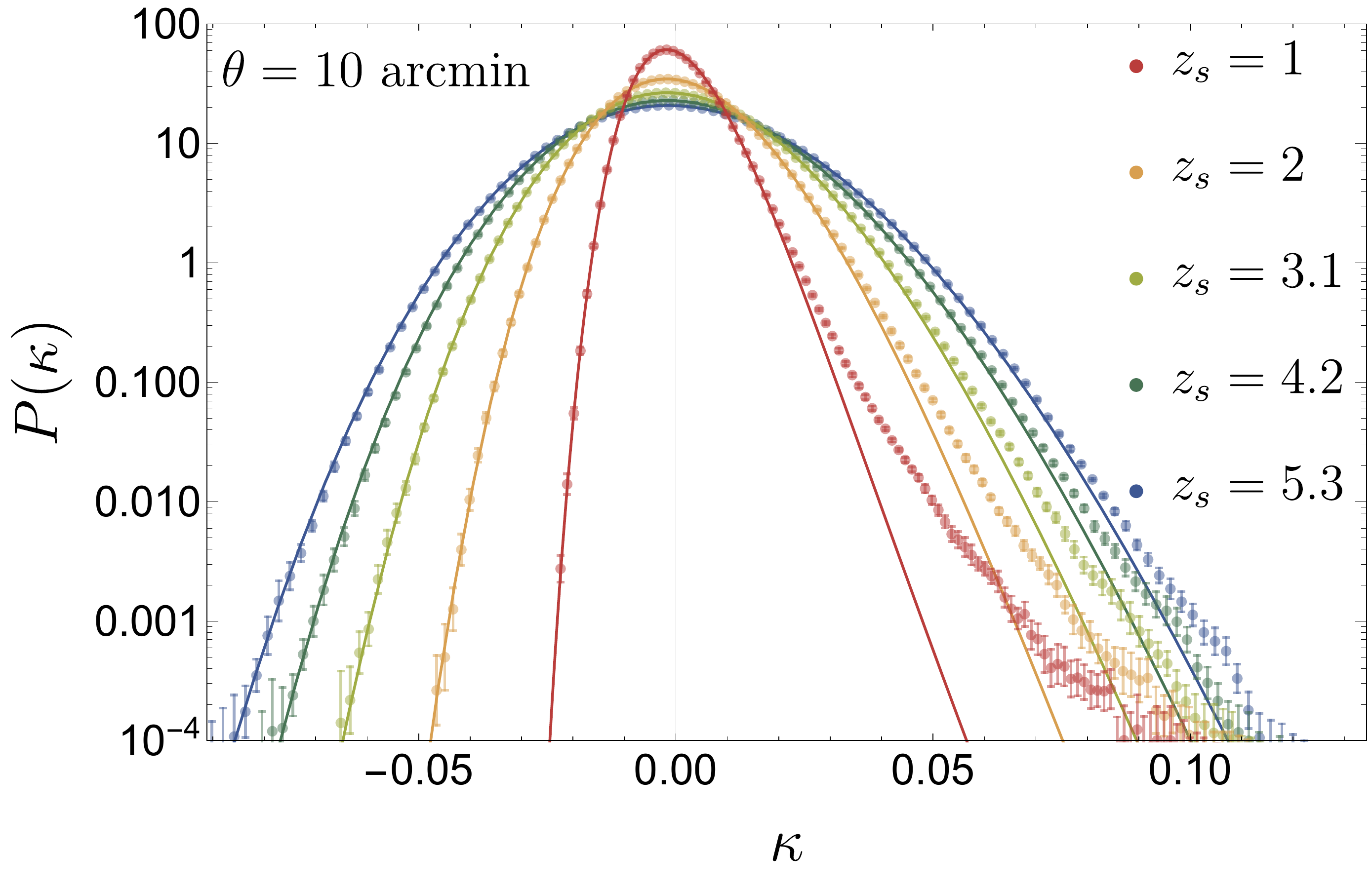}\\
    \includegraphics[width=\columnwidth]{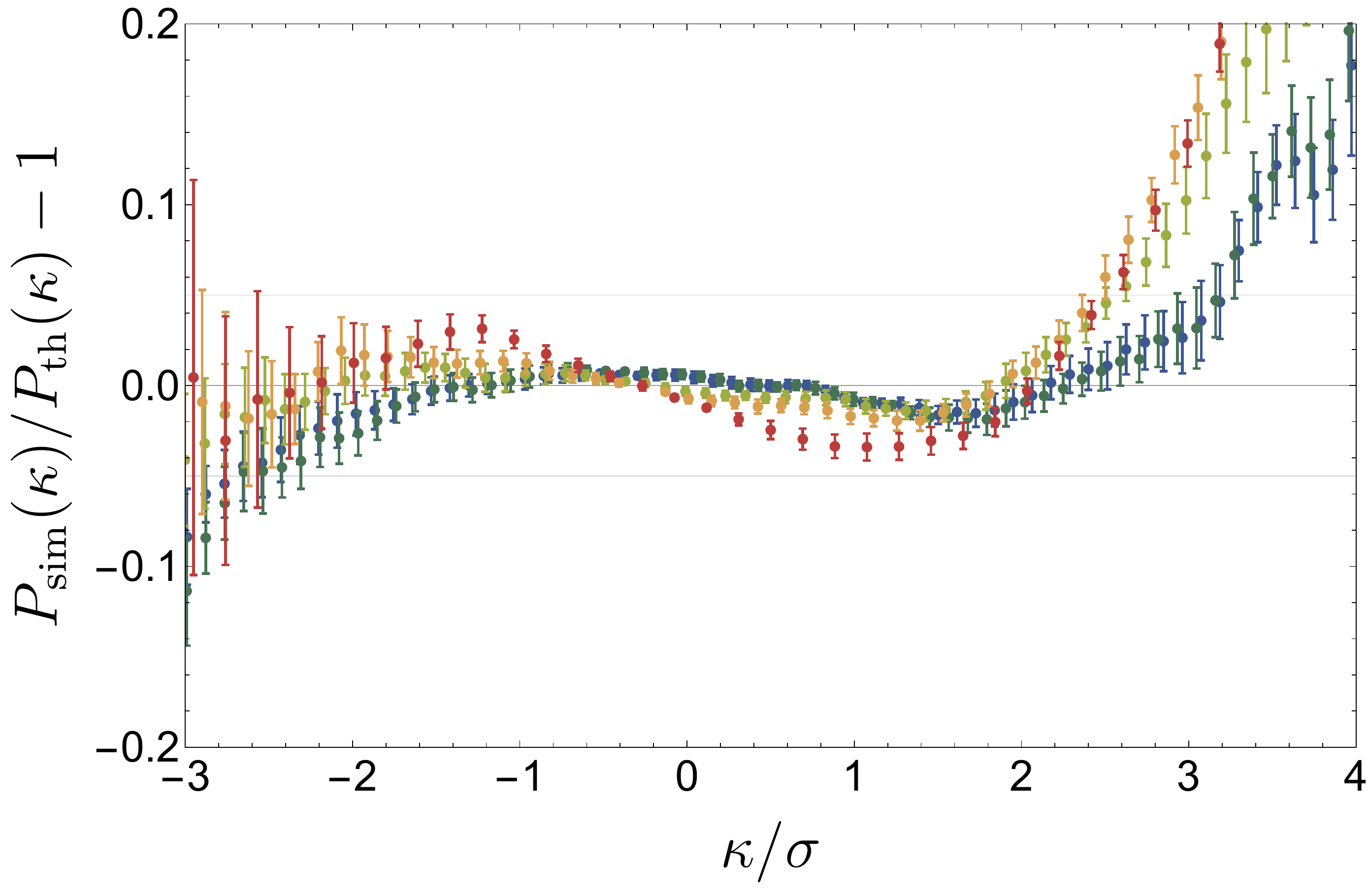}
    \caption{Same as figure~\ref{redshift_pdf} but using the analytical shortcut predictions given by equation~(\ref{PDFanalapprox1}).}
    \label{shortcut_redshift_pdf}
\end{figure}

\begin{figure}
    \centering
    \includegraphics[width=\columnwidth]{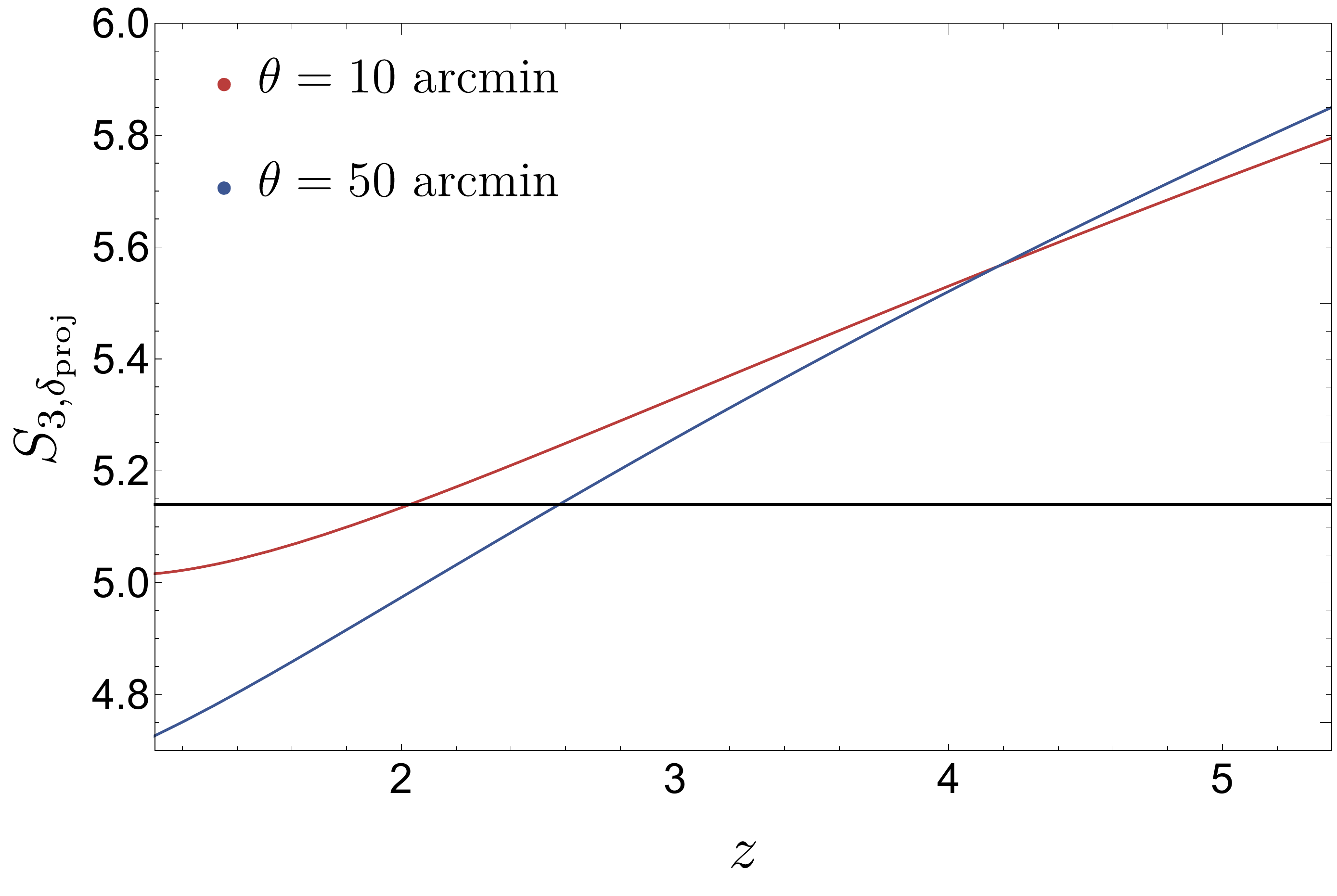}
    \caption{Skewness of the projected density for different source redshifts and opening angles as computed in solid lines with large deviation theory, in dashed line with analytical shortcut and the thick solid line gives the unsmoothed skewness for a cylinder. By construction the analytical shortcut tends to the thick horizontal line.}
    \label{S3proj}
\end{figure}

The PDF of the projected density is determined from an inverse Laplace transformation~\eqref{laplacePDF} of its (numerically determined) CGF.
In practice, it can be useful to approximate this complex integral using an analytical saddle-point technique, as was successfully done for 3D cosmic densities in \cite{saddle}.
The saddle-point approximation relies on identifying the main contribution to the exponent of the integral determining the projected density PDF, which drives the exponential decay of that PDF. 
Similarly to the Legendre transformation relating the CGF to the rate function from equation~\eqref{Legendre}, the projected rate function can be obtained from the projected CGF
\begin{equation}
    \psi_{\rm proj}(\hat\delta_{\rm proj}) = \sup_y \, [y\hat\delta_{\rm proj}-\phi_{proj}(y)]\,.
\end{equation}
Using this relationship, we found empirically, that for source redshifts $z_s$ and angles $\theta$ considered here, the rate function of projected densities is well approximated by
\begin{equation}
\psi_{\rm proj}^{\rm approx}(\hat\delta_{\rm proj})=\frac{\tau_{\rm SC}^2(1+\hat\delta_{\rm proj})}{2\sigma_{\rm proj,nl}^2(\theta)}\,,
    \label{rateanalapprox}
\end{equation}
where $\sigma_{\rm proj}^2(\theta)$ is the non-linear variance of the projected density and $\tau_{\rm SC}$ the inverse cylindrical collapse mapping from equation~\eqref{collapse}. This suggests that the projected density PDF for the weak-lensing convergence can be approximated by assuming a Gaussian distribution for the linear density contrast $\tau$ and a nonlinear mapping to the projected density contrast $\delta_{\rm proj}$ given by the cylindrical collapse model. In this approximation, the projected density PDF is given by
\begin{align}
\hat{\mathcal P}_{\rm proj}^{\rm approx}(\hat\delta_{\rm proj})&=\frac{\langle\zeta(\tau)\rangle}{\langle1\rangle^2} \mathcal P^{\rm approx}_{\rm proj}\left(\frac{\langle\zeta(\tau)\rangle}{\langle1\rangle}\hat\delta_{\rm proj}\right)\,,\\
\mathcal P_{\rm proj}^{\rm approx}(\hat\delta_{\rm proj})\!&=\!\frac{\tau_{\rm SC}'(1+\hat\delta_{\rm proj})}{\sqrt{2\pi}\sigma_{\rm proj,nl}(\theta)}\exp\!{\left[\!-\!\frac{\tau^2_{\rm SC}(1+\hat\delta_{\rm proj})}{2\sigma_{\rm proj,nl}^2(\theta)}\!\right]}\!,
    \label{PDFanalapprox1}
\end{align}
where we used the short-hand notation $\langle\zeta(\tau)\rangle=\int ( 1+\delta_{\rm proj}) \mathcal P(\delta_{\rm proj})\, d\delta_{\rm proj}$ for the mean projected density induced by a zero mean initial density mapped by cylindrical collapse and $\langle1\rangle$ for $\int \mathcal P(\delta_{\rm proj})\, d\delta_{\rm proj}$. Note that this last term is not stricly one -- although often very close to 1 for the variances considered -- because the mapping from $\tau$ to $\hat\delta_{\rm proj}$ is not one-to-one ($\zeta$ tends to infinity for a finite value of $\tau=\nu$).

figure~\ref{shortcut_redshift_pdf} shows that this simplistic approximation  works remarkably well in the 2$\sigma$ region around the peak of the PDF, being well within 5 per cent error for all source redshifts considered at the smallest angular size of $\theta=10$~arcmin. The success of this approximation can be explained when looking at the skewness for projected density for a wide range of opening angles and source redshifts as done in figure~\ref{S3proj}. 
Indeed this approximation is equivalent to considering densities in a long cylinder with a large deviation theory approach but without taking into account the effects of a smoothing scale -- or equivalently using a top-hat window function -- and thus directly implementing all the tree-order unsmoothed cumulants for the variable $\tau$. For a cylindrical collapse the unsmoothed skewness $S_3^{2D} = 36/7 \approx 5.14$ is somehow a good approximation of the real predicted values. Note that on the one hand smoothing diminishes the values of the skewness and on the other hand non-linearities, not included in the tree-order prediction, tend to increase that skewness and consequently the unsmoothed version of a unique cylinder works incidentally best as opposed to introducing scaling terms in equation~(\ref{rateanalapprox}). 

\begin{figure}
    \centering
    \includegraphics[width=\columnwidth]{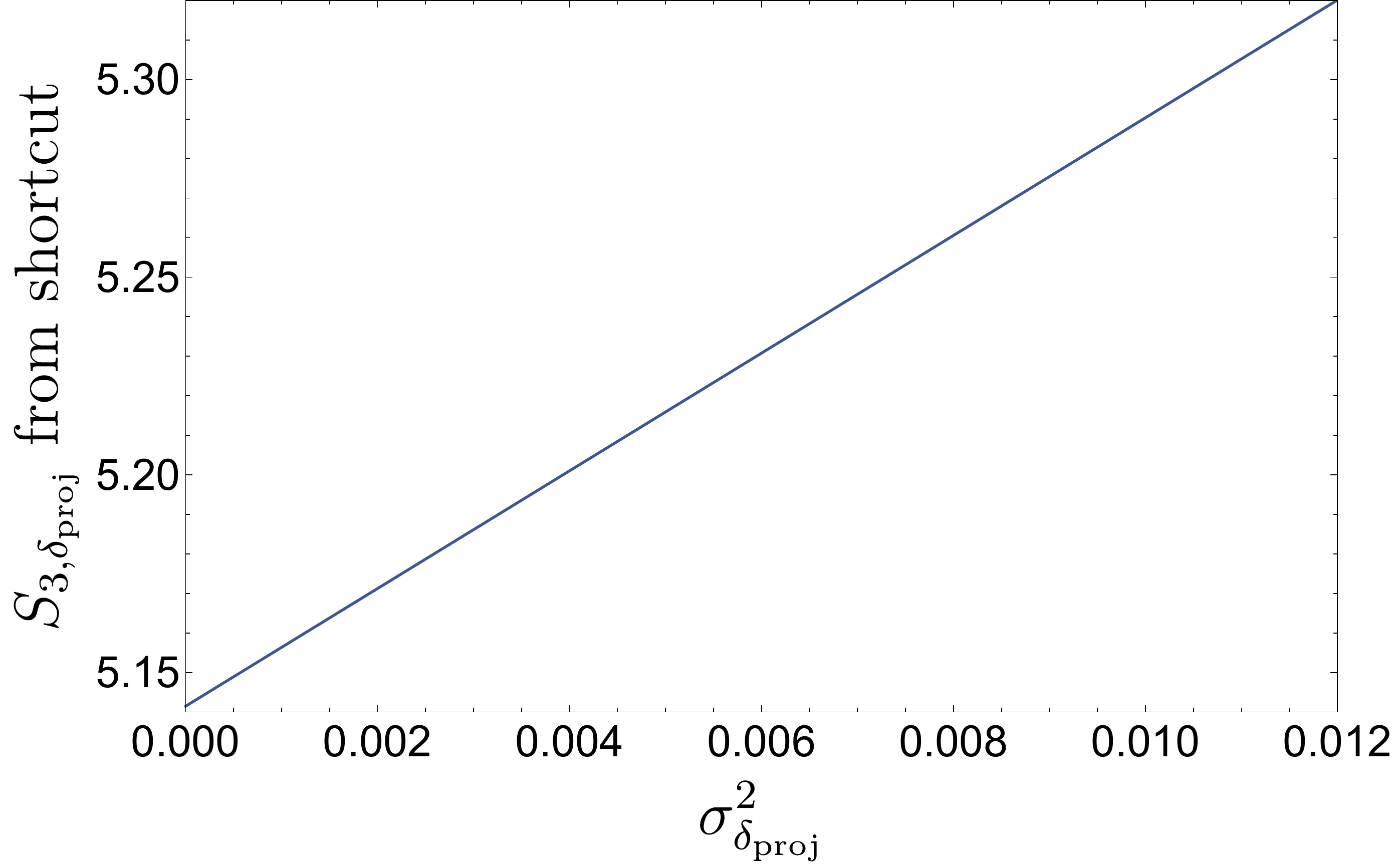}
    \caption{Skewness of the projected density as computed in the analytical approximation in equation~(\ref{PDFanalapprox1}). The linear dependence with the projected variance is a direct consequence of the re-mapping from $\tau$ to $\delta_{\rm proj}.$}
    \label{dependence}
\end{figure}

Now note that, technically, relying on the saddle-point approximation for a nonlinearly transformed variable amounts to making a different hypothesis about the robustness of reduced cumulants $S_p$ with respect to changes in the variance. Assuming constant $S_p$ for the variable $\tau$ as with any new variable is different from assuming constant $S_p$ for the projected densities, the limit for zero variance is the same but the extrapolation to non-zero value of the variance is different (each cumulant differs in the two approaches not at tree order but by their higher order corrections). We refer to \cite{saddle} for a thorough analysis of this effect but still mention the small linear $\sigma^2$ dependence of the skewness going from $\tau$ to $\delta_{\rm proj}$ as is illustrated in figure~\ref{dependence}.


\section{CGF for N bins of nulled convergence}
\label{app:phiN}

There are two ways that we found to derive equation~(\ref{eq:phiN}). The first is purely mathematical and applies to any sets of random variables only correlated to their neighbours, the second one relies more on physical intuition. The two approaches are described below.

\subsection{From the definition of cumulant generating function}

Let us consider N bins of nulled convergence $\tilde \kappa_i$ that are therefore only correlated to their nearest neighbouring bins.
The joint CGF can be written as a generalisation of (\ref{log})
\begin{equation}
\begin{split}
     \phi_{1\!,\!\cdots\!,\! N}(y_1\!,\!\cdots\!,\!y_N)&=\!\! \!\!\!\!\!\!\!\sum_{p_1,\cdots,p_N\leq 0}\! \!\frac{y_1^{p_1}\cdots y_N^{p_N}}{p_1!\cdots p_N!}\left\langle\tilde\kappa_1^{p_1}\cdots\tilde\kappa_N^{p_N}\right\rangle_c\!\!\!-\!\!1\\
   &= \! \!\!\!\!\!\!\!\sum_{p_2,\cdots,p_N\leq 0}\!\!\frac{y_2^{p_2}\cdots y_N^{p_N}}{p_2!\cdots p_N!}\left\langle\tilde\kappa_2^{p_2}\cdots\tilde\kappa_N^{p_N}\right\rangle_c\!\!\!-\!\!1 \\
   &+\!\!\!\!\!\!\!
   \sum_{p_1>1,p_2\leq 0}\!\!\frac{y_1^{p_1} y_2^{p_2}}{p_1!p_2!}\left\langle\tilde\kappa_1^{p_1}\tilde\kappa_2^{p_2}\right\rangle_c
\end{split}
\end{equation}
because $\tilde \kappa_1$ only correlates with $\tilde \kappa_2$ (all the other cumulants are zero).
From there, we get the recursion relation
\begin{multline}
 \phi_{1,\cdots ,N}(y_1,\cdots,y_N)=\phi_{2,\cdots, N}(y_2,\cdots,y_N)\\+\phi_{1,2}(y_1,y_2)-\phi_{2}(y_2)
\end{multline}
which once applied recursively eventually leads to
\begin{equation}
\label{dem}
\phi_{1,\cdots ,N}(y_1,\!\cdots\!,y_N)\!=\!\sum_{i=1}^{N-1}\phi_{i,i+1}(y_i,y_{i+1})
\!-\!\!\sum_{i=2}^{N-1}\!\phi_{i}(y_i).
\end{equation}

\subsection{A more physical approach}

Once one generalises equation~(\ref{eq:phi2}) to N bins of nulled convergence, the integral can be decomposed in subparts where only two $\omega_i(R)$ are non-zero so that 
\begin{equation}
\label{1}
    \phi_{1,\cdots ,N}(y_1,\cdots,y_N)=\sum_{i=1}^{n-1} \hat{\phi}_{i}\left(y_{i}, y_{i+1}\right)+\hat{\phi}\left(y_{n}\right)
\end{equation}
where 
\begin{equation}
\label{2}
    \hat{\phi}_{i}\left(y_{i}, y_{i+1}\right)\!=\!\int_{R_{i-1}}^{R_{i}} \! \! {\rm d} R \,\phi_{\rm{cyl}}\left(\omega_{i}(R) y_{i}+\omega_{i+1}(R) y_{i+1} , R\right)
\end{equation}
and 
\begin{equation}
\label{3}
    \hat{\phi}_{n}\left(y_{n}\right)=\int_{R_{n-1}}^{R_{n}} \mathrm{d} R\, \phi_{\mathrm{cyl}}\left(\omega_{n}(R) y_{n} , R\right).
\end{equation}
Now let us express the $\hat{\phi}$ in terms of $\phi$. From equation~(\ref{1}) we can write
\begin{multline}
\label{5}
    \phi_{i,i+1}(y_i,y_{i+1})= \hat{\phi}_{i,i+1}(y_i,y_{i+1})+ \hat{\phi}_{i+1}(y_{i+1}) \\+ \int_{R_0}^{R_{i-1}}\!\!\!\!\!\!\!\!\! {\rm d}R\, \phi_{\rm cyl}(\omega_i(R)y_i) .
\end{multline}
Eventually plugging equation~(\ref{5}) into (\ref{1}) yields the same result as shown in equation~(\ref{dem}) with the previous approach.

\label{lastpage}
\end{document}